\documentclass[reprint,superscriptaddress,amsmath,amssymb,aps,pre]{revtex4-1}

\usepackage{amsmath}
\usepackage{mathtools}
\usepackage{graphicx}
\usepackage{dcolumn}
\usepackage{bm}
\usepackage{color}
\usepackage{bbold}
\usepackage{soul}
\usepackage{subfigure}
\usepackage{braket}
\usepackage[utf8]{inputenc}
\newcommand{\newc}{\newcommand}
\newc{\beq}{\begin{equation}}
\newc{\eeq}{\end{equation}}
\newc{\kt}{\rangle}
\newc{\br}{\langle}
\newc{\beqa}{\begin{eqnarray}}
\newc{\eeqa}{\end{eqnarray}}
\newc{\longra}{\longrightarrow}

\makeatletter
\let\Hy@backout\@gobble
\makeatother

\begin{document}

\title{Exact decomposition of homoclinic orbit actions in chaotic systems: Information reduction}

\author{Jizhou Li}
\affiliation{Department of Physics and Astronomy, Washington State University, Pullman, Washington 99164-2814, USA}
\author{Steven Tomsovic}
\affiliation{Department of Physics and Astronomy, Washington State University, Pullman, Washington 99164-2814, USA}

\date{\today}

\begin{abstract}
Homoclinic and heteroclinic orbits provide a skeleton of the full dynamics of a chaotic dynamical system and are the foundation of semiclassical sums for quantum wave packet, coherent state, and transport quantities. Here, the homoclinic orbits are organized according to the complexity of their phase-space excursions, and exact relations are derived expressing the relative classical actions of complicated orbits as linear combinations of those with simpler excursions plus phase-space cell areas bounded by stable and unstable manifolds. The total number of homoclinic orbits increases exponentially with excursion complexity, and the corresponding cell areas decrease exponentially in size as well. With the specification of a desired precision, the exponentially proliferating set of homoclinic orbit actions is expressible by a slower-than-exponentially increasing set of cell areas, which may present a means for developing greatly simplified semiclassical formulas. 
\end{abstract}


\maketitle

\section{Introduction}
\label{Introduction}

Specific sets of rare classically chaotic orbits are central ingredients for sum rules in classical and quantum systems~\cite{ChaosBook}.  Classical sum rules over unstable periodic orbits describe various entropies, Lyapunov exponents, escape rates, and the uniformity principle~\cite{So07}.  Gutzwiller's trace formula~\cite{Gutzwiller71} for quantum spectra is over unstable periodic orbits, closed orbit theory of atomic spectra~\cite{Du88a,Du88b} gives the absorption spectrum close to the ionization threshold of atoms placed in magnetic fields, and heteroclinic (homoclinic) orbits arising from intersections between the stable and unstable manifolds of different (same) hyperbolic trajectories describe quantum transport between initial and final localized wave packets~\cite{Tomsovic93}.  

It is often the case that the nonlinear flows of phase-space densities are completely captured by the stable and unstable manifolds of one or just a few short periodic orbits, hence also by the homoclinic and heteroclinic orbits that arise from intersections between these manifolds. These orbits can thus play the important role of providing a ``skeleton" of transport for the system. It is not a unique choice, but each choice provides the same information.  For example, an unstable periodic orbit gives rise to an infinity of homoclinic orbits, but it is also true that families of periodic orbits of arbitrary lengths accumulate on some point along every homoclinic orbit~\cite{Birkhoff27,Moser56,Silva87}, and the periodic orbit points can be viewed as being topologically forced by the homoclinic point on which a particular sequence accumulates~\cite{Ozorio89,Li17a}. 

Two problems are immediately apparent.  The first is the particular importance of having accurate evaluations of classical actions because these quantities are divided by $\hbar$ and play the role of phase factors for the interferences between terms, and their remainder after taking the modulus with respect to $2\pi$ must be $\ll 2\pi$.  A straightforward calculation would proceed with the numerical construction of the actions, which would be plagued by the sensitive dependence on initial conditions for long orbits.  An alternative method has been developed by the authors~\cite{Li17a,Li18}.  That scheme converts the calculation of unstable periodic orbit actions into the evaluation of homoclinic orbit action differences.  The homoclinic orbit actions can then be stably obtained as phase-space areas via the MacKay-Meiss-Percival principle~\cite{MacKay84a,Meiss92}, or directly from the stable constructions of homoclinic orbits \cite{Silva87,Doedel89,Beyn90,Moore95b,Li17}. Beside the action functions, another quantity of the periodic orbits, namely their stability exponents, also play the crucial role of the prefactor in the Gutzwiller's trace formula. In Sec.~\ref{Fast and slow scaling relations}, a new relation (Eq.~\eqref{eq:General scaling}) is introduced that determines the stability exponents of periodic orbits from ratios between areas bounded by stable and unstable manifolds, or equivalently, distribution of homoclinic points on the manifolds. Therefore, both the action and the stability exponent of periodic orbits can be calculated from the knowledge of homoclinic orbits, without the numerical construction of periodic orbits themselves. 

The second problem is more fundamental. Namely, the total number of periodic orbits increases exponentially with increasing period and for the homoclinic orbits with increasingly complicated excursions.  This is a reflection of the non-vanishing rate of information entropy production associated with chaotic dynamics, which in an algorithmic complexity sense has been proven equivalent to the Kolmogorov-Sinai entropy~\cite{Brudno78,Alekseev81,Kolmogorov58,Kolmogorov59,Sinai59}, and hence the Lyapunov exponents via Pesin's theorem~\cite{Pesin77,Gaspard90}.  On the other hand, entropies introduced for quantum systems~\cite{Connes87,Alicki94,Lindblad88} vanish due to the non-zero size of $\hbar$, if these systems are isolated, bounded, and not undergoing a measurement process.  This gives one the intuitive notion and hope that there must be a means to escape the exponential proliferation problem of semiclassical sum rules.  

Therefore, a scheme to replace classical and semiclassical sum rules that from the outset clearly have vanishing information entropy content is highly desirable~\cite{Cvitanovic92}.  The pseudo-orbits of the cycle expansion~\cite{Cvitanovic88,Cvitanovic89,ChaosBook}, the primitive orbits of Bogomolny's surface of section method~\cite{Bogomolny92}, and multiplicative semiclassical propagator~\cite{Kaplan98b}   were steps in this direction.  Building on the methods of~\cite{Li17a}, we develop exact relations for the decomposition of homoclinic orbit relative actions with complicated excursions in terms of multiples of the two primary ones and sets of phase-space areas. Accounting for an error tolerance determined by $\hbar$ reduces the exponentially proliferating set of homoclinic orbit actions to combinations of an input set (i.e., phase-space cell areas) that increase more slowly than exponentially (i.e., algebraically) with time, thus resolving the conflict between the entropies of classical and quantum chaotic systems, and directly linking $\hbar$ to the boundary between surviving and non-surviving information in quantum mechanics. 

This paper is organized as follows. Sec.~\ref{Basic concepts} introduces the basic concept of homoclinic tangle. Sec.~\ref{Relative actions} introduces the relative action functions between homoclinic orbit pairs. Sec.~\ref{Hierarchic structure of homoclinic points} reviews the concepts of winding number and transition time of homoclinic orbits, and introduces a hierarchical ordering of homoclinic points in terms of their winding numbers.  Organizing the homoclinic points using the winding numbers, we identify an asymptotic scaling relation between families of homoclinic points, which puts strong constraints on the distribution of homoclinic points along the manifolds. Sec.~\ref{Homoclinic action formulae} gives two central results of this paper. The first one (Sec.~\ref{Exact decomposition}) is an exact formula for the complete expansion of homoclinic orbit actions in terms of primary homoclinic orbits and phase-space cell areas bounded by the manifolds. The second one  (Sec.~\ref{Information reduction}) is the demonstration that a coarse-grained scale, determined by $\hbar$, allows for an approximation that eliminates exponentially small areas from the complete expansion, which gives an approximate action expansion that requires a subset of cell areas growing sub-exponentially.     

\section{Basic concepts}
\label{Basic concepts}

\begin{figure}[ht]
\centering
{\includegraphics[width=6.5cm]{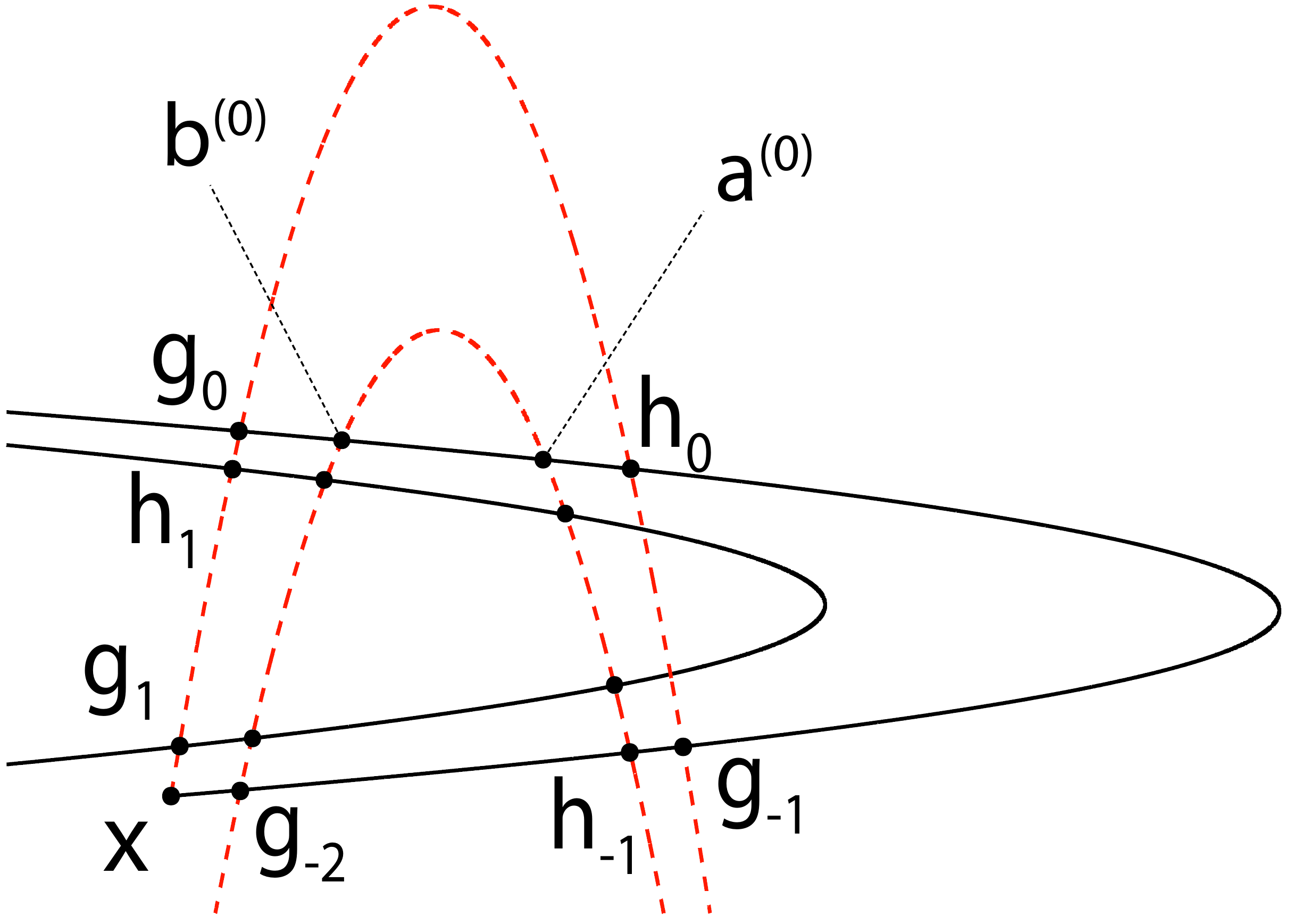}}
 \caption{(Color online) Horseshoe-shaped homoclinic tangle formed by $S(x)$ (red dashed curve) and $U(x)$ (black solid curve), with two primary homoclinic orbits $\lbrace h_0 \rbrace$ and $\lbrace g_0 \rbrace$. Notice that the $U(x)$ segments beyond $g_0$ and $h_1$ are simply connected and omitted from the figure for clarity, and the same for the $S(x)$ segments beyond $g_{-1}$ and $h_{-1}$.   }
\label{fig:Homoclinic_Tangle}
\end{figure}

Consider a two-degree-of-freedom autonomous Hamiltonian system. With energy conservation and applying the standard Poincar\'{e} surface of section technique~\cite{Poincare99}, the continuous flow leads to a discrete area-preserving map $M$ on the two-dimensional phase space $(q,p)$.  Assume the existence of a hyperbolic fixed point $x=(q_x,p_x)$ under $M$: $M(x)=x$.  Associated with it are the one-dimensional stable ($S(x)$) and unstable ($U(x)$) manifolds, which are the collections of phase-space points that approach $x$ under successive forward and inverse iterations of $M$, respectively. Typically, $S(x)$ and $U(x)$ intersect infinitely many times and form a complicated pattern named $\mathit{homoclinic}$ $\mathit{tangle}$ \cite{Poincare99,Easton86,Rom-Kedar90}, as partially illustrated in Fig.~\ref{fig:Homoclinic_Tangle}. This figure demonstrates the simplest but generic type of homoclinic tangle, a ``Smale horseshoe" \cite{Smale63,Smale80}, which results from the exponential stretching along $U(x)$, compressing along $S(x)$, and eventually a binary folding to create mixing dynamics.   
Refer to App.~\ref{Symbolic dynamics} for a detailed introduction of the Smale horseshoe. The area-preserving H\'{e}non map \cite{Henon76} shown by Eq.~\eqref{eq:Henon map} with parameter $a=10$ is used to generate this figure, along with all forthcoming numerical implementations in this article.

The main objects of study in this article, the $\mathit{homoclinic}$ $\mathit{orbits}$, arise from intersections between $S(x)$ and $U(x)$. These are the orbits asymptotic to $x$ under both forward and inverse iterations of $M$. For instance, the point $h_0$ in Fig.~\ref{fig:Homoclinic_Tangle} is a homoclinic intersection between the manifolds, and its orbit $\lbrace \cdots, h_{-1},h_0,h_1,\cdots \rbrace$ approaches $x$ under both forward and inverse iterations. In spite of the infinity of homoclinic orbits arising from the pattern in Fig.~\ref{fig:Homoclinic_Tangle}, for the most part only two of them, $\lbrace h_0 \rbrace$ and $\lbrace g_0 \rbrace$, have a fundamental importance.  They have the special property that the segments $U[x,h_0]$ and $S[h_0,x]$ only intersect at $x$ and $h_0$. Consequently, the loop $US[x,h_0]\equiv U[x,h_0]+S[h_0,x]$ is a single loop, so the orbit $\lbrace h_0 \rbrace$ ``circles" around the loop only once. The same is true for $\lbrace g_0 \rbrace$. A ``$\mathit{winding}$ $\mathit{number}$" of $1$ can thus be associated to both $\lbrace h_0 \rbrace$ and $\lbrace g_0 \rbrace$, and they are commonly referred as the $\mathit{primary}$ $\mathit{homoclinic}$ $\mathit{orbits}$. All other orbits have winding numbers greater than $1$. To be shown later, their classical actions can be built by the two primary orbit actions and certain sets of phase-space areas bounded by $S(x)$ and $U(x)$. More details about the winding numbers will be introduced in Sec.~\ref{Hierarchic structure of homoclinic points}.  

The topological structures of homoclinic tangles are well understood nowadays, and they provide a foundation for our analysis on the homoclinic orbit actions in later sections. With the help of certain generating Markov partitions identified from the homoclinic tangle ($V_0$ and $V_1$ in Fig.~\ref{fig:Horseshoe} in Appendix~\ref{Symbolic dynamics}), the non-wandering orbits of the system can be put into a one-to-one correspondence with bi-infinite strings of integers, i.e., the $\mathit{symbolic}$ $\mathit{dynamics}$ \cite{Hadamard1898,Birkhoff27a,Birkhoff35,Morse38} of chaotic systems. For example, the hyperbolic fixed-point $x$ in Fig.~\ref{fig:Homoclinic_Tangle} is labled by the bi-infinite string $\overline{0}.\overline{0}$, where the overhead bar indicates infinite repetitions of the symoblic string underneath it, and the decimal point indicate the location of the current iteration. This symbolic code reflects the fact that $x$ stays in $V_0$ under all forward and inverse iterations. The primary homoclinic points $h_0$ and $g_0$ are labled by $h_0 \Rightarrow \overline{0}1.1\overline{0}$ and $g_0 \Rightarrow \overline{0}1.\overline{0}$, respectively. Other than the points on $\lbrace h_0 \rbrace$ and $\lbrace g_0 \rbrace$, all homoclinic points $ a $ of $x$ must have a symbolic string of the form
\begin{equation}\label{eq:Homoclinic point symbolic string general form main}
a \Rightarrow \overline{0} 1 s_{-m}\cdots s_{-1}.s_0 s_1 \cdots s_n 1 \overline{0} = \overline{0} 1 \tilde{s}_{-} . \tilde{s}_{+} 1 \overline{0} 
\end{equation}  
along with all possible shifts of the decimal point, where each digit $s_i \in {0,1}$ ($-m \leq i \leq n$). The substrings $\tilde{s}_{-}=s_{-m}\cdots s_{-1}$ and $\tilde{s}_{+}=s_0 s_1 \cdots s_n$. The $\overline{0}$ on both ends means the orbit approaches the fixed point asymptotically. The orbit $\lbrace a \rbrace$ can then be represented by the same symbolic string:
\begin{equation}\label{eq:Homoclinic orbit symbolic string general form main}
\lbrace a \rbrace \Rightarrow \overline{0} 1 \tilde{s}_{-}\tilde{s}_{+} 1 \overline{0}
\end{equation}  
with the decimal point removed, as compared to Eq.~\eqref{eq:Homoclinic point symbolic string general form main}. The finite symbolic segment ``$1 \tilde{s}_{-}\tilde{s}_{+} 1$" is often referred to as the \textit{core} of the symbolic code of $ a $, with its length referred to as the \textit{core length}. 

In the horseshoe map, besides the hyperbolic fixed point $x$, there is another hyperbolic fixed point with reflection, denoted by $x^{\prime}$. This fixed point has symbolic code $x^{\prime} \Rightarrow \overline{1}.\overline{1}$, i.e., it stays in $V_1$ under all forward and inverse iterations. Denote the stability exponents of $x$ and $x^{\prime}$ by $\mu_{0}$ and $\mu_{1}$, respectively, i.e., the subscripts indicate the symbolic code. These two exponents are of special interest later. 

We skip further detailed introduction here and refer the reader to excellent references such as \cite{Easton86,Rom-Kedar90,Wiggins92}, and to App.~\ref{Homoclinic tangle}, \ref{Symbolic dynamics} for the concepts of trellises, symbolic dynamics, and for the definitions of notations adopted throughout this article. The symbolic dynamics will be the main language adapted to identify homoclinic orbits in this study. However, although well-resolved \cite{Sterling99}, the assignment of symbolic codes to homoclinic points is still a non-trivial task in general. The readers are referred to App.~\ref{Systematic assignment of symbolic codes} for a detailed assignment scheme. In the forthcoming contents, the symbolic codes of all homoclinic points is assumed known.  

\section{Relative actions}
\label{Relative actions}

The classical actions of homoclinic orbits are divergent as they come from the infinite sum over the generating functions associated with each iteration along the orbit.  Hence, it is necessary to consider relative actions, which are finite.  For any phase-space point $z_n=(q_n,p_n)$ and its image $M(z_n)=z_{n+1}=(q_{n+1},p_{n+1})$, the mapping $M$ can be viewed as a canonical transformation that maps $z_n$ to $z_{n+1}$ while preserving the symplectic area, therefore a $\mathit{generating}$ ($\mathit{action}$) function $F(q_n,q_{n+1})$ can be associated with this transformation such that \cite{MacKay84a,Meiss92}
\begin{equation}\label{eq:Definition generating function}
\begin{split}
&p_n = - \partial F / \partial q_n \\
&p_{n+1} = \partial F / \partial q_{n+1}\ .
\end{split}
\end{equation}
Despite the fact that $F$ is a function of $q_{n}$ and $q_{n+1}$, it is convenient to denote it as $F(z_n,z_{n+1})$. This should cause no confusion as long as it is kept in mind that it is the $q$ variables of $z_n$ and $z_{n+1}$ that go into the expression of $F$. A special example is the generating function of the fixed point, $F(x,x)$, that maps $x$ into itself under one iteration. For homoclinic orbits $\lbrace h_0 \rbrace$, the $\mathit{classical}$ $\mathit{action}$ is the sum of generating functions between each step
\begin{equation}\label{eq:Definition homoclinic orbit action}
{\cal F}_{\lbrace h_0 \rbrace} \equiv \lim_{N \to \infty} \sum_{n=-N}^{N-1} F(h_n,h_{n+1})
\end{equation} 
However, according to the MacKay-Meiss-Percival action principle \cite{MacKay84a,Meiss92}, convergent relative actions can be obtained by comparing the classical actions of a homoclinic orbit pair: 
\begin{equation}\label{eq:Area-action homoclinic pair}
\begin{split}
& \Delta {\cal F}_{\lbrace h^{\prime}_0 \rbrace \lbrace h_0 \rbrace} \equiv \lim_{N \to \infty} \sum_{n=-N}^{N-1} \left[ F(h^{\prime}_n, h^{\prime}_{n+1}) - F(h_{n}, h_{n+1}) \right] \\
&= \int\limits_{ U[h_0, h^{\prime}_0] } p\mathrm{d}q +\int\limits_{ S[h^{\prime}_0, h_0] } p\mathrm{d}q = {\cal A}^{\circ}_{US[h_0,h^{\prime}_0]}
\end{split}
\end{equation}
where the $\circ$ superscript in the last term indicates that the area evaluated is interior to a path that forms a closed loop, and the subscript indicates the path: $US[h_0,h^{\prime}_0] = U[h_0,h^{\prime}_0]+S[h^{\prime}_0,h_0]$. Such an action difference is referred to as the $\mathit{relative}$ $\mathit{action}$ between $\lbrace h^{\prime}_0 \rbrace$ and $\lbrace h_0 \rbrace$. A special case of interest is the relative action between a homoclinic orbit $\lbrace h_0 \rbrace$ and the fixed point itself $\lbrace x \rbrace$:
\begin{equation}\label{eq:Area-action homoclinic orbit and fixed point orbit}
\begin{split}
\Delta {\cal F}_{\lbrace h_0 \rbrace \lbrace x \rbrace}& =\lim_{N \to \infty} \sum_{n=-N}^{N-1} \left[ F(h_n, h_{n+1}) - F(x,x) \right] \\
&= {\cal A}^{\circ}_{US[x,h_0]}
\end{split}
\end{equation}
which gives the action of $\lbrace h_0 \rbrace$ relative to the fixed point orbit action, and is simply referred to as the relative action of $\lbrace h_0 \rbrace$. An equivalent approach, which makes use of the information about the stable and unstable manifolds of hyperbolic fixed points to obtain convergent expressions of homoclinic and heteroclinic orbit actions as algebraic areas evaluated under these manifolds, were given by Tabacman in \cite{Tabacman95}. There, it was shown that the homoclinic and heteroclinic orbits can be calculated as critical values of certain action functions constructed from the generating function of the system and the local stable and unstable manifolds near the fixed points. However, our goal is to identify hidden relations between the homoclinic orbit actions without numerical constructions of the orbits themselves. As shown ahead, this requires information about the global stable and unstable manifolds. 

A generalization of Eq.~\eqref{eq:Area-action homoclinic orbit and fixed point orbit} applies to four arbitrary homoclinic orbits of $x$, namely $\lbrace a_0 \rbrace$, $\lbrace b_0 \rbrace$, $\lbrace c_0 \rbrace$, and $\lbrace d_0 \rbrace$.  Expressing the relative actions of each of them using Eq.~\eqref{eq:Area-action homoclinic orbit and fixed point orbit}, and calculating the action difference between the following two pairs of orbits gives
\begin{equation}\label{eq:Area-action two homoclinic pairs}
\begin{split}
&( \Delta {\cal F}_{\lbrace a_0 \rbrace \lbrace x \rbrace} - \Delta {\cal F}_{\lbrace b_0 \rbrace \lbrace x \rbrace} ) - ( \Delta {\cal F}_{\lbrace c_0 \rbrace \lbrace x \rbrace} - \Delta {\cal F}_{\lbrace d_0 \rbrace \lbrace x \rbrace} ) \\
& = ( {\cal A}^{\circ}_{US[x,a_0]} -  {\cal A}^{\circ}_{US[x,b_0]} ) -  ( {\cal A}^{\circ}_{US[x,c_0]} -  {\cal A}^{\circ}_{US[x,d_0]} )\\
& = {\cal A}^{\circ}_{SUSU[ a_0, c_0, d_0, b_0 ]}
\end{split}
\end{equation}
where 
\begin{equation}\label{eq:Parallelogram area definition}
\begin{split}
& {\cal A}^{\circ}_{SUSU[ a_0, c_0, d_0, b_0 ]} \equiv \int\limits_{ S[a_0, c_0] } p\mathrm{d}q + \int\limits_{ U[c_0, d_0] } p\mathrm{d}q \\
& \quad + \int\limits_{ S[d_0, b_0] } p\mathrm{d}q + \int\limits_{ U[b_0, a_0] } p\mathrm{d}q
\end{split}
\end{equation}
is the curvy parallelogram area bounded by alternating segments of $S(x)$ and $U(x)$ connecting the four homoclinic points.

\section{Hierarchical structure of homoclinic points}
\label{Hierarchic structure of homoclinic points}

\subsection{Winding numbers and transit times}
\label{Rotary and transition numbers}

The infinite set of homoclinic orbits can be put into a hierarchical structure, organized using a winding number~\cite{Hockett86,Bevilaqua00} that characterizes the complexity of phase-space excursion of each individual orbit. The winding number of a homoclinic point $h$ is defined to be the number of single loops (i.e., loops with no self-intersection) that the loop $US[x,h]$ can be decomposed into~\cite{Bevilaqua00}. The primary homoclinic points $h_0$ and $g_0$ points in Fig.~\ref{fig:Homoclinic_Tangle} are associated with orbits having winding number $1$, since both $US[x,h_0]$ and $US[x,g_0]$ are single loops.  They form the complete first hierarchical family.  

The non-primary homoclinic points $a^{(0)}$ and $b^{(0)}$ in Fig.~\ref{fig:Homoclinic_Tangle} are both associated with winding number $2$; i.e.~the loop $US[x,a^{(0)}]=US[x,h_0]+US[h_0,a^{(0)}]$, both of which are single loops; and similarly for $b^{(0)}$, $US[x,b^{(0)}]=US[x,g_{0}]+US[g_{0},b^{(0)}]$.  All points on a particular orbit are associated with the same winding number.  Roughly speaking, a winding-$n$ orbit ``circles" the complex region $n$ times from the infinite past to the infinite future, and therefore the winding number characterizes the complexity of its phase-space excursion. Figure 1 of Ref.~\cite{Bevilaqua00} has a nice illustration.  

Within each family, the orbits can be further organized by their \textit{transit times}~\cite{Easton86,Rom-Kedar90}, which contains the length of the phase-space excursion of a homoclinic orbit.  With the ``open system" assumption, there are no homoclinic points on segments $U^{\prime}_n$ and $S_n$ (For the definition of fundamental segments $U_{n}$, $U^{\prime}_n$, $S_n$, and $S^{\prime}_n$, see Eq.~\eqref{eq:Definition fundamental segments}). Therefore, any homoclinic point $z_0$ must arise from the intersection between some $U_n$ and $S^{\prime}_m$ segments, with $n$ and $m$ being appropriate integers such that $z_0 \in U_n \cap S^{\prime}_m$. The transit time of $\lbrace z_0 \rbrace$, denoted by $t$, is defined as the difference in the indices of $U_n$ and $S^{\prime}_m$: $t=(n-m)$. Starting from $z_{-n} \in U_0 \cap S^{\prime}_{m-n}$, and mapping $t$ times, $M^{t}(z_{-n})=z_{-m} \in U_{n-m} \cap S^{\prime}_0$. Thus, $t$ is the number of iterations needed to map the orbit from $U_0$ to $S^{\prime}_0$. Note that, excluding the primary homoclinic orbits, $\{g_0\}$ and $\{h_0\}$, all homoclinic orbits have positive definite $t$ since there are no intersections of $U_n$ with $S^{\prime}_0$ with negative integer $n$ or $0$; i.e.~the first intersection of $S^{\prime}_0$ is with $U_1$.

Since the mapping preserves the topology: $M^{k}(z_0)=z_k \in U_{n+k}\cap S^{\prime}_{m+k}$, every orbit $\lbrace z_0 \rbrace$ has one and only one point (which is $z_0$) on $S^{\prime}_m$. Therefore, enumerating homoclinic points on $S^{\prime}_m$ is equivalent to enumerating all distinct homoclinic orbits in the trellis (See Eq.~\eqref{eq:Definition trellis} for the definition of a trellis). In practice, it is convenient to choose $m=-1$.  Equivalently, all homoclinic points on $S^{\prime}_{-1}$ with a maximum $t=n+1$ are intersections with the trellis $T_{-1,n}$ ($=T_{-1,t-1}$). The total number of homoclinic orbits increases exponentially 
\begin{figure}[ht]
\centering
{\includegraphics[width=7cm]{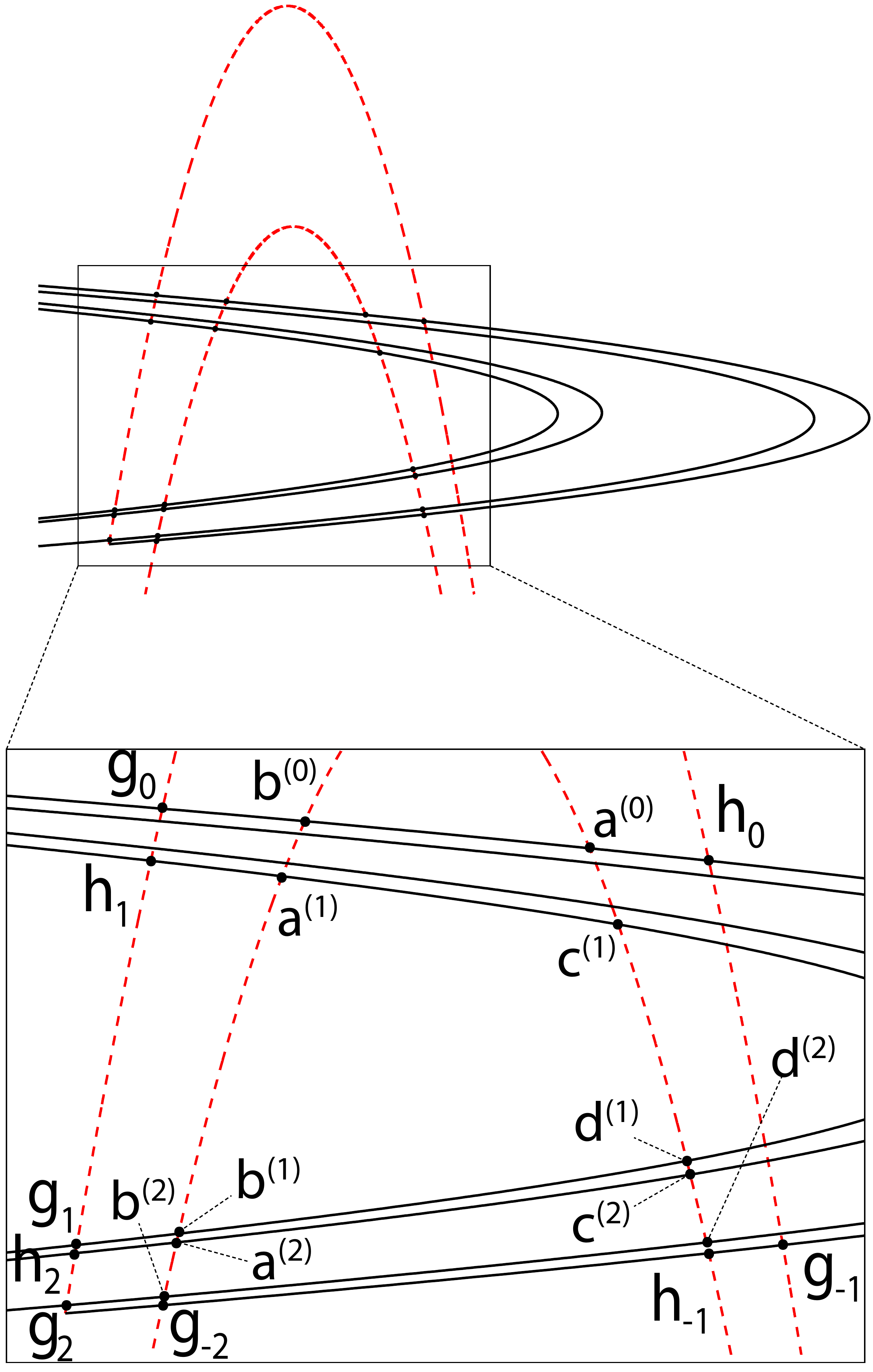}}
 \caption{(Color online) Upper panel: trellis $T_{-1,2}$. Lower panel: (Zoomed-in) The hierarchy of homoclinic points on $S^{\prime}_{-1}$ can be organized as the following: the winding-$1$ points $g_{-2}$ and $h_{-1}$ constitute the first-order family. The sequence of winding-$2$ points $a^{(n)}$ and $b^{(n)}$ ($n \geq 0$) form two second-order families that accumulate on $g_{-2}$ asymptotically under Eqs.~\eqref{eq:Scaling relation 1} and \eqref{eq:Scaling relation 2}. Similarly on the right side of $S^{\prime}_{-1}$, we have the winding-$2$ points $c^{(n)}$ and $d^{(n)}$ ($n \geq 1$) accumulating on $h_{-1}$, which form two second-order families as well. Consequently, three families of areas $[{\cal A}^{\circ}_{SUSU[g_{-2}, b^{(n)}, d^{(n)}, h_{-1}]}]$, $[{\cal A}^{\circ}_{SUSU[b^{(n)}, a^{(n)}, c^{(n)}, d^{(n)}]}]$ and $[{\cal A}^{\circ}_{SUSU[a^{(n)}, b^{(n-1)}, d^{(n-1)}, c^{(n)}]}]$ accumulate on the bottom segment $U[g_{-2},h_{-1}]$ under Eq.~\eqref{eq:Area scaling relation}, with the same asymptotic exponent $\mu_0$.}
\label{fig:Scaling}
\end{figure}  
rapidly with the transit time. For example, in Fig.~\ref{fig:Scaling}, $U_{0}$ intersects $S^{\prime}_{-1}$ at two points: $a^{(0)}$ and $b^{(0)}$. $U_{1}$ intersects $S^{\prime}_{-1}$ at four points $a^{(1)}$, $b^{(1)}$, $c^{(1)}$ and $d^{(1)}$. Furthermore, $U_{2}$ intersects $S^{\prime}_{-1}$ at eight points, where the four points $a^{(2)}$, $b^{(2)}$, $c^{(2)}$ and $d^{(2)}$ are winding-$2$, and the remaining four points, on the upper half of $S^{\prime}_{-1}$ are not explicitly labeled and are winding-$3$. Including $g_{-2}$ and $h_{-1}$, the total number of homoclinic points on $S^{\prime}_{-1}$ is exactly $2^{(t+1)}$.

\subsection{Asymptotic accumulation of homoclinic points}
\label{Asymptotic accumulation}

Although homoclinic tangles create unimaginably complicated phase-space patterns, their behaviors are highly constrained by a few simple rules of Hamiltonian chaos, namely exponential compression and stretching occurs while preserving phase-space areas, and manifolds cannot intersect themselves or other manifolds of the same type. Therefore, locally near any homoclinic point, unstable (stable) manifolds form fine layers of near-parallel curves, with distances in between the curves scaling down exponentially rapidly as they get closer towards that point.  As numerically demonstrated by Eq.~(10) in~\cite{Bevilaqua00}, such asymptotic scaling relations exist inside every family of homoclinic points.  A concrete mathematical description of this phenomenon is given by Lemma 2 in Appendix.~B.~3 of~\cite{Mitchell03a}, which states ``iterates of a curve intersecting the stable manifold approach the unstable manifold." Refer to Appendix.~\ref{ASYMPTOTIC ACCUMULATION EXPONENT} for a brief overview of the lemma.  

The asymptotic scaling ratio of the accumulation is determined by the stability exponent of the hyperbolic fixed point, $\mu_0$, as in Eq.~\eqref{eq:Mitchell scaling relation}. Starting from Eq.~\eqref{eq:Mitchell scaling relation}, let the $z_u$ base point be $g_{-2}$ in Fig.~\ref{fig:Scaling}, and the curve $\overline{{\cal C}}$ that passes through $z_u$ be the stable manifold segment from $S^{\prime}_{-1}$ that passes through $g_{-2}$. Furthermore, choose the ${\cal C}_0$ curve to be $U_0$, which intersects $S^{\prime}_{-1}$ at $a^{(0)}$ and $b^{(0)}$.  The pair of points $a^{(0)}$ and $b^{(0)}$ here play the role of the $z^{(0)}$ point in Fig.~\ref{fig:Mitchell_Theorem}, which are the leading terms of the two families of winding-$2$  homoclinic points $[a^{(n)}]$ and $[b^{(n)}]$, respectively, that accumulate asymptotically on $g_{-2}$. The two families of points $[a^{(n)}]$ and $[b^{(n)}]$ are generated from iterating $U_0$ forward and intersecting the successive images $U_{n}$ ($n \in \mathbb{Z}^{+}$) with $S^{\prime}_{-1}$, and are located on the upper and lower side of $U_{n}$, respectively. The accumulation can be expressed in the asymptotic relation:
\begin{equation}\label{eq:Scaling relation 1}
\begin{split}
& \lim_{n \to \infty} a^{(n)} = g_{-2} \\
& \lim_{n \to \infty} | a^{(n)} - g_{-2} | e^{n\mu_0} = C(g_{-2},a^{(0)})
\end{split}
\end{equation}
where $||$ is the standard Euclidean vector norm, and $C(g_{-2},a^{0})$ is a positive constant depending on the base point $g_{-2}$ and the leading term $a^{(0)}$ in the asymptotic family. Similarly for $b^{(n)}$ we have
\begin{equation}\label{eq:Scaling relation 2}
\begin{split}
& \lim_{n \to \infty} b^{(n)} = g_{-2} \\
& \lim_{n \to \infty} | b^{(n)} - g_{-2} | e^{n\mu_0} = C(g_{-2},b^{(0)})
\end{split}
\end{equation}
Notice that Eqs.~\eqref{eq:Scaling relation 1} and \eqref{eq:Scaling relation 2} are obtained directly from Eq.~\eqref{eq:Mitchell scaling relation}, by the substitutions $z_{u} \to g_{-2}$ and $z^{(n)} \to a^{(n)}/b^{(n)}$. Therefore, the two families of winding-$2$ homoclinic points $[a^{(n)}]$ and $[b^{(n)}]$ accumulate asymptotically onto the winding-$1$ point $g_{-2}$ along the stable manifold, under the scaling relations described by Eqs.~\eqref{eq:Scaling relation 1} and \eqref{eq:Scaling relation 2}. These relations will be denoted symbolically as
\begin{equation}\label{eq:Accumulation along stable}
\begin{split}
& a^{(n)} \xhookrightarrow[S]{n+1} g_{-2} \\
& b^{(n)} \xhookrightarrow[S]{n+1} g_{-2}
\end{split}
\end{equation}
where the $\xhookrightarrow[S]{n+1}$ symbol indicates $a^{(n)}$ and $b^{(n)}$ are the $(n+1)$th member of their respective families, $[a^{(0)},a^{(1)},\cdots]$ and $[b^{(0)},b^{(1)},\cdots]$, that accumulate on $g_{-2}$ along the stable manifold with asymptotic exponent $\mu_0$. 

The asymptotic accumulation relations can be used to infer symbolic dynamics of homoclinic points. Given the symbolic codes of the base point, e.g., $g_{-2}$ from Eq.~\eqref{eq:Scaling relation 1}, the symbolic codes of the entire families of homoclinic points that accumulate on it can be uniquely determined by suitable additions of $110\cdots$ or $100\cdots$ strings to the left side of the core of $g_{-2}$.
Given $g_{-2} \Rightarrow \overline{0}.01\overline{0}$, it can be inferred that (see Fig.~\ref{fig:Trellis_1}):
\begin{equation}\label{eq:Accumulation on g first terms}
  \begin{cases}
       a^{(0)} \Rightarrow \overline{0} 1.11  \overline{0} \\
       b^{(0)} \Rightarrow \overline{0} 1.01  \overline{0}
  \end{cases}
\end{equation}
and
\begin{equation}\label{eq:Accumulation on g general terms}
\begin{cases}
       a^{(n)} \Rightarrow \overline{0} 110^{n-1}.01  \overline{0}  \\
       b^{(n)} \Rightarrow \overline{0} 100^{n-1}.01  \overline{0} 
  \end{cases}
  (n \geq 1)
\end{equation}
where ``$0^{n-1}$" denotes $(n-1)$ repetitions of $0$. The general rule is, the symbolic codes of $a^{(n)}$ and $b^{(n)}$ $(n \geq 0)$ are obtained by adding the substrings ``$110^{n}$" and ``$100^{n}$", respectively, to the left end of the core of $g_{-2}$, keeping the position of the decimal point relative to the right end of the core. 

Following the same pattern, on the right side of $S^{\prime}_{-1}$ (see Fig.~\ref{fig:Trellis_1}), there are two families of winding-$2$ homoclinic points $[c^{(n)}]$ and $[d^{(n)}]$ ($n \geq 1$) that accumulate asymptotically along the stable manifold on the winding-$1$ point $h_{-1}$ under scaling relations similar to Eqs.~\eqref{eq:Scaling relation 1} and \eqref{eq:Scaling relation 2}:
\begin{equation}\label{eq:Accumulation along stable 2}
\begin{split}
& c^{(n)} \xhookrightarrow[S]{n} h_{-1} \\
& d^{(n)} \xhookrightarrow[S]{n} h_{-1}
\end{split}
\end{equation}
and their symbolic codes are determined from that of $h_{-1}$:
\begin{equation}\label{eq:Accumulation on h general terms}
\begin{split}
& ( c^{(n)} \Rightarrow \overline{0} 110^{n-1}.11 \overline{0} ) \xhookrightarrow[S]{n} ( h_{-1} \Rightarrow \overline{0}.11\overline{0} ) \\
& ( d^{(n)} \Rightarrow \overline{0} 100^{n-1}.11 \overline{0} ) \xhookrightarrow[S]{n} ( h_{-1} \Rightarrow \overline{0}.11\overline{0} )
\end{split}
\end{equation}  
with the same rule of adding the ``$110^{n-1}$" and ``$100^{n-1}$" substrings. This assignment rule for the symbolic code is valid for any homoclinic points in the system. As the construction is rather technical, refer to App.~\ref{Systematic assignment of symbolic codes} for the detailed systematic assignments of symbolic dynamics. 

An important consequence of the above asymptotic relations between homoclinic points is that the phase-space areas spanned by them also scale down at the same rate. Using the present example, three families of areas can be easily identified, which are $[{\cal A}^{\circ}_{SUSU[g_{-2}, b^{(n)}, d^{(n)}, h_{-1}]}]$, $[{\cal A}^{\circ}_{SUSU[b^{(n)}, a^{(n)}, c^{(n)}, d^{(n)}]}]$, and $[{\cal A}^{\circ}_{SUSU[a^{(n)}, b^{(n-1)}, d^{(n-1)}, c^{(n)}]}]$ ($n \geq 2$).  Each follows the scaling relation,
\begin{equation}\label{eq:Area scaling relation}
\lim_{n \to \infty} \frac{ {\cal A}^{\circ}_{SUSU[g_{-2}, b^{(n)}, d^{(n)}, h_{-1}]} }{ {\cal A}^{\circ}_{SUSU[g_{-2}, b^{(n+1)}, d^{(n+1)}, h_{-1}]} } = e^{\mu_0},
\end{equation}
and similarly for the $[{\cal A}^{\circ}_{SUSU[b^{(n)}, a^{(n)}, c^{(n)}, d^{(n)}]}]$ and $[{\cal A}^{\circ}_{SUSU[a^{(n)}, b^{(n-1)}, d^{(n-1)}, c^{(n)}]}]$ families as well .  These areas are all from the partition of the lobe $L^{\prime}_{-1}$ using successively propagated lobes $L_{n}$.  Returning to Fig.~\ref{fig:Scaling}, where successive intersections between the fundamental segments $U_n$ and $S^{\prime}_{-1}$ of Eq.~\eqref{eq:Definition fundamental segments} accumulate on $g_{-2}$ and $h_{-1}$, the following three identifications can be made: ${\cal A}^{\circ}_{SUSU[g_{-2}, b^{(n)}, d^{(n)}, h_{-1}]}$ is the area between the lower side of $U_n$ and $U[g_{-2},h_{-1}]$, ${\cal A}^{\circ}_{SUSU[b^{(n)}, a^{(n)}, c^{(n)}, d^{(n)}]}$ is the area between the lower and upper sides of $U_n$, and ${\cal A}^{\circ}_{SUSU[a^{(n)}, b^{(n-1)}, d^{(n-1)}, c^{(n)}]}$ is the area between the upper side of $U_n$ and the lower side of $U_{n-1}$.  As more lobes are added, such areas approach $U[g_{-2},h_{-1}]$, and the ratio tends to $e^{\mu_0}$. Hence Eq.~\eqref{eq:Area scaling relation} can be understood as an asymptotic relation between area partitions of $L^{\prime}_{-1}$ in the neighborhood of $U[g_{-2},h_{-1}]$. 

The above relations are obtained by choosing the $z_u$ base point in Eq.~\eqref{eq:Mitchell scaling relation} to be the winding-$1$ points $g_{-2}$ and $h_{-1}$, and studying the accumulations of winding-$2$ homoclinic points on them. Generally speaking, since the choice of the $z_u$ base point is arbitrary, one can just as well choose $z_u$ to be a winding-$m$ homoclinic point on $U(x)$, and there will always be two families of winding-$(m+1)$ homoclinic points that accumulate on $z_u$ along $S(x)$ under similar relations, with the same scaling ratio $e^{\mu_0}$. Therefore, Eq.\eqref{eq:Area scaling relation} holds for any winding-$m$ homoclinic point and the winding-$(m+1)$ families of areas that accumulate on it. Such relations are true in the neighborhood of any homoclinic point, and they imply that the computation of a few leading area terms in any $[{\cal A}^{\circ}_{SUSU[\cdots]}]$ family suffices to determine the rest of the areas, depending on the desired degree of accuracy.  

An important subtlety in the scaling relations concerns the exponent $\mu_0$.  Due to the exponential compressing and stretching nature of chaotic dynamics, it is well-known that the new cell areas bounded by adjacent stable and unstable segments from a trellis with increasing iteration numbers must become exponentially small.  See Appendix~A of Ref.~\cite{Li18} for a brief review. In particular, one can anticipate that the new cell areas from $T_{-1,n}$ decrease on average similarly to the horizontal strips $H_{ s_{-n} \cdots s_{-1} }$ in Figs.~3 and 4 of Ref.~\cite{Li18}, which scale at the rate $e^{-n\mu}$, where $\mu$ is the system's Lyapunov exponent.  However,  in general $\mu_0 \neq \mu$, $\mu_0$ measuring the stretching rate of the hyperbolic fixed point, which is expected to be $\geq \mu$.  This begs the question as to how this could be consistent.  In Sec.~\ref{Fast and slow scaling relations}, it is shown that this presumably larger exponent $\mu_0$ only applies to calculating the ratios between successive areas within the specific families such as those in Eq~\eqref{eq:Area scaling relation}.  Between different families, the scaling exponents change to smaller values, which is consistent with the Lyapunov exponent being smaller than $\mu_0$.  A shorthand reference to this is to say that Eq.~\eqref{eq:Area scaling relation} is fast scaling relation, in the sense that they happen at faster rates than the average instability of the system as a whole, $\mu$. 

\begin{figure}[ht]
\centering
{\includegraphics[width=7cm]{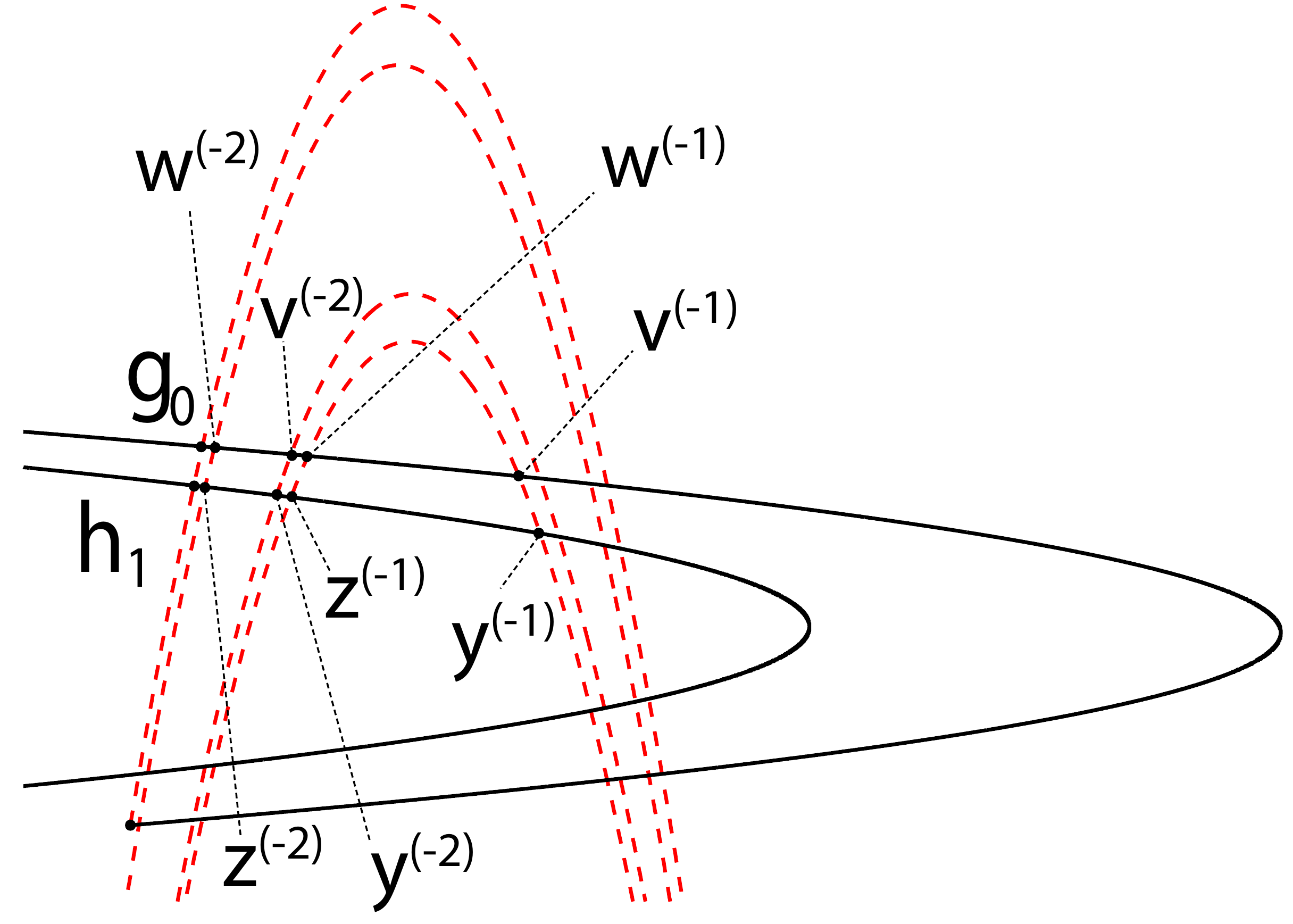}}
 \caption{(Color online) Accumulation of homoclinic points along $U(x)$ under $M^{-1}$. Two families of homoclinic points $[v^{(-n)}]$ and $[w^{(-n)}]$ are created from $S^{\prime}_{-n} \cap U_{0}$, that accumulate on $g_0$ along $U_0$. Notice that only the $n=1, 2$ cases are plotted here. Similarly, the two families $[y^{(-n)}]$ and $[z^{(-n)}]$ are created from $S^{\prime}_{-n} \cap U_1$, and accumulate on $h_1$ along $U_{1}$.    }
\label{fig:Scaling_Inverse}
\end{figure}

Identical scaling results hold under the inverse mapping $M^{-1}$ upon switching the roles of the stable and unstable manifolds.  Shown in Fig.~\ref{fig:Scaling_Inverse} is a simple example of the inverse case,      
where families of homoclinic points accumulate along the unstable manifold. For convenience, the $a^{(0)}$ and $b^{(0)}$ points from Fig.~\ref{fig:Scaling} are relabeled in this figure as $v^{(-1)}$ and $w^{(-1)}$, respectively. Successive inverse mappings of $S^{\prime}_{-1}$ intersect with $U_{0}$ and create two families of winding-$2$ points $[v^{(-n)}]$ and $[w^{(-n)}]$ ($n \geq 1$), which accumulate on the primary point $g_{0}$ along the unstable manifold, under scaling relations similar to Eq.~\eqref{eq:Scaling relation 1}. Similar to Eq.~\eqref{eq:Accumulation along stable}, the accumulation along $U(x)$ is denoted by
\begin{equation}\label{eq:Accumulation along unstable}
\begin{split}
& v^{(-n)} \xhookrightarrow[U]{n} g_{0} \\
& w^{(-n)} \xhookrightarrow[U]{n} g_{0}
\end{split}
\end{equation}
where $\xhookrightarrow[U]{n}$ indicates that $v^{(-n)}$ and $w^{(-n)}$ are the $n$th member of their respective families, $[v^{(-1)},v^{(-2)},\cdots]$ and $[w^{(-1)},w^{(-2)},\cdots]$, that accumulate on $g_0$ along the unstable manifold with asymptotic exponent $\mu_0$. 

Also shown in Fig.~\ref{fig:Scaling_Inverse} are two other families of winding-$2$ points $[y^{(-n)}]$ and $[z^{(-n)}]$ generated from $S^{\prime}_{-n} \cap U_{1}$, which accumulate on $h_1$ along the unstable manifold. Notice that points $y^{(-1)}$ and $z^{(-1)}$ are identical to $c^{(1)}$ and $a^{(1)}$ from Fig.~\ref{fig:Scaling}, respectively. Consequently, three families of areas $[{\cal A}^{\circ}_{SUSU[h_1, g_0, w^{(-n)}, z^{(-n)}]}]$, $[{\cal A}^{\circ}_{SUSU[z^{(-n)}, w^{(-n)}, v^{(-n)}, y^{(-n)}]}]$, and $[{\cal A}^{\circ}_{SUSU[y^{(-n)}, v^{(-n)}, w^{(-n+1)}, z^{(-n+1)}]}]$ ($n \geq 2$) accumulate on $S[h_1,g_0]$ under the asymptotic ratio $\mathrm{e}^{\mu_0}$, similar to Eqs.~\eqref{eq:Area scaling relation}. Therefore, the asymptotic behaviors of the manifolds between $M$ and $M^{-1}$ are identical, upon interchanging the roles of $S(x)$ and $U(x)$. We would like to emphasize that this is a general result that comes from the stability analysis of the system, which holds true whether the system is time-reversal symmetric or not.  

There is an interesting special case of the accumulation relations for which $z_u$ is chosen to be the fixed point $x$ itself.  For this case, the primary orbits $\lbrace g_i \rbrace$ and $\lbrace h_i \rbrace$ themselves become two families of homoclinic points that accumulate on $x$ with asymptotic ratio $\mathrm{e}^{\mu_0}$ under both forward and inverse mappings:
\begin{equation}\label{eq:Scaling relation for primary points along stable}
\begin{split}
& h_i \xhookrightarrow[S]{} x,\\
& g_i \xhookrightarrow[S]{} x,
\end{split}
\end{equation}
and 
\begin{equation}\label{eq:Scaling relation for primary points along unstable}
\begin{split}
& h_{-i} \xhookrightarrow[U]{} x,\\
& g_{-i} \xhookrightarrow[U]{} x
\end{split}
\end{equation}
although the meaning of the order number for each point inside these two families now becomes ambiguous, therefore removed from the top of the ``$\xhookrightarrow{}$" sign. The hyperbolic fixed point $x$ is now viewed as a ``homoclinic point" of winding number $0$, on which the winding-$1$ primaries accumulate. 

\subsection{Partitioning of phase-space areas}
\label{Partitioning of cell areas}

\begin{figure}[ht]
\centering
{\includegraphics[width=7cm]{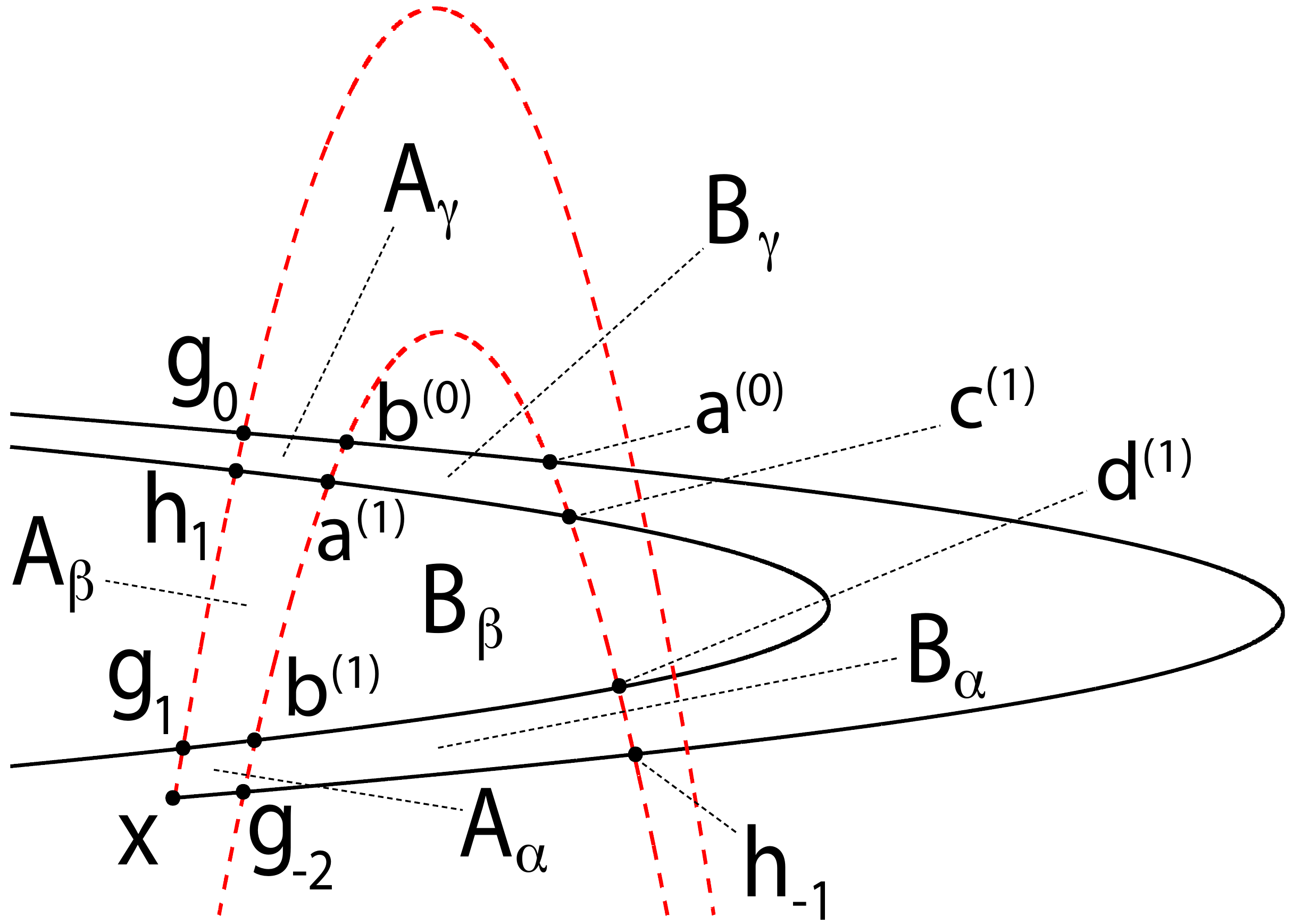}}
 \caption{(Color online) Partitioning of cell areas in $T_{-1,1}$. The three type-I cells are $A_{\alpha}$, $A_{\beta}$, and $A_{\gamma}$. The three type-II cells are $B_{\alpha}$, $B_{\beta}$, and $B_{\gamma}$. The $A$ cell from $T_{-1,0}$ is partitioned into three cell areas in $T_{-1,1}$: $A=A_{\alpha}+A_{\beta}+A_{\gamma}$. Similarly for the type-II cell, $B=B_{\alpha}+B_{\beta}+B_{\gamma}$. }
\label{fig:Trellis_Partition_1}
\end{figure}

Of particular relevance to calculating the homoclinic orbit relative actions is the sequence of trellises $T_{-1,n_u}$, with $n_u=0,1,\cdots,N$. New homoclinic points appear on $S^{\prime}_{-1}$ upon each unit increase of $n_u$, and their relative actions are closely related to certain phase-space areas called $\mathit{cells}$. Given a trellis $T_{n_s,n_u}$ and four homoclinic points $a,b,c,d \in T_{n_s,n_u}$ that form a simple closed region bounded by the loop $SUSU[a,b,c,d]=S[a,b]+U[b,c]+S[c,d]+U[d,a]$, it is called a \textit{cell of $T_{n_s,n_u}$} if there are no stable and unstable manifold segments from $T_{n_s,n_u}$ that enter inside the region. Consequently, there are no homoclinic points other than the four vortices on the boundary of the cell.  For example, both $V_0$ and $V_1$ are cells of $T_{-1,0}$ (Fig.~\ref{fig:Horseshoe}).  However, in $T_{-1,1}$ (Fig.~\ref{fig:Trellis_Partition_1}) they get partitioned by $U_1$ and are not cells anymore since there are unstable segments inside them.  Each trellis gives a specific partition to the phase space. By fixing $n_s=-1$ and increasing the $n_u$ value, the resulting sequence of trellises yields a systematic and ever-finer partition of the phase space, which acts as the skeletal-like structure for the study of homoclinic orbits.

In fact, of all the cell areas of $T_{-1,n_u}$, two subsets are relevant to the action calculations. The first subset, defined as \textit{type-I} cells, are those from the region $V_0$ partition (Fig.~\ref{fig:Horseshoe}). Equivalently, the type-I cells are those with two stable boundary segments located on $S[x,g_0]$ and $S[b^{(0)},g_{-2}]$, respectively. Similarly, the second subset, or the \textit{type-II} cells, are those from the partition of $V^{\prime}$ in Fig.~\ref{fig:Horseshoe}. Equivalently speaking, the type-II cells are those with two stable segments located on $S[b^{(0)},g_{-2}]$ and $S[h_{-1},a^{(0)}]$, respectively. 
Figure~\ref{fig:Trellis_Partition_1} shows the examples of $T_{-1,1}$, three type-I cells $A_{\alpha}$, $A_{\beta}$,$A_{\gamma}$, and three type-II cells $B_{\alpha}$, $B_{\beta}$, $B_{\gamma}$. Section~\ref{Exact decomposition} shows that the knowledge of these types of cell areas is sufficient for the action calculation of all homoclinic orbits. 

\begin{figure}[ht]
\centering
{\includegraphics[width=7cm]{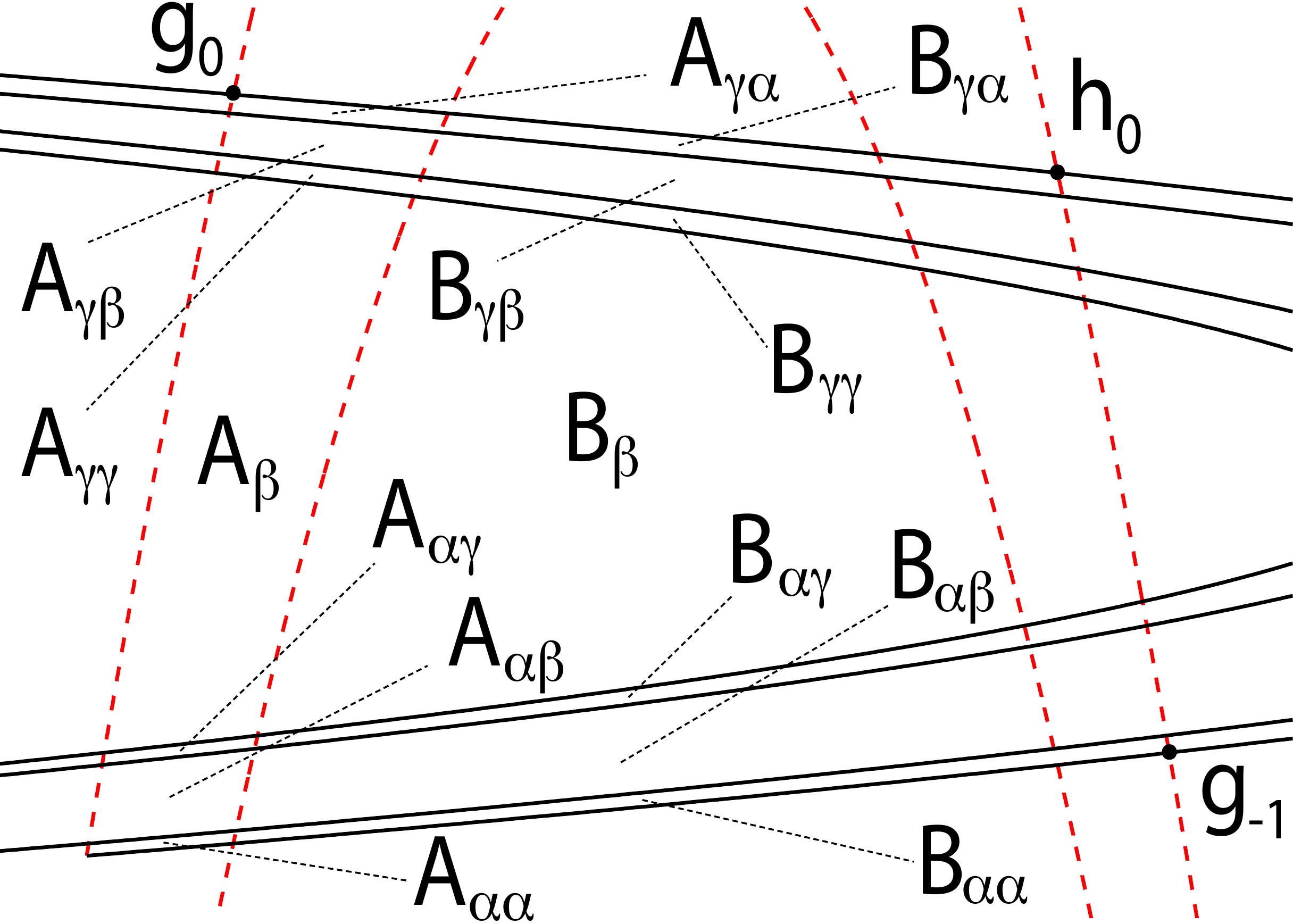}}
 \caption{(Color online) Zoomed-in graph around the complex region of $T_{-1,2}$ (same as the lower panel of Fig.~\ref{fig:Scaling}). The $A_{\alpha}$ and $A_{\gamma}$ areas in Fig.~\ref{fig:Trellis_Partition_1} are partitioned into three sub-areas: $A_{\alpha}=A_{\alpha \alpha}+A_{\alpha \beta}+A_{\alpha \gamma}$ and $A_{\gamma}=A_{\gamma \alpha}+A_{\gamma \beta}+A_{\gamma \gamma}$. The $A_{\beta}$ area does not get partitioned because of the open system assumption, i.e., manifolds outside of the complex region do not revisit the complex region in future iterations. Since the type-I and type-II cells are always partitioned by any lobe $L_n$ simultaneously, the $B_{\alpha}$ and $B_{\gamma}$ cells from Fig.~\ref{fig:Trellis_Partition_1} are partitioned in identical ways with $A_{\alpha}$ and $A_{\gamma}$, respectively. }
\label{fig:Trellis_Partition_2}
\end{figure}

In the partitioning of cell areas from increasing trellises, there are families of areas corresponding to fast and slow scaling relations.  Since the homoclinic orbit actions are ultimately expressed using these areas, an investigation of this kind is crucial for the understanding of asymptotic clustering of homoclinic orbit actions. The partitioning process is recursive in nature, and the partition of the existing cells of $T_{-1,n}$ by $T_{-1,n+1}$ is the critical step. This process eventually leads to an organization of the cells into tree-like structures, and a classification of the scaling rates using the branches of the trees. As introduced in the discussion of Fig.~\ref{fig:Trellis_Partition_2}, these structures are identical for the type-I and type-II cells, so it suffices to concentrate mostly on the type-I cells. 

The partition starts from $T_{-1,0}$, where the only type-I cell is $V_0$.  In order to introduce a partition subscript, $V_0$ is denoted $A$.  In the next iteration, $A$ is partitioned by $T_{-1,1}$, in which the lobe $L_1$ enters $A$ dividing it into three finer cells, namely $A_{\alpha}$, $A_{\beta}$, and $A_{\gamma}$, as shown in Fig.~\ref{fig:Trellis_Partition_1}. Similarly, denote the cell $V^{\prime}$ of $T_{-1,0}$ by $B$. $B$ is partitioned by $T_{-1,1}$ in an identical way: $B=B_{\alpha}+B_{\beta}+B_{\gamma}$ since the unstable lobes always enter the type-I and type-II regions simultaneously for the complete horseshoe map, and also for a large class of incomplete horseshoe maps as well. 

In the next iteration, $T_{-1,2}$ introduces finer partitions in which $L_2$ enters $A_{\alpha}$ and $A_{\gamma}$, dividing both of them into three new cells: $A_{\alpha}=A_{\alpha \alpha}+A_{\alpha \beta}+A_{\alpha \gamma}$ and $A_{\gamma}=A_{\gamma \alpha}+A_{\gamma \beta}+A_{\gamma \gamma}$, as labeled in Fig.~\ref{fig:Trellis_Partition_2}. Therefore, future partitions of a cell correspond to the      
addition of the $\alpha$, $\beta$, and $\gamma$ symbols to the end of its existing subscript, except if its subscript ends in $\beta$ (which terminates that sequence). 

In open systems such as the H\'{e}non map, the $A_{\beta}$ area does not get partitioned by future iterations because points outside the complex region do not re-enter the complex region, therefore no unstable manifolds will extend inside the lobes $L_{i}$ for all $i \in \mathbb{Z}$. Since $A_{\beta}$ belongs to the inside of $L_{1}$, it will not be partitioned by any future trellises. The same are true for $A_{\alpha \beta}$, $A_{\gamma \beta}$, and all areas whose subscript end with $\beta$ in future trellises, which belong to some future lobes $L_{i}$.  

\begin{figure}[ht]
\centering
{\includegraphics[width=6cm]{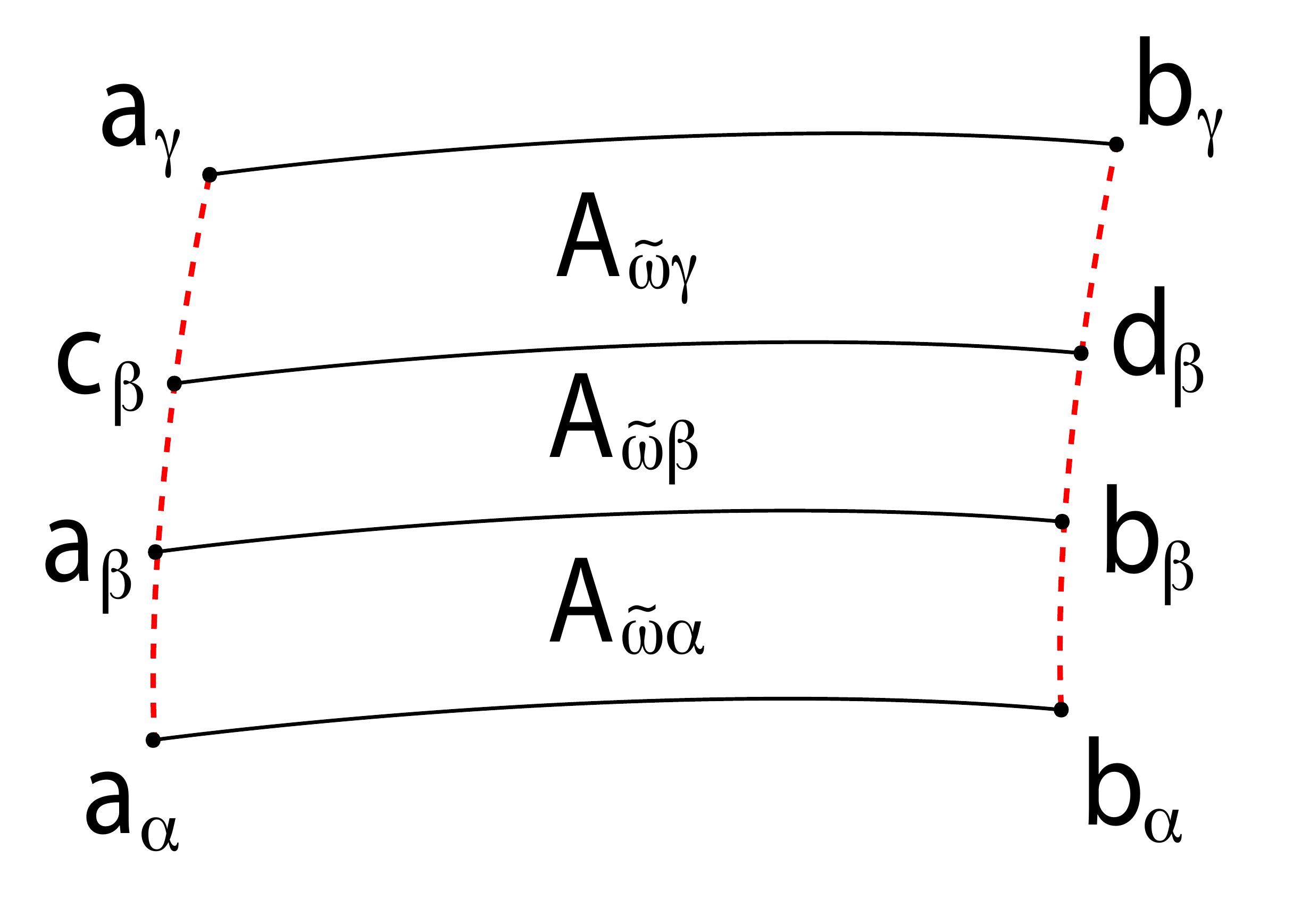}}
 \caption{(Schematic, color online) The cell $A_{\tilde{\omega}}$ in $T_{-1,n}$ is partitioned into three new cells in $T_{-1,n+1}$ by lobe $L_{n+1}$: $A_{\tilde{\omega}} = A_{\tilde{\omega} \alpha} + A_{\tilde{\omega} \beta} + A_{\tilde{\omega} \gamma}$, where $A_{\tilde{\omega} \beta} \subset L_{n+1}$. The rule of assignment is, $A_{\tilde{\omega} \beta} $ is always assigned to the middle cell, and $A_{\tilde{\omega} \alpha} $ is assigned to the cell with the two corners, namely $a_{\alpha}$ and $b_{\alpha}$, such that $a_{\beta},\ c_{\beta}  \xhookrightarrow[S]{k} a_{\alpha} $ and $b_{\beta},\ d_{\beta}  \xhookrightarrow[S]{k} b_{\alpha}$, where the order number $k$ is an appropriate integer that depends on $\tilde{\omega}$. Finally, $A_{\tilde{\omega} \gamma}$ is assigned to the last cell. The same pattern applies to all the $B$ cells as well. }
\label{fig:Cell_Area_Partition}
\end{figure}

The relative position of the new cells is nontrivial. For example, as shown in Fig.~\ref{fig:Trellis_Partition_2}, $A_{\alpha \alpha}$, $A_{\alpha \beta}$, and $A_{\alpha \gamma}$ are positioned from the bottom to the top, while $A_{\gamma \alpha}$, $A_{\gamma \beta}$, $A_{\gamma \gamma}$ are positioned from the top to the bottom, begging the question, how should the order of symbols be assigned for the newly generated cells in a consistent way.  The answer is buried in the scaling relations among homoclinic points.  As shown in Fig.~\ref{fig:Cell_Area_Partition}, consider 
an arbitrary cell area $A_{\tilde{\omega}}$ in $T_{-1,n}$, which is partitioned into three new cells, $A_{\tilde{\omega} \alpha}$, $A_{\tilde{\omega} \beta}$, and $A_{\tilde{\omega} \gamma}$ in $T_{-1,n+1}$ by lobe $L_{n+1}$.  Here $\tilde{\omega}$ denotes a length-$n$ string of symbols composed by arbitrary combinations of $\alpha$ and $\gamma$ (but not $\beta$).  The middle cell is always labeled by $\tilde{\omega} \beta$. Let the four homoclinic points on the corners of this cell be $a_{\beta}$, $b_{\beta}$, $c_{\beta}$, and $d_{\beta}$, respectively, all of which belong to $U_{n+1}$. The $\tilde{\omega} \alpha$ subscript is then assigned to the cell with the two corners on which $a_{\beta}$, $b_{\beta}$, $c_{\beta}$, and $d_{\beta}$ accumulate:
\begin{equation}
\begin{split}
&a_{\beta},\ c_{\beta}  \xhookrightarrow[S]{k} a_{\alpha} \\
&b_{\beta},\ d_{\beta}  \xhookrightarrow[S]{k} b_{\alpha}.
\end{split}
\end{equation}
where the order number $k$ depends on the detailed forms of $\tilde{\omega}$.  The $\tilde{\omega} \gamma$ subscript is assigned to the remaining cell. 

If the symbolic codes of $a_{\alpha}$ and $b_{\alpha}$ are $a_{\alpha} \Rightarrow \overline{0} 1 \tilde{s}_{-} . \tilde{s}_{+} 1 \overline{0}$ and $b_{\alpha} \Rightarrow \overline{0} 1 \tilde{s}^{\prime}_{-}.\tilde{s}^{\prime}_{+} 1 \overline{0}$, where $\tilde{s}_{\pm}$ and $\tilde{s}^{\prime}_{\pm}$ are substrings composed by $0$s and $1$s, then it can be inferred using Eq.~\eqref{eq:Orbit symbolic codes general rules} that
\begin{equation}\label{eq:a_beta b_beta symbolic codes}
\begin{split}
& a_{\beta} \Rightarrow \overline{0} 10 0^{k-1} 1 \tilde{s}_{-} . \tilde{s}_{+} 1 \overline{0} \\
& c_{\beta} \Rightarrow \overline{0} 11 0^{k-1} 1 \tilde{s}_{-} . \tilde{s}_{+} 1 \overline{0}
\end{split}
\end{equation}
and 
\begin{equation}\label{eq:b_beta d_beta symbolic codes}
\begin{split}
&  b_{\beta}  \Rightarrow \overline{0} 10 0^{k-1} 1 \tilde{s}^{\prime}_{-}.\tilde{s}^{\prime}_{+} 1 \overline{0} \\
&  d_{\beta}  \Rightarrow \overline{0} 11 0^{k-1} 1 \tilde{s}^{\prime}_{-}.\tilde{s}^{\prime}_{+} 1 \overline{0}\ .
\end{split}
\end{equation}

\begin{figure}[ht]
\centering
{\includegraphics[width=7cm]{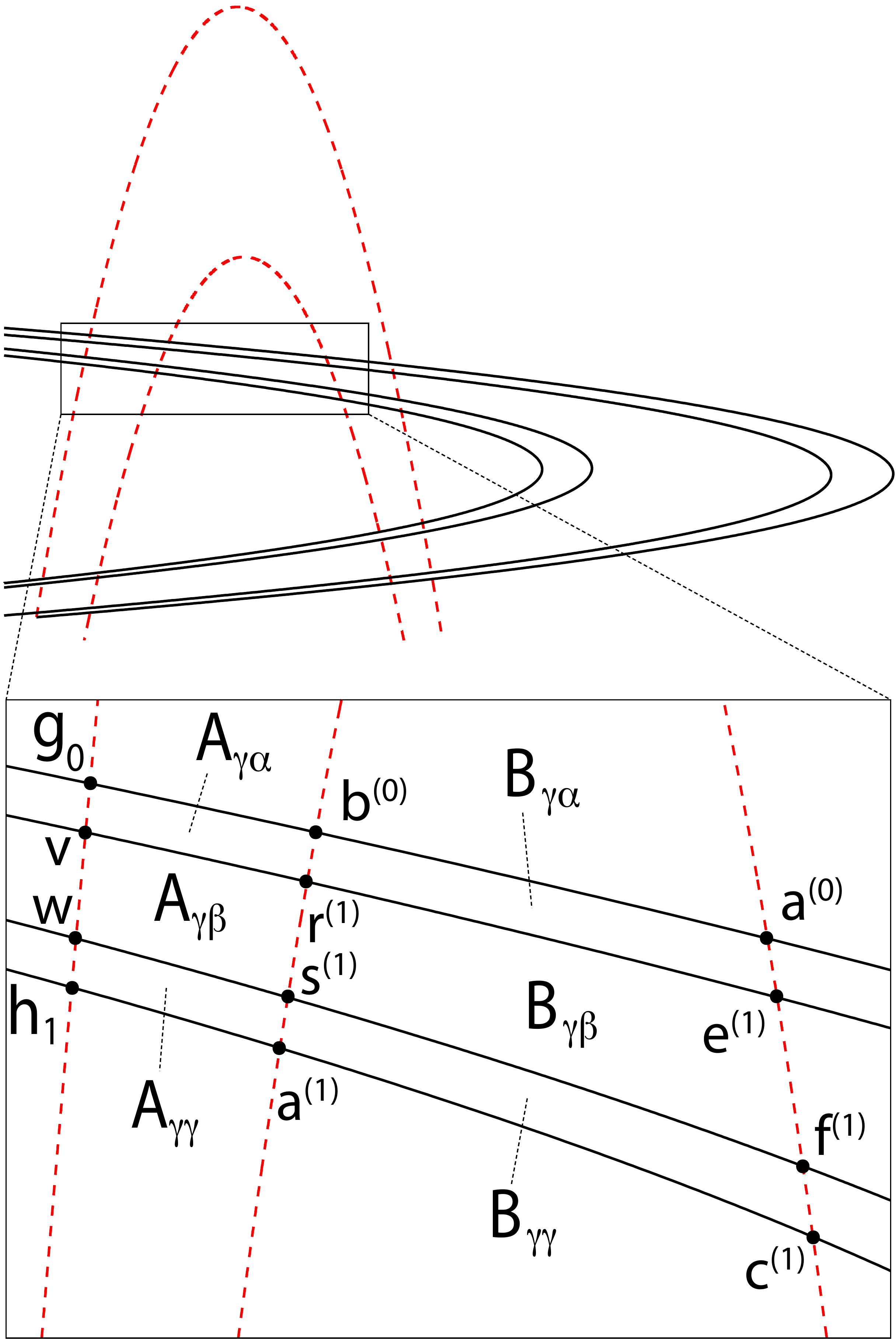}}
 \caption{(Zoomed-in graph, color online) The $A_{\gamma}$ and $B_{\gamma}$ cells from Fig.~\ref{fig:Trellis_Partition_1} are partitioned by $L_2$ into three new cells each. The $A_{\gamma \beta}$ area is assigned to the middle one. Since $v,w \xhookrightarrow[S]{1} g_0$ and $r^{(1)},s^{(1)} \xhookrightarrow[S]{1} b^{(0)}$, $A_{\gamma \alpha}$ is assigned to the top one, leaving $A_{\gamma \gamma}$ to be the bottom one. The same rules apply to the $B$ cells as well. }
\label{fig:Area_Cells_Zoom}
\end{figure}

For a concrete example, consider the partition $A=A_{\alpha}+A_{\beta}+A_{\gamma}$ in Fig.~\ref{fig:Trellis_Partition_1}, where $A_{\tilde{\omega}}=A$ with $\tilde{\omega}$ being an empty string. The $A_{\beta}$ is first identified as the one in the middle. Notice that its corners, $g_1, h_1\xhookrightarrow[S]{} x$, and $a^{(1)},b^{(1)} \xhookrightarrow[S]{2} g_{-2}$, thus $A_{\alpha}$ is assigned to the cell at the bottom; $A_{\gamma}$ is thus the cell at the top.  One can verify that the assignments of cells in Fig.~\ref{fig:Trellis_Partition_2} follow the same pattern. In particular, the relative positions of the $A_{\gamma \alpha}$, $A_{\gamma \beta}$, and $A_{\gamma \gamma}$ cells are indeed reversed.  This can be seen from the zoomed-in Fig.~\ref{fig:Area_Cells_Zoom}, where the four corners of $A_{\gamma \beta}$, namely $v$, $w$, $r^{(1)}$, 
and $s^{(1)}$, accumulate on $g_0$ and $b^{(0)}$: $v,w \xhookrightarrow[S]{1} g_0$ and $r^{(1)},s^{(1)} \xhookrightarrow[S]{1} b^{(0)}$.  Thus, $A_{\gamma \alpha}$ is assigned to the cell on the top of $A_{\gamma \beta}$, and $A_{\gamma \gamma}$ the one at the bottom. The partition of the $B$ cells follow an identical scheme.  

\begin{figure}[ht]
\centering
{\includegraphics[width=6.5cm]{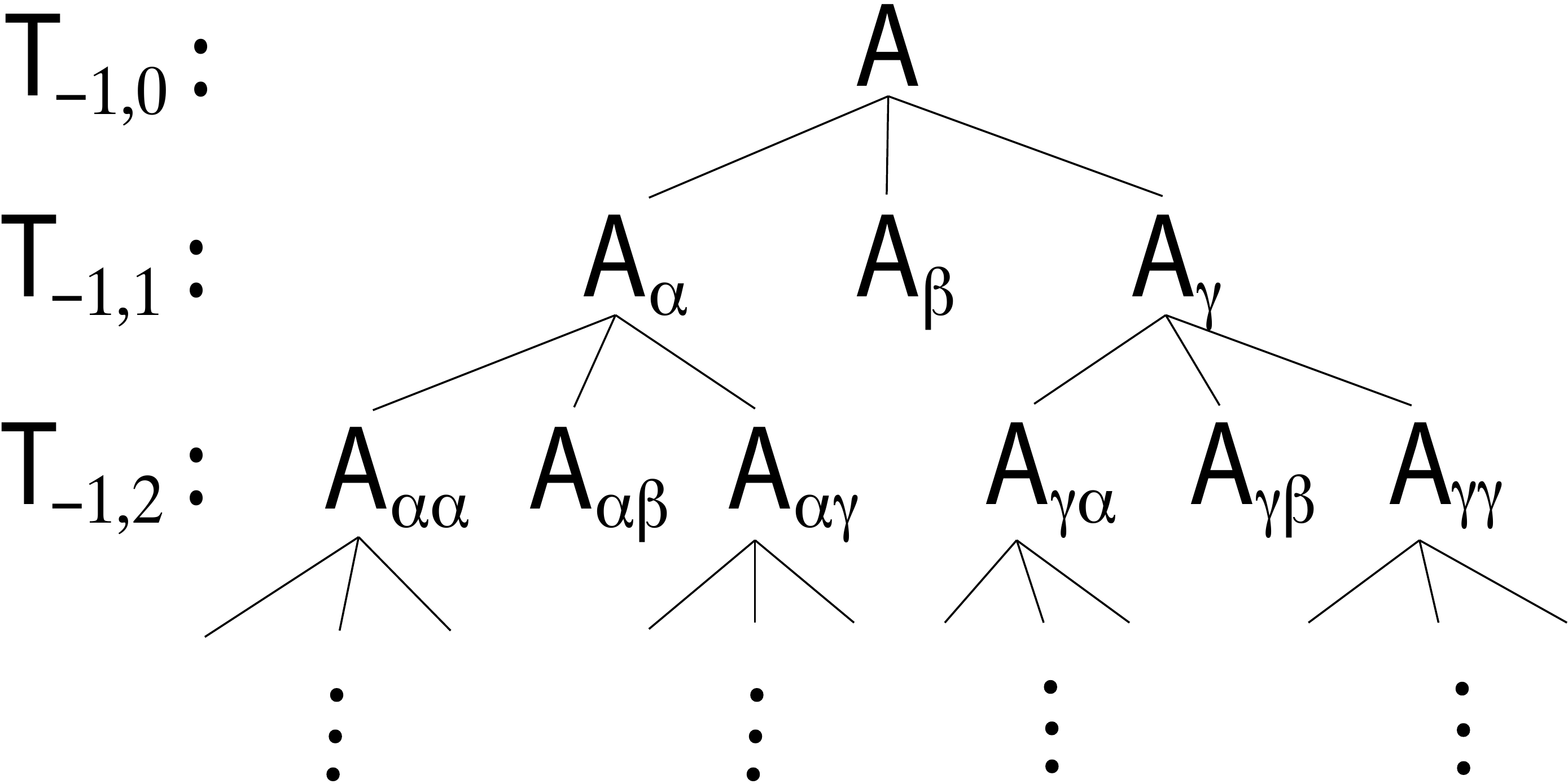}}
 \caption{The partition tree of type-I cell areas. Nodes at the $n$th level along the tree are areas generated from the partition of $T_{-1,n-1}$ by $T_{-1,n}$. Every $\alpha$ and $\gamma$ nodes are partitioned into three new nodes at the next level, while the $\beta$ nodes do not get partitioned any further. The partition tree of type-II areas follows an identical pattern upon changing the symbols $A$ into $B$.  To order the cells as in the trellis, proceeding from the top of the tree, reverse the order for the next level down each time the number of $\gamma$ symbols is odd.}
\label{fig:A_tree}
\end{figure}

A complete assignment of the areas' symbols are determined by the accumulation relations between homoclinic points along $S(x)$, which can be carried on with increasing iterations of $T_{-1,n}$ to obtain ever finer partitions of type-I and type-II cell areas. The progressive partitioning of the type-I cells can be represented by a \textit{partition tree} shown in Fig.~\ref{fig:A_tree}. 
Defining the node $A$ to be the $0$th level of the tree, which is a cell generated by $T_{-1,0}$, then nodes at the $n$th level along the tree represent the cells newly generated by $T_{-1,n}$. Notice the $\beta$ nodes do not get expanded at the next level, because of the open system assumption.   A finite truncation of the partition tree to the $n$th level corresponds to the partition of the type-I areas up to $T_{-1,n}$.  Note that the partition tree of type-II cell areas is identical with the type-I tree upon changing the symbols $A$ into $B$.

\subsection{Scaling relations and periodic orbit exponents}
\label{Fast and slow scaling relations}

In this section we demonstrate numerically a fundamental relation between the stability exponents of periodic orbits and the scaling ratios in certain families of areas of the partition tree. The relation provides an efficient way to compute the stability exponents of periodic orbits from the areas bounded by stable and unstable manifolds, which does not require the numerical construction of periodic orbits. 

The complete and exact decomposition of the homoclinic orbit actions requires only the areas of the partition trees. On the other hand, their areas scale down asymptotically with the tree level exponentially, with the exponents determined by the specific paths that one moves down the trees. The simplest example is a path of consecutive ``$\alpha$"-directions. Starting from any $\alpha$, $\beta$, or $\gamma$ node of the tree, denoted by $A_{\tilde{\omega} \alpha}$, $A_{\tilde{\omega} \beta}$, and $A_{\tilde{\omega} \gamma}$ respectively, and move to deeper levels along the left directions. The successive cells areas visited by such paths form three families: $[A_{\tilde{\omega} \alpha^{n}}]$, $[ A_{\tilde{\omega} \alpha^{n-1} \beta}]$, and $[A_{\tilde{\omega} \alpha^{n-1} \gamma}]$, that scale down with the stability exponent of the fixed point: 
 \begin{equation}
\label{eq:Fast scaling}
\lim_{n \to \infty} \frac{A_{\tilde{\omega} \alpha^{n}}}{A_{\tilde{\omega} \alpha^{n+1}}} = \lim_{n \to \infty} \frac{A_{\tilde{\omega} \alpha^{n-1} \beta}}{ A_{\tilde{\omega} \alpha^{n} \beta}} =  \lim_{n \to \infty} \frac{A_{\tilde{\omega} \alpha^{n-1} \gamma} }{ A_{\tilde{\omega} \alpha^{n} \gamma} }=e^{\mu_0}\ ,
\end{equation}  
where $\alpha^n$ denotes $n$ consecutive $\alpha$ characters in the string. Identical relations hold for the $B$ cells as well. 

\begin{figure}[ht]
\centering
{\includegraphics[width=6.5cm]{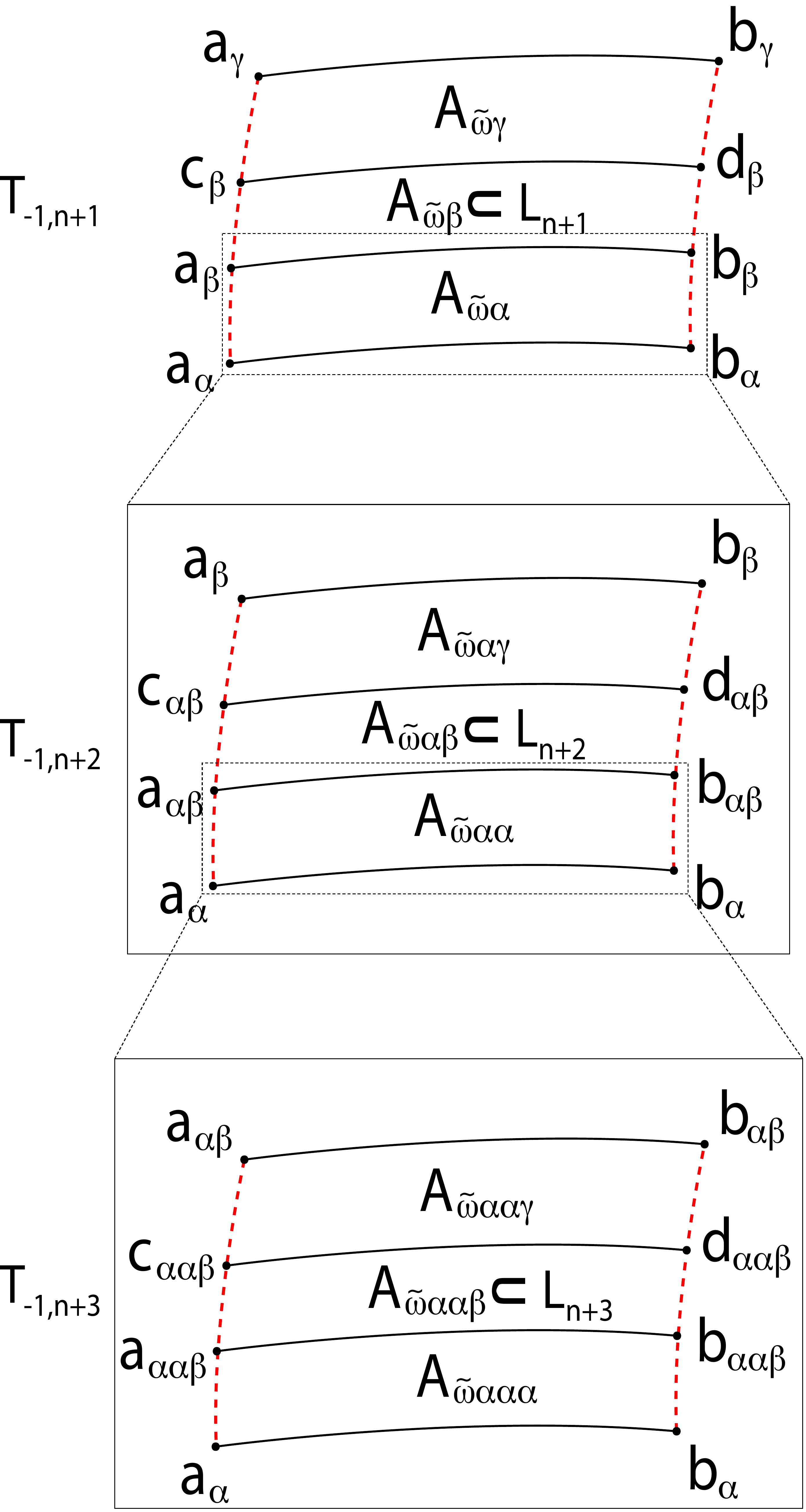}}
 \caption{(Schematic, color online) Successive partitions of $A_{\tilde{\omega}} \subset T_{-1,n}$ in later trellis $T_{-1,n+m}$. Upper panel ($T_{-1,n+1}$): the same as Fig.~\ref{fig:Cell_Area_Partition}, where $A_{\tilde{\omega}}$ is partitioned into three areas by $L_{n+1}$. Middle panel ($T_{-1,n+2}$): zoomed-in graph of $A_{\tilde{\omega} \alpha}$ in $T_{-1,n+2}$, where $A_{\tilde{\omega} \alpha}$ is partitioned by $L_{n+2}$ into three new areas. Lower panel ($T_{-1,n+3}$): zoomed-in graph of $A_{\tilde{\omega} \alpha \alpha}$ in $T_{-1,n+3}$, where $A_{\tilde{\omega} \alpha \alpha}$ is partitioned by $L_{n+3}$ into three new areas. The addition of successive lobes create four families of homoclinic points, $[a_{\alpha^{m-1}\beta}]$, $[c_{\alpha^{m-1}\beta}]$, $[b_{\alpha^{m-1}\beta}]$, and $[d_{\alpha^{m-1}\beta}]$, that accumulate on $a_{\alpha}$ and $b_{\alpha}$ under Eqs.~\eqref{eq:Cell area partition point accumulation 1} and \eqref{eq:Cell area partition point accumulation 2} with exponent $\mu_0$. Therefore, the three families of areas $[A_{\tilde{\omega} \alpha^{m}}]$, $[ A_{\tilde{\omega} \alpha^{m-1} \beta}]$, and $[A_{\tilde{\omega} \alpha^{m-1} \gamma}]$ also converge onto $U[a_{\alpha},b_{\alpha}]$ with exponent $\mu_0$, as described by Eq.~\eqref{eq:Fast scaling}. }
\label{fig:Cell_Area_Partition_Successive}
\end{figure}

The exponents in Eqs.~\eqref{eq:Area scaling relation} and \eqref{eq:Fast scaling} are identical, and this is not a coincidence. Returning to Sec.~\ref{Asymptotic accumulation}, the three families of areas $[{\cal A}^{\circ}_{SUSU[g_{-2}, b^{(n)}, d^{(n)}, h_{-1}]}]$, $[{\cal A}^{\circ}_{SUSU[b^{(n)}, a^{(n)}, c^{(n)}, d^{(n)}]}]$, and $[{\cal A}^{\circ}_{SUSU[a^{(n)}, b^{(n-1)}, d^{(n-1)}, c^{(n)}]}]$ ($n \geq 2$), are just $[B_{\tilde{\omega} \alpha^{n}}]$, $[ B_{\tilde{\omega} \alpha^{n-1} \beta}]$, and $[B_{\tilde{\omega} \alpha^{n-1} \gamma}]$ ($n \geq 2$), respectively, upon letting $\tilde{\omega}=\emptyset$ (null string). Therefore, Eq.~\eqref{eq:Area scaling relation} is just a special case of Eq.~\eqref{eq:Fast scaling}. In fact, just as Eq.~\eqref{eq:Area scaling relation} is a direct consequence of the accumulation relations in Eq.~\eqref{eq:Accumulation along stable} and \eqref{eq:Accumulation along stable 2}, the general formula Eq.~\eqref{eq:Fast scaling} also comes from the accumulation of corresponding homoclinic points at the vertices of the cells. This can be demonstrated by Fig.~\ref{fig:Cell_Area_Partition_Successive}, 
where three families of areas $[A_{\tilde{\omega} \alpha^{n}}]$, $[ A_{\tilde{\omega} \alpha^{n-1} \beta}]$, and $[A_{\tilde{\omega} \alpha^{n-1} \gamma}]$ ($n \geq 2$) accumulate on $U[a_{\alpha},b_{\alpha}]$. Starting from the $A_{\tilde{\omega}}$ cell in $T_{-1,n}$ and mapping to higher iterations, the addition of $L_{n+m}$ ($m=1,2,\cdots$) partitions $A_{\tilde{\omega} \alpha^{m-1} }$ into three new areas: $A_{\tilde{\omega} \alpha^{m} }$, $A_{\tilde{\omega} \alpha^{m-1} \beta}$, and $A_{\tilde{\omega} \alpha^{m-1} \gamma}$, which approach the $U[a_{\alpha},b_{\alpha}]$ segment asymptotically. The two sequences of points $[a_{\alpha^{m-1}\beta}]$ and $[c_{\alpha^{m-1}\beta}]$ ($m \geq 1$), which are created from successive intersections between $U_{n+m}$ and $S[a_{\alpha},a_{\gamma}]$, give rise to two families of points that accumulate on the base point $a_{\alpha}$:
\begin{equation}\label{eq:Cell area partition point accumulation 1}
a_{\alpha^{m-1}\beta},\ c_{\alpha^{m-1}\beta} \xhookrightarrow[S]{k+m-1} a_{\alpha} 
\end{equation} 
with exponent $\mu_0$, where $k$ depends on the detailed form of $\tilde{\omega}$. 

Similarly, the two sequences of points $[b_{\alpha^{m-1}\beta}]$ and $[d_{\alpha^{m-1}\beta}]$ ($m \geq 1$), generated from successive intersections between $U_{n+m}$ and $S[b_{\alpha},b_{\gamma}]$, give rise to two families of points that accumulate on the base point $b_{\alpha}$:
\begin{equation}\label{eq:Cell area partition point accumulation 2}
b_{\alpha^{m-1}\beta},\ d_{\alpha^{m-1}\beta} \xhookrightarrow[S]{k+m-1} b_{\alpha} 
\end{equation} 
with the same exponent $\mu_0$ as well. 

The scaling relations for the cell areas in Eq.~\eqref{eq:Fast scaling} come from the scaling relations of their vertices in Eqs.~\eqref{eq:Cell area partition point accumulation 1} and \eqref{eq:Cell area partition point accumulation 2}. In particular, denote the length of the stable manifold segment $S[a,b]$ by $d_s(a,b)$, then the lengths $d_s(a_{\alpha},a_{\alpha^{m-1}\beta})$, $d_s(a_{\alpha^{m-1}\beta},c_{\alpha^{m-1}\beta})$, and $d_s(c_{\alpha^{m-1}\beta},a_{\alpha^{m-2}\beta})$ scales as (see Fig.~\ref{fig:Cell_Area_Partition_Successive})
\begin{equation}\label{eq:Lengths along stable scaling under mu_0}
\begin{split}
&\lim_{m \to \infty} \frac{d_s(a_{\alpha},a_{\alpha^{m-1}\beta})}{d_s(a_{\alpha},a_{\alpha^{m}\beta})}=\lim_{m\to\infty}\frac{d_s(a_{\alpha^{m-1}\beta},c_{\alpha^{m-1}\beta})}{d_s(a_{\alpha^{m}\beta},c_{\alpha^{m}\beta})}\\
&=\lim_{m \to \infty} \frac{d_s(c_{\alpha^{m-1}\beta},a_{\alpha^{m-2}\beta})}{d_s(c_{\alpha^{m}\beta},a_{\alpha^{m-1}\beta})} = e^{\mu_0}\ .
\end{split}
\end{equation}

Considering that the points in Eq.~\eqref{eq:Lengths along stable scaling under mu_0} are infinitely close under the $m\to\infty$ limit, so the stable manifold segments connecting them are infinitely close to straight-line segments, the distances between homoclinic points can be replaced by the differences in their $p$ (or $q$) coordinates (assuming the generic cases in which the local manifolds do not form caustics):
\begin{equation}\label{eq:Coordinate scaling under mu_0}
\begin{split}
& \lim_{m\to\infty} \frac{p(a_{\alpha^{m-1}\beta})-p(a_{\alpha}) } {p(a_{\alpha^{m}\beta})-p(a_{\alpha})} = \lim_{m\to\infty} \frac{p(c_{\alpha^{m-1}\beta})-p(a_{\alpha^{m-1}\beta}) } {p(c_{\alpha^{m}\beta}) - p(a_{\alpha^{m}\beta}) } \\
& = \lim_{m\to\infty} \frac{p(a_{\alpha^{m-2}\beta})-p(c_{\alpha^{m-1}\beta}) } {p(a_{\alpha^{m-1}\beta}) - p(c_{\alpha^{m}\beta}) } = e^{\mu_0}
\end{split}
\end{equation}
where $p(a)$ denotes the $p$-coordinate value of $a$. The same relations hold for the $q$-coordinate values as well. The leading terms of the homoclinic families in Eq.~\eqref{eq:Coordinate scaling under mu_0} are shown in Fig.~\ref{fig:Cell_Area_Partition_Successive}. 

Thus, the asymptotic area scaling relations originate from the asymptotic relations between the positions of homoclinic points on the invariant manifolds. Furthermore, the scaling relations between the phase-space positions of certain homoclinic points give rise to the stability exponent of the fixed point $x$. In fact, the same relations exist for the stability exponent of any unstable periodic orbit in general~\cite{Li19b}.

As an example of Eq.~\eqref{eq:Fast scaling}, the three families of areas $[{\cal A}^{\circ}_{SUSU[g_{-2}, b^{(n)}, d^{(n)}, h_{-1}]}]$, $[{\cal A}^{\circ}_{SUSU[b^{(n)}, a^{(n)}, c^{(n)}, d^{(n)}]}]$, and $[{\cal A}^{\circ}_{SUSU[a^{(n)}, b^{(n-1)}, d^{(n-1)}, c^{(n)}]}]$ ($n \geq 2$) from Eq.~\eqref{eq:Area scaling relation} can be identified as $[B_{\tilde{\omega} \alpha^n }]$, $[B_{\tilde{\omega} \alpha^{n-1} \beta }]$, and $[B_{\tilde{\omega} \alpha^{n-1} \gamma }]$, respectively, by letting $\tilde{\omega}$ be an empty string.  Comparing the areas in Fig.~\ref{fig:Scaling} and Fig.~\ref{fig:Trellis_Partition_2}, the leading terms in the tree families are identified as ${\cal A}^{\circ}_{SUSU[g_{-2}, b^{(2)}, d^{(2)}, h_{-1}]}=B_{\alpha \alpha}$, ${\cal A}^{\circ}_{SUSU[b^{(2)}, a^{(2)}, c^{(2)}, d^{(2)}]}=B_{\alpha \beta}$, and ${\cal A}^{\circ}_{SUSU[a^{(2)}, b^{(1)}, d^{(1)}, c^{(2)}]}=B_{\alpha \gamma}$. Although not plotted in the figure, future lobes partition $B_{\alpha \alpha}$ into every-finer areas and create the three infinite families of areas that converge to the bottom segment $U[g_{-2},h_{-1}]$.  

To check the accuracy of Eq.~\eqref{eq:Fast scaling}, the first seven areas of the three families $ [A_{\alpha^n}] $, $[ A_{\alpha^{n-1} \beta} ]$, and $[ A_{\alpha^{n-1} \gamma} ]$ are given in Table~\ref{tab:Fast_scaling_numerics}.  The three columns give the scaling exponents obtained from $[A_{\alpha}, A_{\alpha \alpha}, A_{\alpha \alpha \alpha}, \cdots]$, $[ A_{\beta}, A_{\alpha \beta}, A_{\alpha \alpha \beta},\cdots ]$, and $[ A_{\gamma}, A_{\alpha \gamma}, A_{\alpha \alpha \gamma},\cdots ]$, respectively. 
\begin{table}[h!]
  \begin{center}
    \begin{tabular}{l|c|c|c}
      \textbf{n} & \boldmath$\log( \frac { A_{\alpha^n} }{ A_{\alpha^{n+1}} } )$  & \boldmath$\log( \frac {A_{\alpha^{n-1} \beta} }{ A_{\alpha^{n} \beta} } )$ & \boldmath$\log( \frac {A_{\alpha^{n-1} \gamma} }{ A_{\alpha^{n} \gamma} } )$  \\
      \hline
      1 & 2.144099 &  2.103342  & 2.197343  \\
      2 & 2.142725 &  2.142323  & 2.156467   \\
      3 & 2.142084 & 2.142521  & 2.144631    \\
      4 & 2.141952 &  2.142060  & 2.142364    \\
      5 & 2.141929 &  2.141949  & 2.141991    \\
      6 & 2.141927 &  2.141929  &  2.141933   \\
      \hline
      \boldmath$\mu_0$ &  2.141926 &  2.141926 & 2.141926\\
    \end{tabular}
    \caption{$e^{\mu_{0}}$ scaling: The scaling exponents in the ``$\alpha$"-direction starting from $A_{\alpha}$, $A_{\beta}$, and $A_{\gamma}$, are listed in the three columns, respectively. Clearly, they all converge to $\mu_{0}$ asymptotically.}
    \label{tab:Fast_scaling_numerics}
  \end{center}
\end{table}
Even for the first ratio (worst case), the predicted exponent is good to better than two decimal places.  By the bottom of each column, the distinction first appears only in the sixth digit.

\begin{table}[h!]
  \begin{center}
    \begin{tabular}{l|c|c|c}
      \textbf{n} & \boldmath$\log( \frac { A_{\gamma^{n-1} \alpha} }{ A_{\gamma^{n} \alpha} } )$  & \boldmath$\log( \frac {A_{\gamma^{n-1} \beta} }{ A_{\gamma^{n} \beta} } )$ & \boldmath$\log( \frac {A_{\gamma^{n} } }{ A_{\gamma^{n+1} } } )$  \\
      \hline
      1 & 1.320085   & 2.365152  &  1.468471   \\
      2 & 1.707766  &  1.384612  &  1.446403   \\
      3 & 1.343392 &  1.460855  &  1.500372   \\
      4 & 1.535619  &  1.496668  &  1.477362   \\
      5 & 1.467206  &  1.478053  &  1.484760   \\
      6 & 1.487618  & 1.484611   &  1.482549   \\
      7 & 1.481780  &  1.482579  &  1.483168   \\
      8 & 1.483367  &  1.483164  &  1.482999   \\
      \hline
      \boldmath$\mu_1$ &  1.483036 &  1.483036  & 1.483036 \\
    \end{tabular}
    \caption{$e^{\mu_{1}}$ scaling: The scaling exponents in the ``$\gamma$"-direction starting from $A_{\alpha}$, $A_{\beta}$, and $A_{\gamma}$, are listed in the three columns, respectively. Clearly, they all converge to $\mu_{1}$, the stability exponent of the periodic orbit $\overline{1}$.}
    \label{tab:Slow_scaling_numerics}
  \end{center}
\end{table}

The opposite direction down the tree follows increasing repetitions of $\gamma$ leading to the families, $[A_{\tilde{\omega} \gamma^{n-1} \alpha}]$, $[A_{\tilde{\omega} \gamma^{n-1} \beta}]$, and $[A_{\tilde{\omega} \gamma^{n} }]$ ($n \geq 1$), respectively. The exponential shrinking rate is much slower, and numerical evidence with specific families of cells shown in Tables~\ref{tab:Slow_scaling_numerics} and \ref{tab:Slow_scaling_numerics_more} indicate that the scaling along such ``$\gamma$" directions converge to the stability exponent $\mu_{1}$ of $x^{\prime}$, i.e., the hyperbolic fixed point with reflection:
\begin{equation}
\label{eq:Slow scaling}
 \lim_{n \to \infty} \frac{A_{\tilde{\omega} \gamma^{n-1} \alpha} }{A_{\tilde{\omega} \gamma^{n} \alpha}} =  \lim_{n \to \infty} \frac{A_{\tilde{\omega} \gamma^{n-1} \beta}}{ A_{\tilde{\omega} \gamma^{n} \beta} }= \lim_{n \to \infty} \frac{A_{\tilde{\omega} \gamma^{n} } }{A_{\tilde{\omega} \gamma^{n+1} }} = e^{\mu_1}
\end{equation}
which is in complete analogy to Eq.~\eqref{eq:Fast scaling}, except for a different direction along the tree, and with a different scaling exponent. 

\begin{table}[h!]
  \begin{center}
    \begin{tabular}{l|c|c|c}
      \textbf{n} & \boldmath$\log( \frac { A_{\alpha \gamma^{n-1} \alpha} }{ A_{\alpha \gamma^{n} \alpha} } )$  & \boldmath$\log( \frac {A_{\alpha \gamma^{n-1} \beta} }{ A_{\alpha \gamma^{n} \beta} } )$ & \boldmath$\log( \frac {A_{\alpha \gamma^{n} } }{ A_{\alpha \gamma^{n+1} } } )$  \\
      \hline
      1 &  1.364533  & 2.471588  &  1.449553   \\
      2 & 1.703491  & 1.352048   &  1.444654   \\
      3 & 1.332763  & 1.460781   & 1.502057    \\
      4 & 1.541193  & 1.497815   &  1.476780   \\
      5 & 1.465512  &  1.477561  &  1.484950   \\
      6 & 1.488134  & 1.484780  & 1.482495   \\
      7 & 1.481634  & 1.482527   &  1.483189   \\
      \hline
      \boldmath$\mu_1$ &  1.483036 &  1.483036  & 1.483036 \\
    \end{tabular}
    \caption{$e^{\mu_{1}}$ scaling: The scaling exponents in the ``$\gamma$"-direction starting from $A_{\alpha \alpha}$, $A_{\alpha \beta}$, and $A_{\alpha \gamma}$, are listed in the three columns, respectively. They all converge to $\mu_{1}$.}
    \label{tab:Slow_scaling_numerics_more}
  \end{center}
\end{table}

The above tables indicate that the scaling of cells along consecutive ``$\alpha$" directions yield the exponent $\mu_0$, and cells along consecutive ``$\gamma$" directions yield the exponent $\mu_1$. Such phenomena are still just special cases of a general relation that links the scaling exponents along different directions to the symbolic codes of periodic orbits. The association is simple: a scaling step in the ``$\alpha$"-direction contributes a symbolic digit ``$0$", and a scaling step in the ``$\gamma$"-direction contributes a digit ``$1$".  To formulate this process, define a mapping $\Psi$ that maps a string of Greek letters ``$\alpha$'' and ``$\gamma$" to a string of symbolic codes of ``$0$" and ``$1$", with the grammar $\alpha \mapsto 0$ and $\gamma \mapsto 1$. For example, $\Psi(\gamma \alpha \gamma)=101$, and the asymptotic scaling exponent in successive ``$\gamma \alpha \gamma$"-directions is the stability exponent of the $\overline{101}$ periodic orbit, $\mu_{101}$.  

In the most general case, consider beginning with an arbitrary node (denoted by either $A_{\tilde{\omega} \alpha}$, $A_{\tilde{\omega} \beta}$, or $A_{\tilde{\omega} \gamma}$, depending on its location) in the type-I partition tree, and study the scaling exponent in an arbitrary direction $\tilde{\eta}$ deepening along the tree. Here $\tilde{\eta}$ is a Greek letter string composed by ``$\alpha$"s and ``$\gamma$"s that specifies the scaling path. The scaling exponent along $\tilde{\eta}$ is determined by the stability exponent of the periodic orbit $\overline{\Psi( \tilde{\eta} )}$, $\mu_{\Psi(\tilde{\eta})}$:
 \begin{equation}
\label{eq:General scaling}
\begin{split}
&\lim_{n \to \infty} \frac{A_{\tilde{\omega} \tilde{\eta}^{n-1} \alpha} }{A_{\tilde{\omega} \tilde{\eta}^{n} \alpha}} =  \lim_{n \to \infty} \frac{A_{\tilde{\omega} \tilde{\eta}^{n-1} \beta}}{ A_{\tilde{\omega} \tilde{\eta}^{n} \beta} }= \lim_{n \to \infty} \frac{A_{\tilde{\omega} \tilde{\eta}^{n-1} \gamma}}{ A_{\tilde{\omega} \tilde{\eta}^{n} \gamma} } \\
&= e^{\mu_{\Psi(\tilde{\eta})}} 
\end{split}
\end{equation}
which is in complete analogy to Eqs.~\eqref{eq:Fast scaling} and \eqref{eq:Slow scaling}. Notice the relations are independent of $\tilde{\omega}$, i.e., any node of the tree can be used as a starting node (the $n=1$ terms) of the scaling. Identical relations hold for $B$ cells in the type-II partition tree as well. 

\begin{table}[h!]
  \begin{center}
    \begin{tabular}{l|c|c|c}
      \textbf{n} & \boldmath$\log( \frac { A_{ \alpha (\alpha\gamma)^{n-1}  \alpha } }{ A_{ \alpha (\alpha\gamma)^{n}  \alpha } } )$  & \boldmath$\log( \frac { A_{ \alpha (\alpha\gamma)^{n-1} \beta } }{ A_{ \alpha (\alpha\gamma)^{n} \beta } } )$ & \boldmath$\log( \frac { A_{\alpha (\alpha\gamma)^{n-1} \gamma} }{ A_{\alpha (\alpha\gamma)^{n} \gamma} } )$   \\
      \hline
      1  &  3.520098  & 4.629501 &  3.603747   \\ 
      2 & 3.226675  &3.202485 & 3.292394   \\
      3 & 3.259026   &3.255664  & 3.248603   \\
      4 & 3.256531  &3.256733 & 3.257234   \\     
      \hline
      \boldmath$\mu_{01}$ &  3.256614 &  3.256614 &  3.256614 \\
    \end{tabular}
    \caption{$e^{\mu_{01}}$ scaling: The scaling exponents in the ``$\alpha\gamma$"-direction starting from $A_{\alpha \alpha}$, $A_{\alpha \beta}$, and $A_{\alpha \gamma}$, are listed in the three columns, respectively. Clearly, they all converge to $\mu_{01}$, the stability exponent of the periodic orbit $\overline{01}$. }
    \label{tab:01_scaling_numerics}
  \end{center}
\end{table}

Similar to Eq.~\eqref{eq:Fast scaling}, the origin of Eq.~\eqref{eq:General scaling} comes from a fundamental relation linking the stability exponents of unstable periodic orbits to the distribution of certain families of homoclinic points (which can be identified as the vertices of the cell areas) on the invariant manifolds. 

\section{Homoclinic action formulas}
\label{Homoclinic action formulae}

All the tools are now in place to develop exact relations expressing the classical actions of any homoclinic orbit in $T_{-1,N}$ (therefore up to transition time $N+1$), in terms of the type-I and type-II cell areas of $T_{-1,N}$.  In this method, the calculation of numerical orbits, which suffers from sensitive dependence on initial errors and unstable in nature, are converted into the calculation of areas bounded by $S(x)$ and $U(x)$, which can be evaluated in stable ways. The exact relations of Sec.~\ref{Exact decomposition} are perfectly adapted for the development of approximations in Sec.~\ref{Information reduction} that make use of the asymptotic scaling relations among the areas, and that leads to approximate expressions for the homoclinic orbit actions in $T_{-1,N}$ using only the type-I and type-II cell areas from $T_{-1,d(N)}$, where $d(N)$ is an integer much smaller than $N$. Consequently, it is possible to express the exponentially increasing set of homoclinic orbit actions using a set of areas that is increasing at a much slower rate (e.g., algebraic or linear).  

\subsection{Projection operations}
\label{Projection operations}

The main process leading to the homoclinic action formulas in this section is to express the actions of the homoclinic orbits with large winding numbers in terms of those with small winding numbers, i.e., the decomposition of orbits according to their hierarchical structure. To accomplish this, there are some projection operations to be defined which establish mappings between orbits with different winding numbers. 

Given a winding-$n$ ($n \geq 1$) homoclinic point $y$ and two winding-$(n+1)$ points $z$ and $w$ such that $z \xhookrightarrow[S]{k} y$ and $w \xhookrightarrow[S]{k} y$ ($\forall k \geq 1$) and $S[y,w] \subset S[y,z]$, define the \textit{projection operation along the stable manifold}, denoted by $P_S$, to be the mapping that maps $z$ and $w$ into the base point $y$:
\begin{equation}\label{eq:P_S definition}
 P_S (z) = P_S (w) = y.
\end{equation}

The corresponding operation on the symbolic strings, denoted by $\pi_S$, can be readily obtained by working backward from Eq.~\eqref{eq:Orbit symbolic codes general rules}. Namely, given the symbolic codes of $z$ and $w$, the $\pi_S$ operation deletes the substrings ``$110^{k-1}$" and ``$100^{k-1}$", respectively, from the left ends of the cores of $z$ and $w$, while maintaining the position of the decimal point relative to the right end of the core. The resulting symbolic code is then $y$. Take the points $a^{(0)} \Rightarrow \overline{0} 1.11 \overline{0}$, $b^{(0)} \Rightarrow \overline{0} 1.01 \overline{0} $ and $g_{-2} \Rightarrow \overline{0} .01 \overline{0}$ in Fig.~\ref{fig:Homoclinic_Tangle} as examples, we know $a^{(0)},b^{(0)} \xhookrightarrow[S]{1} g_{-2}$, thus $P_S(a^{(0)})=P_S(b^{(0)})=g_{-2}$. Correspondingly for the symbolic codes
\begin{equation}\label{eq:pi_S definition special cases}
\begin{split}
& \pi_S ( \overline{0} 1.11 \overline{0} ) =  \overline{0} .01 \overline{0} \\
& \pi_S ( \overline{0} 1.01 \overline{0} ) =  \overline{0} .01 \overline{0} 
\end{split}
\end{equation}
where the $\pi_S$ operation deletes either the ``$11$" (for $a^{(0)}$) or ``$10$" (for $b^{(0)}$) substring from the left of the cores while keeping the position of the decimal points relative to the right end of the core unchanged. 

Similar operations can be defined for the accumulating homoclinic families along the unstable manifold under the inverse mappings as well. Given a winding-$n$ homoclinic point $y^{\prime} $, and the winding-$(n+1)$ points $z^{\prime}$ and $w^{\prime}$ such that $z^{\prime} \xhookrightarrow[U]{k} y^{\prime}$ and $w^{\prime} \xhookrightarrow[U]{k} y^{\prime}$ and $U[y^{\prime},w^{\prime}] \subset U[y^{\prime},z^{\prime}]$, define the \textit{projection operation along the unstable manifold}, denoted by $P_U$, to be the mapping:
\begin{equation}\label{eq:P_U definition}
 P_U (z^{\prime}) = P_U (w^{\prime}) = y^{\prime}.
\end{equation}
The corresponding operation $\pi_U$ on the symbolic codes is then defined by working backward from Eq.~\eqref{eq:Orbit symbolic codes general rules inverse mapping}. Namely, given the symbolic codes of $z^{\prime}$ and $w^{\prime}$, the $\pi_U$ operation deletes the substrings ``$0^{k-1}11$" and ``$0^{k-1}01$", respectively, from the right ends of the cores of $z^{\prime}$ and $w^{\prime}$, while maintaining the position of the decimal point relative to the left end of the core. The resulting symbolic code then gives $y^{\prime}$. 

In the preceding definitions, the projection operations must be applied to homoclinic points with winding numbers $\geq 2$. However, they can be naturally extended to apply to the primary (winding-$1$) points as well. The extension is straightforward: for any primary homoclinic point $g_i$ or $h_i $, define
\begin{equation}\label{eq:P_S P_U definition on primary points}
P_S(g_i)=P_U(g_i)=P_S(h_i)=P_U(h_i)=x  
\end{equation}
with corresponding $\pi_S$ and $\pi_U$ operations mapping the symbolic codes of $h_i$ and $g_i$ into $\overline{0}.\overline{0}$, i.e., that of the hyperbolic fixed point $x$. This is consistent with the scaling relations of Eqs.~\eqref{eq:Scaling relation for primary points along stable} and \eqref{eq:Scaling relation for primary points along unstable} as well.

Since $\pi_S$ and $\pi_U$ operate on different sides of the cores, it is easy to see that they commute: $\pi_S \pi_U = \pi_U \pi_S$.  Since the symbolic codes are in one-to-one correspondences with the phase space points, the projection operations $P_S$ and $P_U$ also commute: $P_S P_U =P_U P_S$. Therefore, a mixed string of operations consisting of $n$ applications of $P_S$ and $m$ applications of $P_U$, disregarding their relative orders, can always be written as $P_S ^{n} P_U^{m}$, and similarly for the mixed string of operations of $\pi_S$ and $\pi_U$ as well. Such operations are extensively used in the decomposition scheme in Sec.~\ref{Exact decomposition}.

As an example, consider the $c^{(1)} \Rightarrow \overline{0} 11.11 \overline{0}$, $h_{-1} \Rightarrow \overline{0} .11 \overline{0} $, and $h_1 \Rightarrow \overline{0} 11. \overline{0} $ points from Fig.~\ref{fig:Decomposition_Example}. The accumulation relations are $c^{(1)} \xhookrightarrow[S]{1} h_{-1} \xhookrightarrow[U]{} x$ and $c^{(1)} \xhookrightarrow[U]{1} h_{1} \xhookrightarrow[S]{} x$, thus $P_U P_S (c^{(1)})=P_U (h_{-1})=x $ and $P_S P_U (c^{(1)})=P_S (h_{1})=x$. On the other hand, using the symbolic dynamics we have $\pi_U \pi_S (\overline{0} 11.11 \overline{0}) = \pi_U (\overline{0} .11 \overline{0}) = \overline{0} .\overline{0}$ and $\pi_S \pi_U (\overline{0} 11.11 \overline{0}) = \pi_S (\overline{0} 11. \overline{0}) = \overline{0} .\overline{0}$, consistent with the results from the accumulation relations. 

\subsection{Exact decomposition}
\label{Exact decomposition}

The derivation of the exact formula makes repeated use of the MacKay-Meiss-Percival action principle described by Eqs.~\eqref{eq:Area-action homoclinic pair} and \eqref{eq:Area-action homoclinic orbit and fixed point orbit}, and expresses the relative classical actions of homoclinic orbits as sums of phase-space areas bounded by $S(x)$ and $U(x)$.  The fixed-point orbit $\lbrace x \rbrace$ becomes a natural candidate for a reference orbit, and the actions of all homoclinic orbits $\lbrace h \rbrace$ can be expressed relative to $\lbrace x \rbrace$ in the form of $\Delta {\cal F}_{\lbrace h \rbrace \lbrace x \rbrace}$, as shown by Eq.~\eqref{eq:Area-action homoclinic orbit and fixed point orbit}.  

Start by calculating the actions of the two primary orbits $\lbrace g_0 \rbrace$ and $\lbrace h_0 \rbrace$, which readily follow from Eq.~\eqref{eq:Area-action homoclinic orbit and fixed point orbit}.  The two areas ${\cal A}^{\circ}_{US[x,h_0]}$ and ${\cal A}^{\circ}_{US[x,g_0]}$ are straightforward to evaluate since only short segments of $S(x)$ and $U(x)$ are required.  Having the primary relative orbit actions available, the actions of all winding-$n$ orbits ($n \geq 2$) can be determined recursively from the actions of the winding- ($n-1$) and winding-($n-2$) orbits. In particular, given any winding-$n$ ($n \geq 2$) homoclinic point $y \in S^{\prime}_{-1} \cap U_{m}$, the action of $\lbrace y \rbrace$ can be expressed using three auxiliary orbits: $\lbrace P_S(y) \rbrace$, $\lbrace P_U(y) \rbrace$, and $\lbrace P_S P_U (y) \rbrace$. Substituting $\lbrace y \rbrace$, $\lbrace P_U (y) \rbrace$, $\lbrace P_S (y) \rbrace$, and $\lbrace P_S P_U (y) \rbrace$ into Eq.~\eqref{eq:Area-action two homoclinic pairs} gives
\begin{equation}\label{eq:Action formulae once reduction action difference}
\begin{split}
&( \Delta {\cal F}_{\lbrace y \rbrace \lbrace x \rbrace} - \Delta {\cal F}_{\lbrace P_U(y) \rbrace \lbrace x \rbrace} ) \\
&\qquad  - ( \Delta {\cal F}_{\lbrace P_S(y) \rbrace \lbrace x \rbrace} - \Delta {\cal F}_{\lbrace P_S P_U(y) \rbrace \lbrace x \rbrace} )\\
&\qquad \qquad = {\cal A}^{\circ}_{SUSU[ y, P_S(y), P_S P_U(y), P_U(y) ]}
\end{split}
\end{equation}
and therefore
\begin{equation}\label{eq:Action formulae once reduction}
\begin{split}
& \Delta {\cal F}_{\lbrace y \rbrace \lbrace x \rbrace} = \Delta {\cal F}_{\lbrace P_S(y) \rbrace \lbrace x \rbrace} + \Delta {\cal F}_{\lbrace P_U(y) \rbrace \lbrace x \rbrace}\\
& - \Delta {\cal F}_{\lbrace P_S P_U(y) \rbrace \lbrace x \rbrace} + {\cal A}^{\circ}_{SUSU[ y, P_S(y), P_S P_U(y), P_U(y) ]}\ .
\end{split}
\end{equation}
Notice that the $P_S$ and $P_U$ operations reduce the winding number of $y$ by $1$.   Similarly, from Eqs.~\eqref{eq:Orbit symbolic codes general rules} and \eqref{eq:Orbit symbolic codes general rules inverse mapping} the core length is reduced by at least $2$, since their effect is to delete substrings of a minimum of two digits from the original core (``$110^{k-1}$" or ``$100^{k-1}$" for $P_S$, ``$0^{k-1}11$" or ``$0^{k-1}01$" for $P_U$). Therefore, the three auxiliary orbits are guaranteed to have simpler and shorter phase-space excursions than $\lbrace y \rbrace$. In this sense, Eq.~\eqref{eq:Action formulae once reduction} provides a decomposition of the relative action of any arbitrary homoclinic orbit into the relative actions of three simpler auxiliary homoclinic orbits, plus a phase-space area bounded by the manifolds. By repeated contractions, the decomposition could be pushed to involving only the primary homoclinic orbits, the fixed point, and a set of ${\cal A}^{\circ}_{SUSU[\cdots]}$ areas. Implied by this process is that the inverse sequences could be used beginning with the two primary homoclinic orbits, fixed point, and a set of areas to construct the relative actions of all the homoclinic orbits.

\begin{figure}[ht]
\centering
{\includegraphics[width=6.5cm]{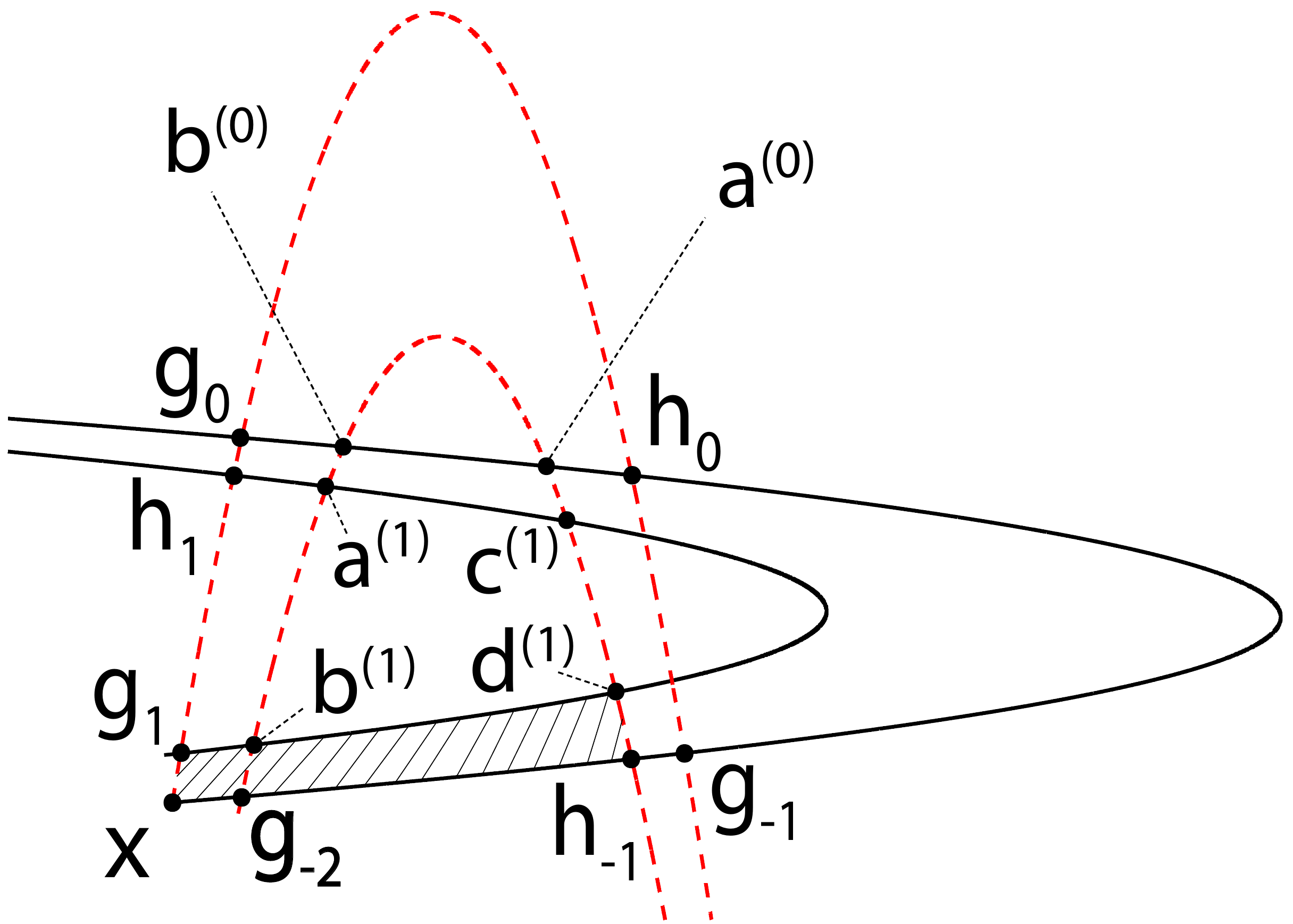}}
 \caption{(Color online) An example of the homoclinic orbit action decomposition. As shown by Eq.~\eqref{eq:Action formulae once reduction example with d1}, the relative action of the winding-$2$ orbit $\lbrace d^{(1)} \rbrace$ is decomposed into the sum of the relative actions of the winding-$1$ orbits $\lbrace h_{-1} \rbrace$ and $\lbrace g_1 \rbrace$, and a phase-space area ${\cal A}^{\circ}(d^{(1)})={\cal A}^{\circ}_{SUSU[ d^{(1)}, h_{-1}, x, g_1 ]}$ marked by the hatched region in the figure. Similar decomposition can be done for any homoclinic point on $S^{\prime}_{-1}$.}
\label{fig:Decomposition_Example}
\end{figure}  

The particular form of ${\cal A}^{\circ}_{SUSU[ y, P_S(y), P_S P_U(y), P_U(y) ]}$ indicates that the area depends only on the homoclinic point $y$. Once $y$ is chosen, the uniqueness of $P_S (y)$, $P_U(y)$, and $P_S P_U (y)$ means that the area is uniquely calculated. Thus, in the forthcoming contents the short-handed notation
\begin{equation}\label{eq:A_SUSU area short notation}
{\cal A}^{\circ}(y) \equiv {\cal A}^{\circ}_{SUSU[ y, P_S(y), P_S P_U(y), P_U(y) ]}
\end{equation}  
will be used frequently to simplify the notation. 

An important outcome, buried in Eq.~\eqref{eq:Action formulae once reduction}, relates to the particular form of ${\cal A}^{\circ}(y)$.  For any $y \in S^{\prime}_{-1} \cap U_m$, the locations of its projections are highly constrained: $P_S(y) \in S^{\prime}_{-1}$, $P_U(y) \in S[x, g_0]$, and $P_S P_U (y) \in S[x,g_0]$.  As a consequence, ${\cal A}^{\circ}(y)$ is always expressible by the type-I and -II cell areas of $T_{-1,m}$. Consider $d^{(1)} \in ( S^{\prime}_{-1} \cap U_1 )$ from Fig.~\ref{fig:Decomposition_Example} for example, the use of Eq.~\eqref{eq:Action formulae once reduction} yields:
\begin{equation}
\label{eq:Action formulae once reduction example with d1}
\begin{split}
\Delta {\cal F}_{\lbrace d^{(1)} \rbrace \lbrace x \rbrace} & = \Delta {\cal F}_{\lbrace h_{-1} \rbrace \lbrace x \rbrace} + \Delta {\cal F}_{\lbrace g_1 \rbrace \lbrace x \rbrace}\\
& - \Delta {\cal F}_{\lbrace x \rbrace \lbrace x \rbrace} + {\cal A}^{\circ}(d^{(1)})
\end{split}
\end{equation} 
where $\Delta {\cal F}_{\lbrace x \rbrace \lbrace x \rbrace}=0$ gives zero contributions. Comparing Fig.~\ref{fig:Decomposition_Example} with Fig.~\ref{fig:Trellis_Partition_1}, the ${\cal A}^{\circ}(d^{(1)})$ term (hatched region in Fig.~\ref{fig:Decomposition_Example}) is expressible by two cell areas from the type-I and type-II partition trees of $T_{-1,1}$:
\begin{equation}\label{eq:Action formulae once reduction  example with c1 area expression}
{\cal A}^{\circ}(d^{(1)}) = A_{\alpha} + B_{\alpha} 
\end{equation}
both of which are finite curvy trapezoids bounded by the manifolds that can be evaluated simply. The same results hold for all homoclinic points on $S^{\prime}_{-1}$ with a single exception---$a^{(0)}$. The use of Eq.~\eqref{eq:Action formulae once reduction} on $a^{(0)}$ gives
\begin{equation}\label{eq:Action formulae once reduction example with a0}
\Delta {\cal F}_{\lbrace a^{(0)} \rbrace \lbrace x \rbrace} = \Delta {\cal F}_{\lbrace g_{-2} \rbrace \lbrace x \rbrace} + \Delta {\cal F}_{\lbrace g_0 \rbrace \lbrace x \rbrace}+ {\cal A}^{\circ}(a^{(0)}) \nonumber
\end{equation} 
where the evaluation of ${\cal A}^{\circ}(a^{(0)})={\cal A}^{\circ}_{SUSU[ a^{(0)}, g_{-2}, x, g_0 ]}$ requires the additional area ${\cal A}^{\circ}_{SU[ a^{(0)}, b^{(0)} ]}$ that is not part of the partition tree areas. Although the calculation of ${\cal A}^{\circ}_{SU[ a^{(0)}, b^{(0)} ]}$ is not difficult, to make the scheme consistent for all homoclinic points, an alternate form of Eq.~\eqref{eq:Action formulae once reduction} is used for $a^{(0)}$ only:
\begin{equation}\label{eq:Action formulae once reduction alternative for a0}
\begin{split}
\Delta {\cal F}_{\lbrace a^{(0)} \rbrace \lbrace x \rbrace} & = \Delta {\cal F}_{\lbrace h_{-1} \rbrace \lbrace x \rbrace} + \Delta {\cal F}_{\lbrace g_0 \rbrace \lbrace x \rbrace}\\
&+ {\cal A}^{\circ}_{SUSU[ a^{(0)}, h_{-1}, x, g_0 ]}
\end{split}
\end{equation} 
so ${\cal A}^{\circ}_{SUSU[ a^{(0)}, h_{-1}, x, g_0 ]}$ is expressible by cell areas $A+B$.  

Although ${\cal A}^{\circ}(y)$ is expressible by linear combinations of type-I and type-II partition tree areas, $A_{\tilde{\omega}}$ and $B_{\tilde{\omega}}$, the precise mapping between this area and the tree area symbols must be determined.  Given the symbolic code of any homoclinic point $y \in ( S^{\prime}_{-1} \cap U_{m} )$, the explicit mapping links ${\cal A}^{\circ}(y)$ with specific linear combinations of cell areas from the type-I and type-II partition trees of $T_{-1,m}$.  Since the transition time of $y$ is $m+1$, according to Eq.~\eqref{eq:transition time core length relation}, its core length is $m+3$. Let $\tilde{s} = s_1 s_2 \cdots s_{m+2} s_{m+3} $ ($s_i \in \lbrace 0,1 \rbrace$, $s_1=s_{m+3}=1$) be the core of the symbolic code of $y$, then the linear combination of cell areas depends solely on $\tilde{s}$.  As the association is rather technical, the details are given in App.~\ref{Area correspondence relations}.  The correspondence is given by Eq.~\eqref{eq:A_SUSU in terms of partition tree cells} using the notation and other relations also defined in the appendix.

Even though the actions of individual homoclinic orbits can always be calculated directly with the MacKay-Meiss-Percival action principle: $\Delta {\cal F}_{\lbrace y \rbrace \lbrace x \rbrace}={\cal A}^{\circ}_{US[x,y]}$, for those orbits with large transit times, the integration path $US[x,y]$ will be stretched exponentially long and extend far from the fixed point. Accurate interpolation of the path will require an exponentially growing set of points on the manifolds to maintain a reasonable density, an impractical task given the formidable computation time and memory space. On the other hand, using Eqs.~\eqref{eq:Action formulae once reduction} and \eqref{eq:A_SUSU in terms of partition tree cells}, the entire set of the homoclinic orbit actions arising from any trellis $T_{-1,N}$, can be calculated with the two primary orbit actions, $ \Delta {\cal F}_{\lbrace h_{0} \rbrace \lbrace x \rbrace}$ and $ \Delta {\cal F}_{\lbrace g_{0} \rbrace \lbrace x \rbrace}$, and the areas of the cells of the type-I and type-II partition trees of $T_{-1,N}$.  These areas are confined to a finite region of the phase space, and bounded by stable and unstable manifolds with small curvatures, which are far easier to compute.  Notice that both the symbolic codes of homoclinic points and the numerical areas in the partition trees can be generated with straightforward computer algorithms, so the recursive use of Eqs.~\eqref{eq:Action formulae once reduction} and \eqref{eq:A_SUSU in terms of partition tree cells} give rise to an automated computational scheme for the exact calculation of homoclinic orbit actions.

Equivalently, one may carry out the recursive process explicitly, which leads to an expression of the homoclinic orbit action as a cell-area expansion. This is done by expanding the three auxiliary homoclinic orbit actions in Eq.~\eqref{eq:Action formulae once reduction} using the equation itself, repeatedly, until all auxiliary orbits reduce to the primary homoclinic orbits. However, there is a technical difficulty of Eq.~\eqref{eq:Action formulae once reduction} to take into account: the point $P_{U}(y)$ is no longer on $S^{\prime}_{-1}$, so the area term in its own expansion, ${\cal A}^{\circ}(P_U(y))$, is no longer being expressed by the type-I and type-II cell areas. Consequently, Eq.~\eqref{eq:A_SUSU in terms of partition tree cells} breaks down for $P_U (y)$. The same is true for point $P_S P_U (y)$ as well. To adjust for this problem, all that is needed is to identify the representative point of the orbit $\lbrace P_U (y)\rbrace$ on $S^{\prime}_{-1}$, denoted by $P^{\prime}_U (y)$.  In fact, $P^{\prime}_U (y)$ is just an image of $P_U (y)$ under several inverse mappings. The number of inverse mappings is straightforwardly identified.  All homoclinic points on $S^{\prime}_{-1}$ have symbolic codes of the form $\overline{0} \tilde{\zeta}.01 \overline{0}$ (if they are located on $S[b^{(0)},g_{-2}]$) or $\overline{0} \tilde{\zeta}.11 \overline{0}$  (if they are located on $S[h_{-1},a^{(0)}]$), where $\tilde{\zeta}$ denotes an arbitrary symbolic string of binary digits. Equivalently stated, the decimal point in the symbolic code of any homoclinic point on $S^{\prime}_{-1}$ is always two digits left of the right end of its core. Hence, the resultant shift of the decimal point of $P_U (y)$ yields $P^{\prime}_U (y)$. Suppose the decimal point of $P_U (y)$ is $n^{\prime}$ digits to the right side of the right end of its core, then the $P^{\prime}_U$ operation can be defined as
\begin{equation}\label{eq:P_prime_U definition}
P^{\prime}_U (y) \equiv M^{-(n^{\prime}+2)} P_U (y).
\end{equation}

\begin{figure}[ht]
\centering
{\includegraphics[width=6.5cm]{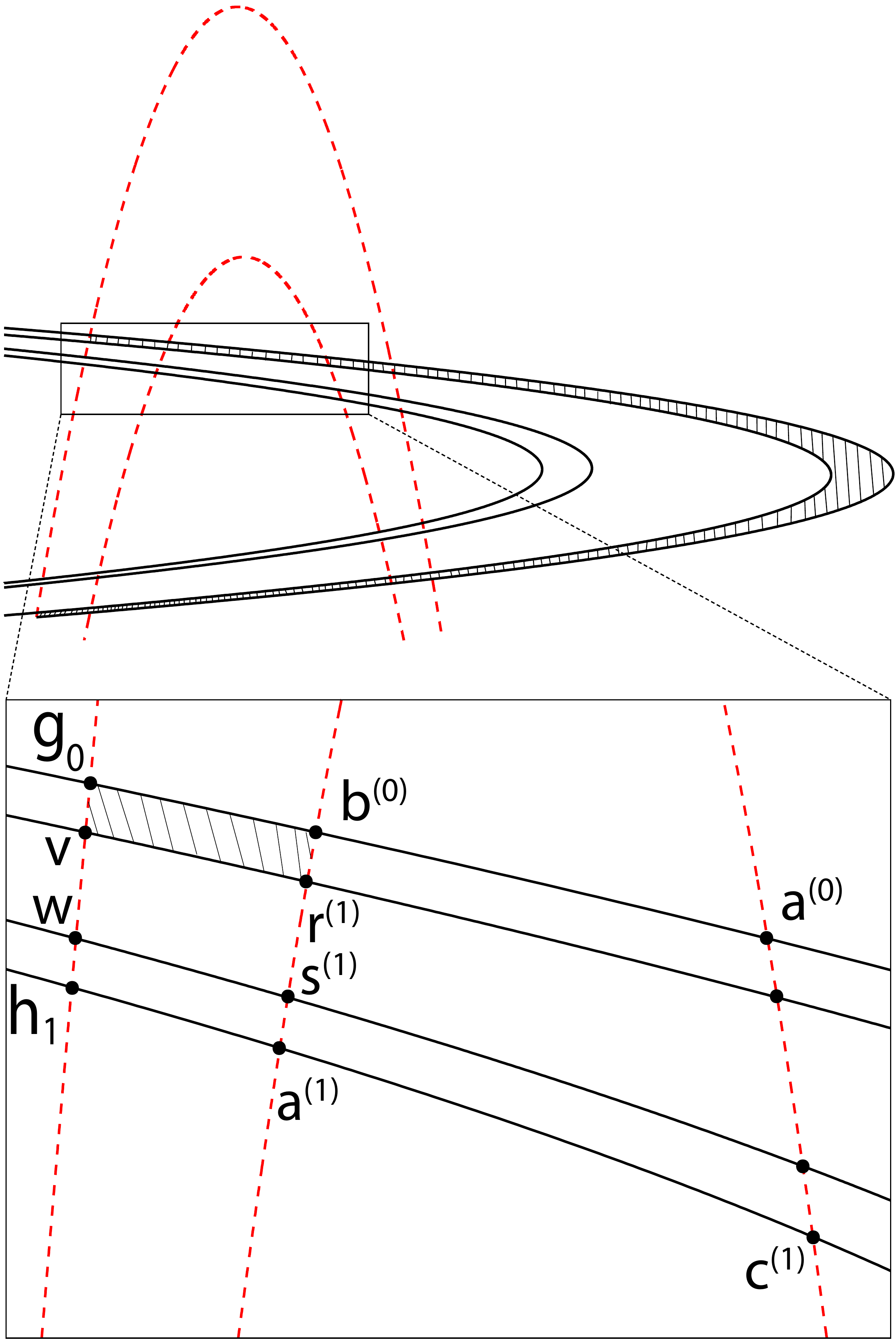}}
 \caption{(Color online) Relative areas for the decomposition of the winding-$3$ orbit $\lbrace r^{(1)} \rbrace$. The ${\cal A}^{\circ}(r^{(1)})$ term in Eq.~\eqref{eq:Action formulae once reduction for r1} is marked as the hatched region in the lower panel, which is just $-A_{\gamma\alpha}$. The long and curvy, hatched region in the upper panel is the ${\cal A}^{\circ}(v)$ term in Eq.~\eqref{eq:Action formulae once reduction for v}. Areas like this may not be expressible by the type-I and type-II cell areas.  }
\label{fig:Projection_Shift_Example}
\end{figure}  

The corresponding symbolic operation $\pi^{\prime}_U$ can be defined as a shift of the decimal point for $n^{\prime}+2$ digits towards the left, after the operation $\pi_U$. 

For the special cases of $y=h_i$ or $y=g_i$, i.e., a primary homoclinic point, $P_U(y)$ reduces to $x$, and $n^{\prime}$ loses its meaning.  For those cases, define
\begin{equation}
\label{eq:P_prime_U definition on primaries}
P^{\prime}_U(h_i)=P^{\prime}_U(g_i)=x
\end{equation}
and the corresponding $\pi^{\prime}_U$ operation maps the symbolic codes of the primary homoclinic points into $\overline{0}.\overline{0}$, i.e., that of the hyperbolic fixed point. 

The commutative relations hold for both the projection operations and their symbolic counterparts: $P_S P^{\prime}_U = P^{\prime}_U P_S$ and $\pi_S \pi^{\prime}_U =\pi^{\prime}_U \pi_S$.  Using the $P^{\prime}_U$ operation, Eq.~\eqref{eq:Action formulae once reduction} can be written alternatively as
\begin{equation}\label{eq:Action formulae once reduction alternative}
\begin{split}
& \Delta {\cal F}_{\lbrace y \rbrace \lbrace x \rbrace} = \Delta {\cal F}_{\lbrace P_S(y) \rbrace \lbrace x \rbrace} + \Delta {\cal F}_{\lbrace P^{\prime}_U(y) \rbrace \lbrace x \rbrace}\\
& - \Delta {\cal F}_{\lbrace P_S P^{\prime}_U(y) \rbrace \lbrace x \rbrace} + {\cal A}^{\circ}(y),
\end{split}
\end{equation}
in which the representative points $P^{\prime}_U (y)$ and $P_S P^{\prime}_U (y)$ of the auxiliary homoclinic orbits $\lbrace P^{\prime}_U (y) \rbrace$ and $\lbrace P_S P^{\prime}_U (y) \rbrace$ both locate on $S^{\prime}_{-1}$ now. Therefore, the recursive expansion of Eq.~\eqref{eq:Action formulae once reduction alternative} can be continued until all auxiliary orbits involved are primary homoclinic orbits.

The above motivation for introducing this extra $P^{\prime}_U$ operation is better demonstrated with the example in Fig.~\ref{fig:Projection_Shift_Example}. For the winding-$3$ homoclinic point $r^{(1)}$, we have: $r^{(1)} \xhookrightarrow[S]{1} b^{(0)}$ and $r^{(1)} \xhookrightarrow[U]{1} v$, therefore the projection operations on it give: $P_S ( r^{ (1) } ) = b^{(0)}$, $P_U ( r^{(1)} ) = v$, and $P_S P_U (r^{(1)}) = g_0$. Thus, Eq.~\eqref{eq:Action formulae once reduction}, when applied to $r^{(1)}$, reads:
\begin{equation}\label{eq:Action formulae once reduction for r1}
\begin{split}
\Delta {\cal F}_{\lbrace r^{(1)} \rbrace \lbrace x \rbrace} & = \Delta {\cal F}_{\lbrace b^{(0)} \rbrace \lbrace x \rbrace} + \Delta {\cal F}_{\lbrace v \rbrace \lbrace x \rbrace}\\
 & - \Delta {\cal F}_{\lbrace g_0 \rbrace \lbrace x \rbrace} + {\cal A}^{\circ}(r^{(1)})
\end{split}
\end{equation} 
where ${\cal A}^{\circ}(r^{(1)})={\cal A}^{\circ}_{SUSU[ r^{(1)}, b^{(0)}, g_0, v ]}$ is the negative area of the hatched region ($-A_{\gamma \alpha}$) from the lower panel of Fig.~\ref{fig:Projection_Shift_Example}. Among the three auxiliary orbit actions in the above expression, $ \Delta {\cal F}_{\lbrace g_0 \rbrace \lbrace x \rbrace}$ is already a primary orbit action, therefore no further decomposition is needed for it. The other two, $\Delta {\cal F}_{\lbrace b^{(0)} \rbrace \lbrace x \rbrace}$ and $\Delta {\cal F}_{\lbrace v \rbrace \lbrace x \rbrace}$, are both winding-$2$ orbits, and thus need to be further decomposed via Eq.~\eqref{eq:Action formulae once reduction} again. This is fine for $\Delta {\cal F}_{\lbrace b^{(0)} \rbrace \lbrace x \rbrace}$, since $b^{(0)}$ is already on $S^{\prime}_{-1}$, and thus:
\begin{equation}\label{eq:Action formulae once reduction for b0}
\begin{split}
\Delta {\cal F}_{\lbrace b^{(0)} \rbrace \lbrace x \rbrace} & = \Delta {\cal F}_{\lbrace g_{-2} \rbrace \lbrace x \rbrace} + \Delta {\cal F}_{\lbrace g_0 \rbrace \lbrace x \rbrace}\\
& - \Delta {\cal F}_{\lbrace x \rbrace \lbrace x \rbrace} + {\cal A}^{\circ}(b^{(0)})
\end{split}
\end{equation}
where ${\cal A}^{\circ}(b^{(0)})={\cal A}^{\circ}_{SUSU[ b^{(0)}, g_{-2}, x, g_0 ]}=A_{\alpha}+A_{\beta}+A_{\gamma}$. However, the same procedure, when applied to $\Delta {\cal F}_{\lbrace v \rbrace \lbrace x \rbrace}$, gives rise to undesired subtleties. Notice that $v \not\in S^{\prime}_{-1}$, $P_S (v) = g_0$, and $P_U (v) = g_2$, which lead to the expansion
\begin{equation}\label{eq:Action formulae once reduction for v}
\begin{split}
\Delta {\cal F}_{\lbrace v \rbrace \lbrace x \rbrace} & = \Delta {\cal F}_{\lbrace g_{0} \rbrace \lbrace x \rbrace} + \Delta {\cal F}_{\lbrace g_2 \rbrace \lbrace x \rbrace}\\
& - \Delta {\cal F}_{\lbrace x \rbrace \lbrace x \rbrace} + {\cal A}^{\circ}(v)
\end{split}
\end{equation}
where ${\cal A}^{\circ}(v)={\cal A}^{\circ}_{SUSU[ v, g_{0}, x, g_2 ]}$ is a long, thin, and folded area indicated by the hatched region in the upper panel of Fig.~\ref{fig:Projection_Shift_Example}. The expressions of such areas in terms of the type-I and -II cells are not immediately apparent, and the correspondence relation Eq.~\eqref{eq:A_SUSU in terms of partition tree cells} will fail. The fix, however, is simple and straightforward: use the representative point of $\lbrace v \rbrace$ on $S^{\prime}_{-1}$. This point can be easily identified from the symbolic dynamics. Given that $b^{(0)} \Rightarrow \overline{0} 1.01 \overline{0}$ and $r^{(1)} \xhookrightarrow[S]{1} b^{(0)}$, we know from Eq.~\eqref{eq:Orbit symbolic codes general rules} that $r^{(1)} \Rightarrow \overline{0} 101.01 \overline{0}$. Since $v=P_U (r^{(1)})$, its symbolic code is then $v \Rightarrow \pi_U ( \overline{0} 101.01 \overline{0}) = \overline{0} 101. \overline{0}$, which indicates that $v= M^{2} (b^{(0)})$. Therefore, the representative point of $\lbrace v \rbrace$ on $S^{\prime}_{-1}$ is identified to be $b^{(0)}$. Correspondingly, one can verify the validity of Eq.~\eqref{eq:P_prime_U definition} since $P^{\prime}_U (r^{(1)}) = b^{(0)}$, i.e., $P^{\prime}_U(y)$ indeed yields the correct representative point of $\lbrace P_U (y) \rbrace$ on $S^{\prime}_{-1}$. Therefore, $\Delta {\cal F}_{\lbrace v \rbrace \lbrace x \rbrace}=\Delta {\cal F}_{\lbrace b^{(0)} \rbrace \lbrace x \rbrace}$, which is expressible via Eq.~\eqref{eq:Action formulae once reduction for b0} again. The final expression for $\lbrace r^{(1)} \rbrace$ is then
\begin{equation}\label{eq:Action formulae complete reduction for r1}
\begin{split}
\Delta {\cal F}_{\lbrace r^{(1)} \rbrace \lbrace x \rbrace} & =3 \Delta {\cal F}_{\lbrace g_0 \rbrace \lbrace x \rbrace} + 2 {\cal A}^{\circ}(b^{(0)})+ {\cal A}^{\circ}(r^{(1)})\\
 & = 3 \Delta {\cal F}_{\lbrace g_0 \rbrace \lbrace x \rbrace} + 2A - A_{\gamma \alpha}
\end{split}
\end{equation} 
which only involves $\Delta {\cal F}_{\lbrace g_0 \rbrace \lbrace x \rbrace}$ and several type-I cell areas. As shown by Eq.~\eqref{eq:Action formulae complete reduction} later, similar decomposition can be written for any homoclinic orbit, and the resulting expansions will only involve the two primary orbit actions, $\Delta {\cal F}_{\lbrace g_0 \rbrace \lbrace x \rbrace}$ and $\Delta {\cal F}_{\lbrace h_0 \rbrace \lbrace x \rbrace}$, plus a linear combination of some type-I and type-II cell areas.

The general process proceeds as follows. Consider the case of $\lbrace y \rbrace $ with winding-$2$. Then $P_S P_U (y) = x$, thus $ \Delta {\cal F}_{\lbrace P_S P_U(y) \rbrace \lbrace x \rbrace} = 0$. The two non-vanishing auxiliary orbits are $\lbrace P_S (y) \rbrace$ and $\lbrace P^{\prime}_U(y) \rbrace$, both of which are primary orbits, so Eq.~\eqref{eq:Action formulae once reduction alternative} is already a complete expansion. For all higher winding cases, $n \geq 3$, it is possible to expand the $ \Delta {\cal F}_{\lbrace P_S(y) \rbrace \lbrace x \rbrace} $ and $ \Delta {\cal F}_{\lbrace P^{\prime}_U(y) \rbrace \lbrace x \rbrace} $ terms in Eq.~\eqref{eq:Action formulae once reduction alternative} using the equation itself to obtain a twice-iterated formula
\begin{equation}\label{eq:Action formulae twice reduction}
\begin{split}
& \Delta {\cal F}_{\lbrace y \rbrace \lbrace x \rbrace} = \Delta {\cal F}_{\lbrace P_{S}^{2}(y) \rbrace \lbrace x \rbrace} + \Delta {\cal F}_{\lbrace P_S P^{\prime}_U(y) \rbrace \lbrace x \rbrace}\\
&+ \Delta {\cal F}_{\lbrace P^{\prime 2}_{U} (y) \rbrace \lbrace x \rbrace} -  \Delta {\cal F}_{\lbrace P_{S}^{2} P^{\prime}_U (y) \rbrace \lbrace x \rbrace} - \Delta {\cal F}_{\lbrace P_{S} P^{\prime 2}_{U} (y) \rbrace \lbrace x \rbrace}\\
&   + {\cal A}^{\circ}(y) +  {\cal A}^{\circ}( P_S (y) ) +  {\cal A}^{\circ}( P^{\prime}_U(y) )\ .
\end{split}
\end{equation}
Since $y$, $P_S(y)$, and $P^{\prime}_U(y)$ are all located on $S^{\prime}_{-1}$, with the help of Eq.~\eqref{eq:A_SUSU in terms of partition tree cells}, the three ${\cal A}^{\circ}$ areas in the above formula are all expressible using type-I and type-II areas. For the orbits with $n=3$, both $ \Delta {\cal F}_{\lbrace P_{S}^{2} P^{\prime}_U (y) \rbrace \lbrace x \rbrace}$ and $\Delta {\cal F}_{\lbrace P_{S} P^{\prime 2}_{U} (y) \rbrace \lbrace x \rbrace}$ vanish, so no more expansions are needed. An example of this is already provided by Eq.~\eqref{eq:Action formulae complete reduction for r1} previously. For the $n \geq 4$ cases, the above procedure can be carried on repeatedly, until the $P_{S}^{n-i} P_U^{\prime i}(y)$ ($1 \leq i \leq n-1$) action terms are present, which reduce $y$ into $x$. To further simplify the notations, define the mixed projections of $P_S$ and $P^{\prime}_U$ on $y$ as
\begin{equation}\label{eq:Definition mixed projections}
P(y;i;j) \equiv P_S^{i-j} P^{\prime j}_U (y),\ \ (i \geq j).
\end{equation}
Then, a general formula for the complete action decomposition of any winding-$n$ homoclinic orbit $\lbrace y \rbrace$ (where $y\in S^{\prime}_{-1}\cap U_m $) can be written as
\begin{equation}\label{eq:Action formulae complete reduction}
 \Delta {\cal F}_{\lbrace y \rbrace \lbrace x \rbrace}  = \sum_{i=0}^{n-1}  \Delta {\cal F}_{\lbrace P(y;n-1;i) \rbrace \lbrace x \rbrace}+ \sum_{i=0}^{n-2} \sum_{j=0}^{i} {\cal A}^{\circ}\big( P(y;i;j) \big) 
\end{equation}
where $ \Delta {\cal F}_{\lbrace P(y;n-1;i) \rbrace \lbrace x \rbrace}= \Delta {\cal F}_{\lbrace P_{S}^{n-1-i} P^{\prime i}_U (y) \rbrace \lbrace x \rbrace}$ are relative actions of the primary homoclinic orbits, therefore either $\Delta {\cal F}_{\lbrace h_0 \rbrace \lbrace x \rbrace}$ or $\Delta {\cal F}_{\lbrace g_0 \rbrace \lbrace x \rbrace}$.  The ${\cal A}^{\circ}(P(y;i;j))$ terms in the double sum are areas of the curvy parallelograms spanned by four homoclinic points of various winding numbers, generated from mixed projections of $y$.  By design, all $P(y;i;j)$ points in these areas are located on $S^{\prime}_{-1}$, thus the ${\cal A}^{\circ}(P(y;i;j))$ terms are expressible using the type-I and type-II cells via Eq.~\eqref{eq:A_SUSU in terms of partition tree cells}. 

Eq.~\eqref{eq:Action formulae complete reduction} gives a complete expansion of the homoclinic orbit actions in terms of the primary homoclinic orbit actions plus the cell areas of type-I and type-II partition trees. It converts the determinations of numerical orbits into area calculations in a finite region of the phase space, and avoids exponentially extending integration paths associated with complicated orbits. Furthermore, the two types of cells come from a nearly parallel and linear foliated phase-space region with relatively small curvature along the manifolds, so the numerical interpolation of the manifolds does not require a very dense set of points, and therefore renders the calculations practical. 
    
Nevertheless, the total number of the cell areas proliferates with the same rate as the homoclinic points on $S^{\prime}_{-1}$.  This is because the cells can be put into an one-to-one correspondence with the non-primary homoclinic points on $S^{\prime}_{-1}$, such that each cell corresponds to the homoclinic point at its upper right corner. For example, in Fig.~\ref{fig:Trellis_Partition_1}, the cells $A_{\alpha}$, $A_{\beta}$, $A_{\gamma}$, $B_{\alpha}$, $B_{\beta}$, and $B_{\gamma}$ correspond to points $b^{(1)}$, $a^{(1)}$, $b^{(0)}$, $d^{(1)}$, $c^{(1)}$, and $a^{(0)}$, respectively. As we increase the integer $N$ of the trellis $T_{-1,N}$, new cells emerge at an identical rate with new homoclinic points on $S^{\prime}_{-1}$, both of which proliferate as $2^{N+2}=\mathrm{e}^{ \alpha (N+2)}$, where $\alpha = \log 2$ is the topological entropy of the system. Therefore, the exact evaluation of homoclinic orbit actions, Eq.~\eqref{eq:Action formulae complete reduction} requires an exponentially increasing set of areas for its input, as must happen. 

\begin{figure}[ht]
\centering
{\includegraphics[width=6.5cm]{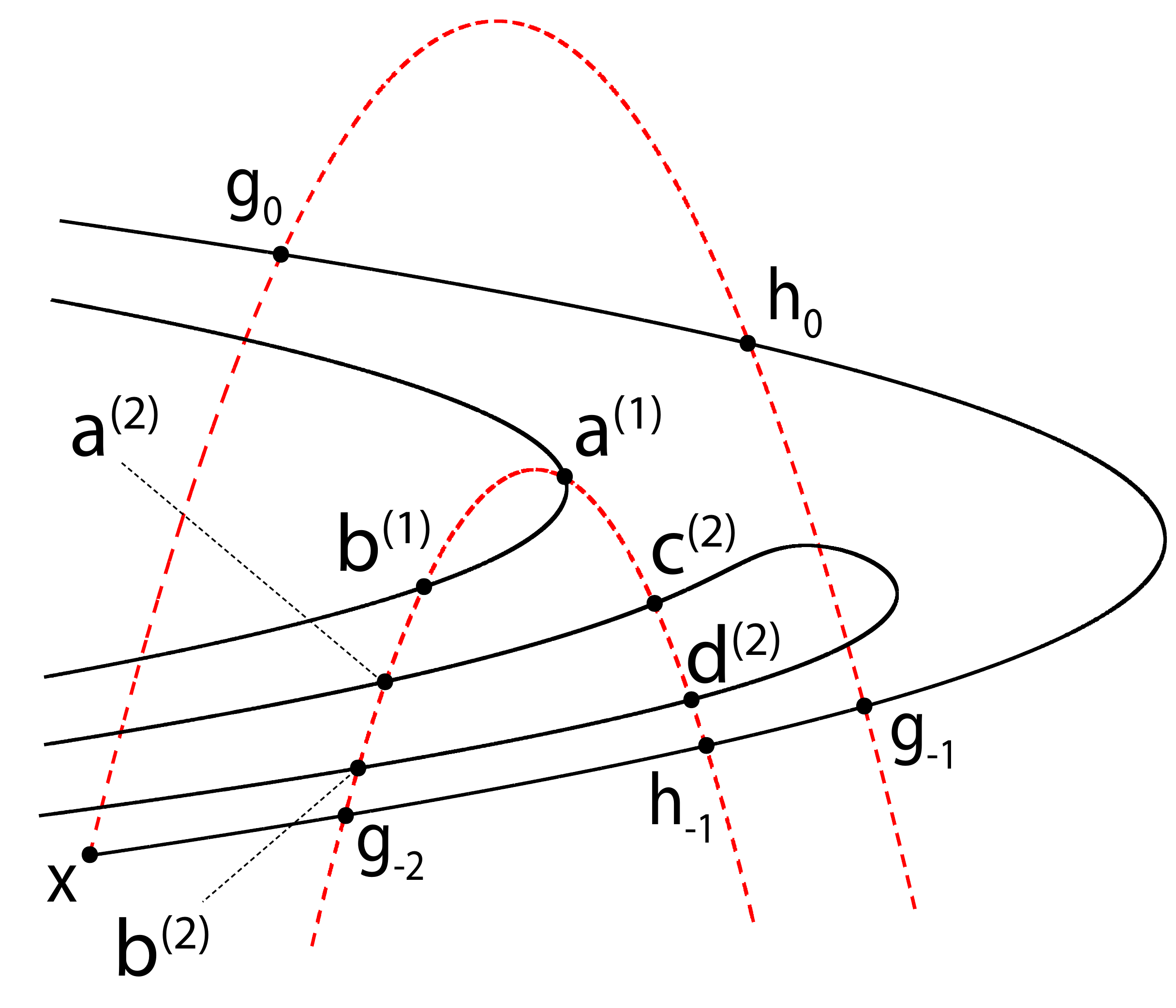}}
 \caption{(Schematic, color online) Homoclinic tangle forming an incomplete horseshoe. Comparing to the complete horseshoe case (Fig.~\ref{fig:Scaling}), the points $a^{(0)}$, $b^{(0)}$, $c^{(1)}$, and $d^{(1)}$ are pruned. However, the accumulation relations (thus the projection operations) for the unpruned homoclinic points remain the same. Therefore, for the unpruned homoclinic points, Eqs.~\eqref{eq:Action formulae once reduction} and \eqref{eq:Action formulae complete reduction} remain valid. }
\label{fig:Incomplete_Horseshoe}
\end{figure}  

A few words are in order for the symbolic dynamics. In all the derivations up till now, we have assumed the homoclinic tangle forms a complete horseshoe structure, which allows all possible sequences of binary digits. Although this is often true for highly chaotic systems, for other types of systems with mixed dynamics, the homoclinic tangles will in general form incomplete horseshoe structures that coexist with stability islands in phase space. A simple kind of incomplete horseshoe is shown by Fig.~\ref{fig:Incomplete_Horseshoe}. The symbolic dynamics of such systems are more complicated as certain substrings are not admissible by the dynamics and therefore ``pruned" from the symbol plane \cite{Cvitanovic88a,Cvitanovic91}. Therefore, not all symbolic strings may exist, and their very existence are determined by a ``pruning front" \cite{Cvitanovic88a} which separates the allowed and disallowed orbits in the symbol plane. In spite of this apparent complication, the foundations of our final result Eq.~\eqref{eq:Action formulae complete reduction} hold true in general, even for incomplete horseshoes. Namely, the accumulation of homoclinic points along the manifolds, and the projection operations defined accordingly, remain valid for all types of horseshoe structures. For instance, in Fig.~\ref{fig:Incomplete_Horseshoe}, although $a^{(0)}$ and $b^{(0)}$ are pruned, we still have: $a^{(n)} \xhookrightarrow[S]{n} g_{-2} $ and $b^{(n)} \xhookrightarrow[S]{n} g_{-2} $ (where $n \geq 1$). As compared to Eq.~\eqref{eq:Accumulation along stable}, the pruning removes the first members ($a^{(0)}$ and $b^{(0)}$) of the two accumulating families, but leaves the rest unchanged. Therefore, as long as the pruning front (or a finite approximation of it) has been established by methods such as \cite{Hagiwara04}, Eqs.~\eqref{eq:Action formulae once reduction} and \eqref{eq:Action formulae complete reduction} will be applicable to any admissible homoclinic orbit $\lbrace y \rbrace$ since all the projections involved are admissible as well. Therefore, their range of applicability is not limited to the complete horseshoes. 

A complication that does arise in the incomplete horseshoe cases is the pruning of the area partition trees. Depending on the complexities of the horseshoes, more types of trees might be needed, and their structures will not be as simple as the one in Fig.~\ref{fig:A_tree}. Certain nodes will be pruned away, and there may not exist a finite grammar rule. Just like the pruning fronts, the partition trees are also system-specific, and we anticipate that the numerical algorithms for generating the pruning front should already contain adequate information for generating the partitions trees as well, although more sophisticated investigations along this direction are needed. 

\subsection{Information reduction}
\label{Information reduction}

In semiclassical approximations, the classical actions divided by $\hbar$ determine phase angles, and as it is already an approximation to begin with, it is possible to tolerate small errors, say $\epsilon=\delta {\cal F}/\hbar$, measured in radians.  As a practical matter, once this ratio is $\lesssim 0.1$ or some similar scale, constructive and destructive interferences are properly predicted, and much greater precision becomes increasingly irrelevant.  Given that the areas in Eq.~\eqref{eq:Action formulae complete reduction}, or similarly of the partition tree cells, shrink exponentially rapidly, most of these corrections can be dropped or ignored. 

Identifying the necessary information begins with an estimate of orders of magnitudes of the areas terms in Eq.~\eqref{eq:Action formulae complete reduction}.  Given any trellis $T_{-1,N}$, the maximum winding number of a homoclinic orbit is $n_{\mathrm{max}}=N/2+2$. Due to the slow scaling direction of the tree structure, the orbit $y \Rightarrow \overline{0}1^{N+1}.11 \overline{0}$ yields an expansion with the largest possible number of significant ${\cal A}^{\circ}(P(y;i;j))$ terms, and hence an upper bound on the number of necessary areas.  

It is reasonable to assume the cell areas $A$ and $B$ of $T_{-1,0}$ are of the same magnitude, and it is sufficient to consider the ratios ${\cal R}={\cal A}^{\circ}(P(y;i;j))/A$. Via Eq.~\eqref{eq:A_SUSU in terms of partition tree cells}, ${\cal A}^{\circ}(P(y;i;j))$ is expressible as a linear combination of cell areas of partition trees of $T_{-1,N-2i}$. These cell areas are at the $(N-2i)$th level of the partition trees, hence the scaling relation, Eq.~\eqref{eq:Slow scaling}, gives ratio estimates $\sim \mathrm{e}^{-\mu_1 (N-2i)}$.  As a result, the inner area sum of Eq.~\eqref{eq:Action formulae complete reduction} gives
\begin{equation}\label{eq:Action formulae area sum scaling estimate}
 \sum_{j=0}^{i} {\cal A}^{\circ}\big( P(y;i;j) \big) \sim A \cdot O \left( (i+1) e^{-\mu_1 (N-2i)} \right).
\end{equation}
Comparing this estimate with the threshold $\delta {\cal F}$ yields a maximum value of the depth $d\equiv N-2i$ of the tree needed:
\begin{equation}\label{eq:Compare area with tolerance}
A (i+1) e^{-\mu_1 d} \geq \delta  {\cal F},
\end{equation}
therefore 
\begin{equation}\label{eq:Compare area with tolerance 2}
e^{-\mu_1 d} \geq \frac{\epsilon \hbar}{A(i+1)}.
\end{equation}
A slightly more conservative bound replaces $i+1$ with $n_{max}\approx N/2$ and gives after some algebra
\begin{equation}
\label{eq:i minimum determination}
d \leq \frac{1}{\mu_1 }\log \frac{NA}{2\epsilon\hbar}.
\end{equation}
Therefore, in order to calculate all homoclinic orbit actions arising from $T_{-1,N}$ within the error tolerance $\epsilon \hbar$, we only need to determine numerically the type-I and type-II cell areas of the partition trees of $T_{-1,d}$.  Recall that the number of cell areas in $T_{-1,d}$ is estimated by 
\begin{equation}\label{eq:Number of cell areas after reduction}
\mathrm{e}^{\alpha d} \sim \left( \frac{N A}{2\epsilon \hbar} \right)^{\frac{\alpha}{\mu_1}} ,
\end{equation}
whereas the number of homoclinic orbits in $T_{-1,N}$ is $ \propto \mathrm{e}^{\alpha N}$, where $\alpha = \log 2$ is the topological entropy of the system. Thus, the exponentially proliferating homoclinic orbit actions in $T_{-1,N}$ is expressible by the algebraically proliferating cell areas from $T_{-1,d}$, a significant information reduction. 

In practice, the use of $T_{-1,d}$ to construct the relative actions of $T_{-1,N}$ alters the area sum in Eq.~\eqref{eq:Action formulae complete reduction}, such that any ${\cal A}^{\circ}(P(y;i;j))$ terms with $P(y;i;j) \not\in T_{-1,d}$ will be excluded from the double sum, leading to the reduced action formula:
\begin{equation}
\begin{split}
\label{eq:Action formulae complete reduction approximate}
 \Delta {\cal F}_{\lbrace y \rbrace \lbrace x \rbrace} & = \sum_{i=0}^{n-1}  \Delta {\cal F}_{\lbrace P(y;n-1;i) \rbrace \lbrace x \rbrace}\\
 & + \sum_{i=0}^{n-2} \sum_{\substack{
   j=0 \\
   P(y;i;j) \in T_{-1,d}
  }}^{i} {\cal A}^{\circ}\big( P(y;i;j) \big) +O(\delta {\cal F}),
  \end{split}
\end{equation}
where the constraint $P(y;i;j)\in T_{-1,d}$ is imposed, therefore eliminating the (exponentially many) cell areas smaller than the error threshold $\delta {\cal F}$. 

\subsection{Numerical example}
\label{Numerical results}

\begin{figure}[ht]
\centering
{\includegraphics[width=8cm]{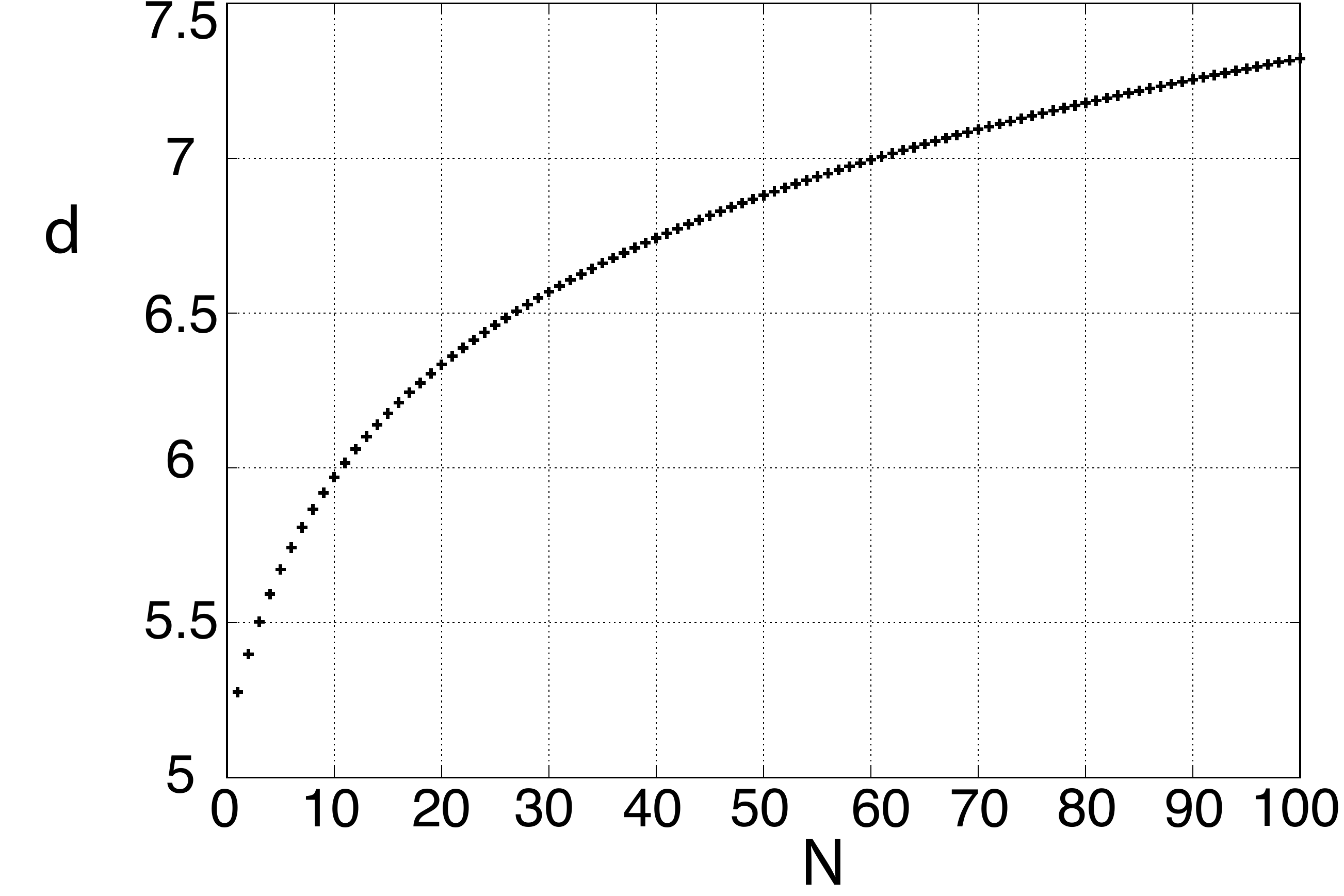}}
 \caption{The logarithmic dependence of $d$ with respect to $N$ using the slow scaling exponent $\mu_s = 1.483$, error tolerance $\delta {\cal F}/A =0.001$ and $a=10$ for the H\'{e}non map.  For the computation of large trellises such as $N=100$, the approximate formula of Eq.~\eqref{eq:Action formulae complete reduction approximate} only requires the computation of cell areas up to trellis number $d=8$. }
\label{fig:t_N_graph}
\end{figure}

For the H\'{e}non map in Eq.~\eqref{eq:Henon map} with $a=10$, and an error tolerance $\delta {\cal F}/A = 0.001$, the natural logarithmic dependence of $d$ on $N$ is shown in Fig.~\ref{fig:t_N_graph}.  The information reduction is significant: even for the calculation of homoclinic orbit actions of $T_{-1,100}$, which is obviously impossible via traditional methods, our scheme only requires the numerical computation of cell areas up to $T_{-1,8}$, an effortless task for personal computers.  

For the numerical verification of Eqs.~\eqref{eq:Action formulae complete reduction} and \eqref{eq:Action formulae complete reduction approximate}, we calculate the relative actions of the homoclinic orbits of $T_{-1,N}$ in three different ways. The first method is to implement the orbit finder method introduced in our previous work \cite{Li17}, which determines the numerical orbits $\lbrace y \rbrace$ and thus their relative actions, $ \Delta {\cal F}^{(\mathrm{ref.})}_{\lbrace y \rbrace \lbrace x \rbrace}$. These actions are the standard reference actions for comparison.  The second method is to calculate the cell areas in the partition trees of $T_{-1,N}$, and evaluate the actions $ \Delta {\cal F}^{(\mathrm{exact})}_{\lbrace y \rbrace \lbrace x \rbrace}$  using Eqs.~\eqref{eq:Action formulae complete reduction} and \eqref{eq:A_SUSU in terms of partition tree cells}.  These should only differ from $ \Delta {\cal F}^{\mathrm{(ref.)}}_{\lbrace y \rbrace \lbrace x \rbrace}$ due to relying on double precision computation since both are exact evaluations with no approximations involved.  On the contrary, in the third method the tolerance is $\delta {\cal F}/A = 0.001$ (where $A \approx 10.973$ for the current case of $a=10$), and only cell areas of the partition trees up to the reduced trellis $T_{-1,d}$ are used with Eqs.~\eqref{eq:Action formulae complete reduction approximate} and \eqref{eq:A_SUSU in terms of partition tree cells} to obtain the approximate actions,  $ \Delta {\cal F}^{(\mathrm{approx.})}_{\lbrace y \rbrace \lbrace x \rbrace}$. 

Every homoclinic orbit up to iteration number $N=10$ is constructed, which corresponds to trellis $T_{-1,10}$. The total number of orbits is $2^{12}=4096$.  The reduced iteration number for this case is $d=6$, i.e., the relative homoclinic orbit actions in $T_{-1,10}$ should be given to an accuracy $A\times O(10^{-3})\sim 1 \times 10^{-2}$ or better using only the cell areas from $T_{-1,6}$. 

Due to the large number of orbits, it is impractical to list the results for $\Delta {\cal F}^{(\mathrm{exact})}_{\lbrace y \rbrace \lbrace x \rbrace}$ and $\Delta {\cal F}^{(\mathrm{approx.})}_{\lbrace y \rbrace \lbrace x \rbrace}$ for every orbit. Instead, we show the two orbits that yield the maximum errors. The homoclinic orbit that leads to the maximum error in $\Delta {\cal F}^{(\mathrm{exact})}_{\lbrace y \rbrace \lbrace x \rbrace}$ out of all $4096$ orbits is $\lbrace y \rbrace \Rightarrow \overline{0}1010001100011\overline{0}$, for which 
\begin{equation}\label{eq:Numerical verification maximum error exact formula}
\Delta {\cal F}^{(\mathrm{exact})}_{\lbrace y \rbrace \lbrace x \rbrace} -   \Delta {\cal F}^{(\mathrm{ref.})}_{\lbrace y \rbrace \lbrace x \rbrace} = 8.08 \times 10^{-8}.
\end{equation}
Compared to the orbit action itself, $\Delta {\cal F}^{(\mathrm{ref.})}_{\lbrace y \rbrace \lbrace x \rbrace} = -466.602\ 850\ 894\ 90$, the relative error is around $1.7 \times 10^{-10}$, almost as good as possible due to the presence of interpolation error. This demonstrates the accuracy of Eq.~\eqref{eq:Action formulae complete reduction}.   

As for $\Delta {\cal F}^{(\mathrm{approx.})}_{\lbrace y \rbrace \lbrace x \rbrace}$, the maximum error emerges for the orbit $\lbrace y \rbrace \Rightarrow \overline{0}111111111111 \overline{0}$, for which
\begin{equation}\label{eq:Numerical verification maximum error approximate formula}
\Delta {\cal F}^{(\mathrm{approx.})}_{\lbrace y \rbrace \lbrace x \rbrace} -   \Delta {\cal F}^{(\mathrm{ref.})}_{\lbrace y \rbrace \lbrace x \rbrace} = -5.453 \times 10^{-3}
\end{equation}
which is well below the error tolerance $1 \times 10^{-2}$. Compared to the orbit action itself, $\Delta {\cal F}^{(\mathrm{ref.})}_{\lbrace y \rbrace \lbrace x \rbrace} = -628.514\ 708\ 240\ 16$, the relative error is around $8.7 \times 10^{-6}$. 
  
\section{Conclusions}
\label{Conclusions}

It is possible to construct the complete set of homoclinic orbit relative actions arising from horseshoe-shaped homoclinic tangles in terms of the primitive orbits' relative actions and an exponentially decreasing set of parallelogram-like areas bounded by stable and unstable manifolds.  Important constraints exist on the distribution of homoclinic points~\cite{Bevilaqua00,Mitchell03a}, which are imposed by the topology of the homoclinic tangle. This enables an organizational scheme for the orbits by their winding numbers and assigns binary symbolic codes to each of them.  The projection operations, $P_S$ and $P_U$, together with the corresponding symbolic operations, $\pi_S$ and $\pi_U$, link homoclinic points of different winding numbers.  Based on a judicious use of the MacKay-Meiss-Percival action principle and mixed projections of all degrees, an exact geometric formula [Eq.~\eqref{eq:Action formulae complete reduction}] emerges that determines their relative actions in terms of cell areas from a finite region of phase space, which are bounded by manifolds with low curvatures.  However, these areas still proliferate at the same rate as the homoclinic points, which become exponentially hard to compute for large iterations numbers $N$.  To overcome this, we made use of the exponential decay of cell areas in the partition trees, and eliminated all small areas that are asymptotically negligible.  The exponentially shrinking areas have their origins in the asymptotic foliations of stable and unstable manifolds, and are thus generic to all chaotic systems. The resulting approximate expression [Eq.~\eqref{eq:Action formulae complete reduction approximate}] relies on a logarithmically reduced amount of information relative to the exact Eq.~\eqref{eq:Action formulae complete reduction}.  It gives the relative actions or orbits in $T_{-1,N}$ using only the areas from $T_{-1,d}$, in exchange for comprising the accuracy by a designated order of magnitude $O(\delta {\cal F}=\epsilon \hbar)$.   

For semiclassical trace formulas, once the actions are determined to within an appropriate tolerance level such as mentioned above, additional accuracy becomes irrelevant and of no consequence.   Straightforward computations of the actions rely on the numerical constructions of orbits, for which the difficulties are twofold. First, in highly chaotic systems, numerical determination of individual long orbits suffer from sensitive dependence on initial errors. Second, the total number of orbits proliferates exponentially rapidly with relevant time scales (the trellis number $N$ in our case). For homoclinic, heteroclinic, and periodic orbits in Hamiltonian chaos with two degrees of freedom, the first difficulty is not fundamental, and solvable in many ways.  The second difficulty, addressed in the present article, illustrates in great detail how information entropy vanishes for quantum systems (isolated, bounded, non-measured) from the perspective of semiclassical theory.  The reduction of information implied by $\hbar$ or any error tolerance criterion produces an exponentially increasing set of output calculations using a slower-than-exponentially (i.e., algebraically) increasing set of input information. 

This method has the potential to serve as a generic paradigm for the information reduction of semiclassical calculations of chaotic systems.  Although the present work is focused on homoclinic orbit actions, the results can be immediately generalized into broader contexts, such as the evaluation of unstable periodic orbit actions.  Such connections are given by Eqs.~(27), (38), and (45) in Ref.~\cite{Li18}. These equations convert the evaluation of periodic orbit actions into the calculation of action differences between certain auxiliary homoclinic orbits constructed from the symbolic codes of the periodic orbit. Therefore, upon the determination of homoclinic orbit actions, the determination of periodic orbit actions becomes a simple manipulation of symbolic strings and subtractions within the homoclinic action set, a trivial task that poses no serious difficulties.  Therefore, just like the homoclinic orbit actions, the exponentially increasing set of periodic orbit actions is expressible with the same reduced set of cell areas as well.  Further extension of the current method concerns the stability exponents of unstable periodic orbits, which is a topic under current investigation. 

\appendix

\section{Homoclinic tangle}
\label{Homoclinic tangle}

In this appendix we illustrate the fundamental concepts and definitions related to homoclinic tangles that are used throughout this article. Consider a two-degree-of-freedom autonomous Hamiltonian system. With energy conservation and applying the standard Poincar\'{e} surface of section technique~\cite{Poincare99}, the continuous flow leads to a discrete area-preserving map $M$ on the two-dimensional phase space $(q,p)$.  Assume the existence of a hyperbolic fixed point $x=(q_x,p_x)$ under $M$: $M(x)=x$.  Associated with it are the one-dimensional stable ($S(x)$) and unstable ($U(x)$) manifolds, which are the collections of phase-space points that approach $x$ under successive forward and inverse iterations of $M$, respectively. Typically, $S(x)$ and $U(x)$ intersect infinitely many times and form a complicated pattern named $\mathit{homoclinic}$ $\mathit{tangle}$ \cite{Poincare99,Easton86,Rom-Kedar90}, as partially illustrated in Fig.~\ref{fig:Homoclinic_Tangle}.   

Homoclinic tangles have been extensively studied as the organizing structures for classical transport and escape problems \cite{Easton86,MacKay84a,Rom-Kedar90,Wiggins92,Mitchell03a,Mitchell03b,Mitchell06,Novick12a,Novick12b}. Of particular interest are the homoclinic orbits, which lie along intersections between $S(x)$ and $U(x)$
\begin{equation}
\label{eq:Definition homoclinic points}
h_{0} = S(x) \bigcap U(x)
\end{equation}
whose images under both $M$ and $M^{-1}$ approach $x$ asymptotically: $M^{\pm\infty}(h_0)=h_{\pm\infty}=x$. The bi-infinite collection of images $M^{n}(h_0)=h_{n}$, is often referred to as a $\mathit{homoclinic}$ $\mathit{orbit}$
\begin{equation}\label{eq:Definition homoclinic orbit}
\lbrace h_0 \rbrace = \lbrace  \cdots, h_{-1}, h_0, h_{1}, \cdots \rbrace\ . 
\end{equation}
A $\mathit{primary}$ $\mathit{homoclinic}$ $\mathit{point}$, $h_0$, arises if the stable and unstable segments, $S[h_0,x]$ and $U[x,h_0]$, intersect only at $x$ and $h_0$. The resulting closed loop $US[x,h_0]=U[x,h_0]+S[h_0,x]$ is topologically equivalent to a circle.  As a result, the phase space excursions of the $\mathit{primary}$ $\mathit{homoclinic}$ $\mathit{orbit}$ $\lbrace h_0 \rbrace$ takes the simplest possible form.  It ``circles" around the loop once from infinite past to infinite future.  Figure~\ref{fig:Homoclinic_Tangle} shows the simplest kind of homoclinic tangle having only two primary homoclinic orbits, $\lbrace h_0 \rbrace$ and $\lbrace g_0 \rbrace$. In practice, more complicated homoclinic tangles are possible. However, generalizations are straightforward and not considered here. 

\begin{figure}[ht]
\centering
{\includegraphics[width=6.5cm]{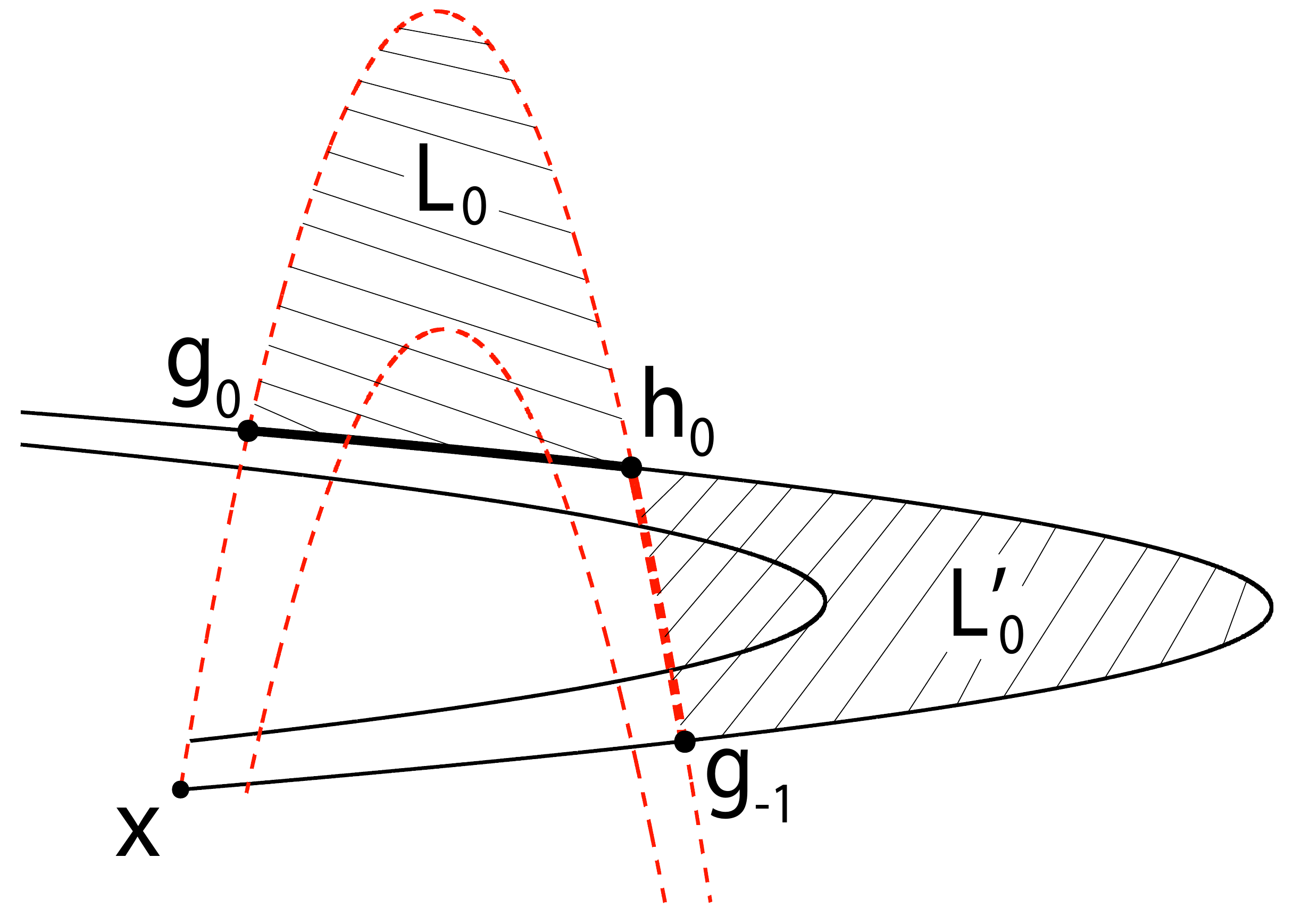}}
 \caption{(Color online) Trellis $T_{-1,1}$. The fundamental segments $U_0$ and $S^{\prime}_0$ are indicated by thick solid and thick dashed curve segments, respectively. The lobes $L_0$ and $L^{\prime}_0$ (hatched regions) form the turnstile which governs the phase-space transport.}
\label{fig:Fundamental_Segments}
\end{figure}     

The entire homoclinic tangle, as an infinite entity, can be constructed from iterations of finite segments on $S(x)$ and $U(x)$. Identifying the $\mathit{fundamental}$ $\mathit{segments}$ as:
\begin{equation}\label{eq:Definition fundamental segments}
\begin{split}
&U_n \equiv U[h_n,g_n]\quad \quad U^{\prime}_n \equiv U(g_{n-1},h_n) \\
&S_n \equiv S(g_{n},h_n)\quad \quad   S^{\prime}_n \equiv S[h_n,g_{n-1}]
\end{split}
\end{equation}
$U_{n+k}=M^{k}(U_n)$ and similarly for $U^{\prime}_n$, $S_n$, and $S^{\prime}_n$. Shown in Fig.~\ref{fig:Fundamental_Segments} are examples of $U_0$ (thick solid segment) and $S^{\prime}_0$ (thick dashed segment). The manifolds can be built as non-overlapping unions of the respective fundamental segments:
\begin{equation}
\begin{split}
&U(x)=\bigcup_{n=-\infty}^{\infty} ( U_n \cup U^{\prime}_n ) \\
&S(x)=\bigcup_{n=-\infty}^{\infty} ( S_n \cup S^{\prime}_n )
\end{split}
\end{equation}
and likewise for the homoclinic tangle. The topology of a homoclinic tangle contains important dynamical information, and is often studied over its truncations, namely a \textit{trellis} \cite{Easton86,Rom-Kedar90} defined as
\begin{equation}\label{eq:Definition trellis}
T_{n_s,n_u} \equiv \left( \bigcup_{i=n_s}^{\infty} (S_i \cup S^{\prime}_i )  \right) \bigcup \left( \bigcup_{i=-\infty}^{n_u} (U_i \cup U^{\prime}_i )  \right)
\end{equation}
where the integers $n_s$ and $n_u$ give the lower and upper bounds for the indices of the stable and unstable fundamental segments, respectively. For example, the pattern shown in Fig.~\ref{fig:Fundamental_Segments} is $T_{-1,1}$. 

For the study of chaotic transport, it is customary to define some special regions inside the homoclinic tangle, which govern the flux in and out of the tangle. Following the conventions~\cite{MacKay84a,Rom-Kedar90}, the phase-space region bounded by loop $US[x,g_0]$ is the $\mathit{complex}$ (also referred to as the $\mathit{resonance}$ $\mathit{zone}$ by Easton \cite{Easton86}), and the regions bounded by the loops $US[h_n,g_n]$ and $US[g_{n-1},h_n]$ are $\mathit{lobes}$ denoted by $L_n$ and $L^{\prime}_n$, respectively. The union of lobes $L_0$ and $L^{\prime}_0$ is often called a $\mathit{turnstile}$~\cite{MacKay84a}, as demonstrated by the hatched regions in Fig.~\ref{fig:Fundamental_Segments}. 

A simplifying assumption adopted here is the ``open system" condition~\cite{Rom-Kedar90,Mitchell03a}, which assumes the lobes $L^{\prime}_n$ and $L_{-n}$ with $n\geq 1$ extend out to infinity as $n$ increases and never enter the complex region.  Consequently, there are no homoclinic points distributed on the segments, $S(g_n,h_n)$ and $U(g_{n-1},h_n)$, which simplifies addressing the homoclinic orbits. However, this restriction is not essential and can be removed to accommodate closed systems as well.

\section{Symbolic dynamics}
\label{Symbolic dynamics}
Symbolic dynamics \cite{Hadamard1898,Birkhoff27a,Birkhoff35,Morse38} is a powerful construct that characterizes the topology of orbits in chaotic systems. In essence, it encodes the trajectories of various initial conditions under the mapping into infinite strings of alphabets, assigned using their phase space itineraries with respect to a generating Markov partition \cite{Bowen75,Gaspard98}. Constructions of exact generating partitions for general mixed systems, if possible, still remain challenging. However, finite approximations  can be obtained via efficient techniques introduced in \cite{Grassberger85a,Christiansen95,Christiansen96,Christiansen97,Rubido18}. 

\begin{figure}[ht]
\centering
{\includegraphics[width=6.5cm]{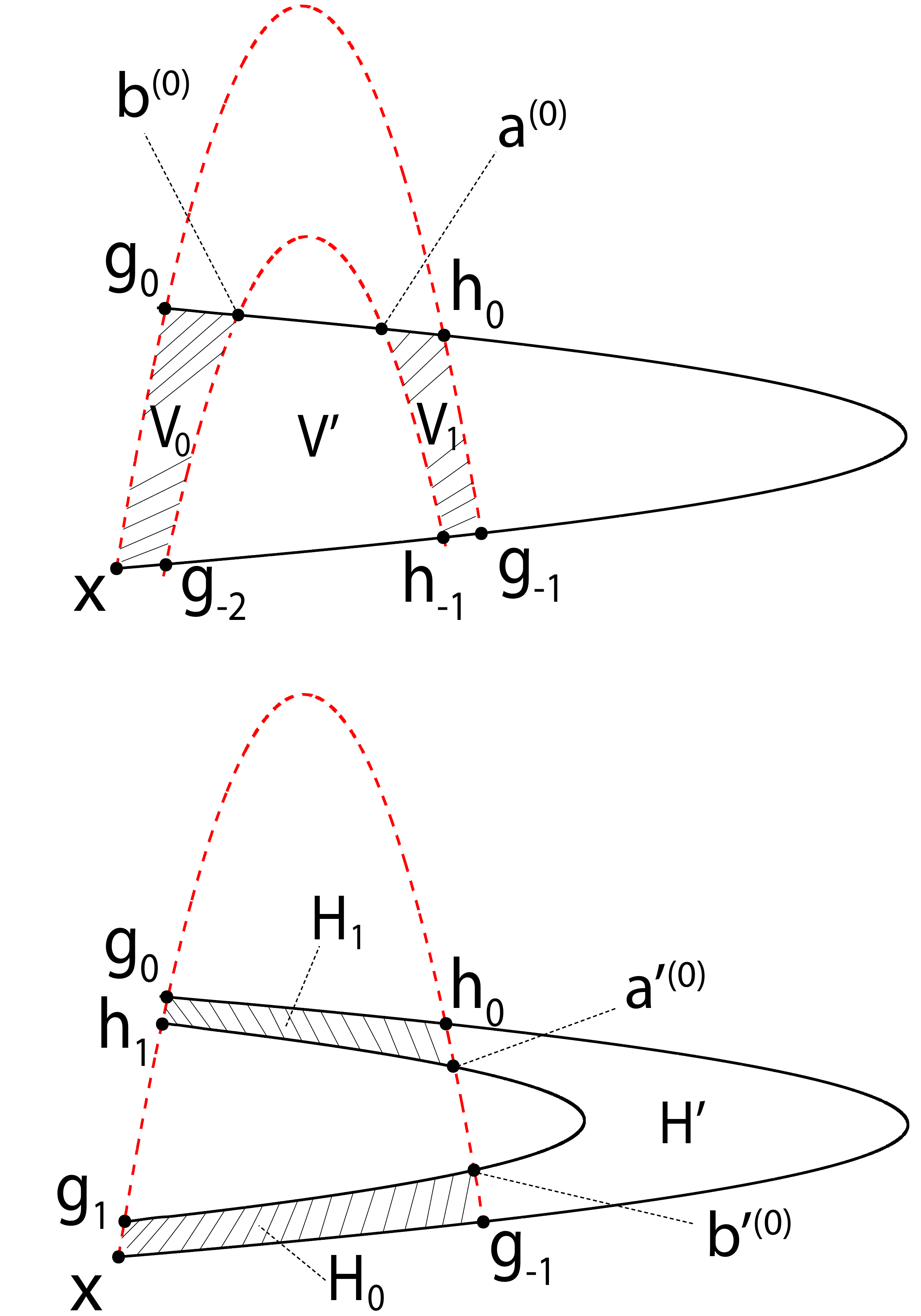}}
 \caption{(Color online) Smale horseshoe formed by $S(x)$ (red dashed curve) and $U(x)$ (black solid curve). The vertical strips $V_0$ and $V_1$ in the upper panel (hatched regions) are the generating partitions of the symbolic dynamics, and are mapped into the horizontal strips $H_0$ and $H_1$ in the lower panel (hatched regions), respectively, under one iteration. The fixed point $x$ has symbolic string $\overline{0}.\overline{0}$, and the primary homoclinic points $h_0$ and $g_0$ have symbolic strings $\overline{0}1.1\overline{0}$ and $\overline{0}1.\overline{0}$, respectively. }
\label{fig:Horseshoe}
\end{figure}  

Assume that the system is highly chaotic and the homoclinic tangle forms a complete Smale horseshoe \cite{Smale63,Smale80}, as the one depicted in Fig.~\ref{fig:Horseshoe}. The generating partition is then the collection of two regions $[V_0,V_1]$ (marked as hatched regions in the upper panel of the figure) where $V_0$ is the closed region bounded by $USUS[x,g_{-2},b^{(0)},g_0]=U[x,g_{-2}]+S[g_{-2},b^{(0)}]+U[b^{(0)},g_0]+S[g_0,x]$, and $V_1$ is the closed region bounded by $USUS[h_{-1},g_{-1},h_0,a^{(0)}]$. Note that the curvy-trapezoid region between $V_0$ and $V_1$ is also labeled in the figure as $V^{\prime}$, which is bounded by loop $USUS[g_{-2},h_{-1},a^{(0)},b^{(0)}]$. The deformation of these regions under the dynamics can be visualized in a simple way: under one iteration of $M$, the curvy-trapezoid region bounded by $USUS[x,g_{-1},h_0,g_0]$ (the union of $V_0$, $V^{\prime}$, and $V_1$) from the upper panel of Fig.~\ref{fig:Horseshoe} is compressed along its stable boundary and stretched along its unstable boundary while preserving the total area, folded into an U-shaped region bounded by $USUS[x,g_0,h_1,g_1]$ in the lower panel, which is the union of $H_0$, $H_1$, and $H^{\prime}$. During this process, the vertical strips $V_0$ and $V_1$ are mapped into the horizontal strips $H_0$ and $H_1$, respectively, marked by the hatched regions in the lower panel of Fig.~\ref{fig:Horseshoe}. In the meantime, $V^{\prime}$ is mapped into the U-shaped region $H^{\prime}$ bounded by $USUS[g_{-1},h_{0},a^{\prime(0)},b^{\prime(0)}]$ and will escape the complex region under further iterations. The inverse mapping of $M^{-1}$ has similar but reversed effects, with $M^{-1}(H_i)=V_{i}$ ($i=0,1$). 

Under the symbolic dynamics, each point $z_0$ inside the complex that never escapes under forward and inverse mappings can be put into an one-to-one correspondence with a bi-infinite symbolic string
\begin{equation}\label{eq:Symbolic string}
z_0 \Rightarrow \cdots s_{-2}s_{-1}.s_0 s_1 s_2 \cdots
\end{equation}
where each digit $s_n$ indicates the region that $M^{n}(z_0)$ lies in: $M^{n}(z_0)=z_n\in V_{s_n}$, where $s_n \in \lbrace 0,1 \rbrace$. The position of the decimal point indicates the present location of $z_0$ since $z_0 \in V_{s_0}$. The symbolic string gives an ``itinerary" of $z_0$ under successive forward and inverse iterations, in terms of the regions $V_0$ and $V_1$ in which each iteration lies. The mapping $M$ then corresponds to a Bernoulli shift on symbolic strings composed by ``$0$"s and ``$1$"s
\begin{equation}\label{eq:Bernoulli shift}
M^{n}(z_0) \Rightarrow \cdots s_{n-2}s_{n-1}.s_n s_{n+1} s_{n+2} \cdots
\end{equation}
therefore encoding the dynamics with simple strings of integers.  Assume a complete horseshoe structure here in which all possible combinations of substrings exist, i.e.~no ``pruning"~\cite{Cvitanovic88a,Cvitanovic91} is needed.  

The area-preserving H\'{e}non map \cite{Henon76} is used as a confirmation of the theory and its approximations:
\begin{equation}\label{eq:Henon map}
p_{n+1} = q_n, \qquad q_{n+1}=a-q_{n}^{2}-p_n.
\end{equation}
With parameter $a=10$, it gives rise to a complete horseshoe-shaped homoclinic tangle; see Fig.~\ref{fig:Horseshoe}.  As it satisfies both the complete horseshoe and open system assumptions, the theory is directly applicable.  Nevertheless, the results derived mostly carry over into more complicated systems possessing incomplete horseshoes \cite{Hagiwara04}, or systems with more than binary symbolic codes, though more work is needed to address such complications.  

The fixed point $x$ has the symbolic string $x\Rightarrow \cdots 0.0 \cdots = \overline{0}.\overline{0}$ where the overhead bar denotes infinite repetitions of ``$0$"s since it stays in (on the boundary of) $V_0$ forever. Consequently, other than the orbit containing the point $\overline{0}1.\overline{0}$, any homoclinic point $ h $ of $x$ must have a symbolic string of the form
\begin{equation}\label{eq:Homoclinic point symbolic string general form}
h \Rightarrow \overline{0} 1 s_{-m}\cdots s_{-1}.s_0 s_1 \cdots s_n 1 \overline{0}
\end{equation}  
along with all possible shifts of the decimal point. The $\overline{0}$ on both ends means the orbit approaches the fixed point asymptotically. The orbit $\lbrace h \rbrace$ can then be represented by the same symbolic string:
\begin{equation}\label{eq:Homoclinic orbit symbolic string general form}
\lbrace h \rbrace \Rightarrow \overline{0} 1 s_{-m}\cdots s_{-1} s_0 s_1 \cdots s_n 1 \overline{0}
\end{equation}  
with the decimal point removed, as compared to Eq.~\eqref{eq:Homoclinic point symbolic string general form}. The finite symbolic segment ``$1 s_{-m}\cdots s_{-1}s_0 s_1 \cdots s_n 1$" is often referred to as the \textit{core} of the symbolic code of $ h $, with its length referred to as the \textit{core length}. To be discussed in Appendix~\ref{Systematic assignment of symbolic codes}, the core length is a measure of the length of the phase-space excursion of $\lbrace h \rbrace$.  

The identification of symbolic strings associated with arbitrary homoclinic points, as well as the ordering of homoclinic points on the fundamental segments $S^{\prime}_n$ or $U_n$, are non-trivial tasks in general. Pioneering works along this line can be found in \cite{Sterling99}, where the symbolic assignment and relative ordering of homoclinic points on $S^{\prime}_0$ were explicitly given for the H\'{e}non map. Refer to Fig.~3 of \cite{Sterling99} for a nice pictorial demonstration. However, \cite{Sterling99} starts from the anti-integrable limit \cite{Aubry90,Aubry95} and derives the results as continuations of the limit. In Appendix~\ref{Systematic assignment of symbolic codes}, we introduce a different analytic scheme, which makes use of the hierarchical structure of the homoclinic tangle (see Sec.~\ref{Hierarchic structure of homoclinic points}) to provide the ordering of homoclinic points on $S^{\prime}_{-1}$ in terms of their symbolic codes. Based on the symbolic codes of the two primary homoclinic points on $S^{\prime}_{-1}$, which are $h_{-1} \Rightarrow \overline{0}.11\overline{0}$ and $g_{-2} \Rightarrow \overline{0}.01\overline{0}$, it recursively builds up the codes of the more complicated homoclinic orbits by adding certain symbolic strings of finite lengths to the primaries, according to their positions in the hierarchic strucutre. The results are equivalent to those of \cite{Sterling99} upon changing the alphabets ``$0$'' $\to$ ``$+$" and ``$1$'' $\to$ ``$-$". This approach naturally facilitates an important accumulation relation (introduced in Sec.~\ref{Asymptotic accumulation}) and thus better integrates into the scheme of the present work.

\section{Systematic assignments of symbolic codes}
\label{Systematic assignment of symbolic codes}

Although the symbolic codes of some simple homoclinic orbits, such as the primary ones, can be easily determined by following the numerical orbits, such tasks become prohibitive for the exponentially proliferating ensemble of more complicated, non-primary orbits.  In addition, a computational method does not reveal the patterns and structural relations buried in substrings of the symbolic codes. In fact, as shown by \cite{Sterling99}, symbolic codes provide a natural ordering of homoclinic points along the fundamental segments, which is otherwise unattainable from numerical methods. Although this problem is essentially solved by \cite{Sterling99} for the H\'{e}non maps in the complete horseshoe region, their approach starts from the anti-intergable limit \cite{Aubry90,Aubry95}, and identifies each homoclinic orbit near the limit as continuations from the anti-integrable limit. Although exact and efficient, it does not make use of the accumulation relations (Sec.~\ref{Asymptotic accumulation}) which are the theoretical foundations of the present paper. This appendix introduces a different approach. Taking advantage of the hierarchical structure of the homoclinic orbits (see Sec.~\ref{Hierarchic structure of homoclinic points}), a recursive scheme is introduced that systematically determines the symbolic codes of the families of winding-$(n+1)$ homoclinic orbits based on the symbolic code of the winding-$n$ orbit on which they accumulate. It results in an ordering of homoclinic points on the fundamental segment $S^{\prime}_{-1}$ in terms of their symbolic codes, which is equivalent to Lemma 7 of \cite{Sterling99} upon switching the alphabets ``$0$'' $\to$ ``$+$" and ``$1$'' $\to$ ``$-$".  This provides a foundation for the exact relations and approximations of Sec.~\ref{Homoclinic action formulae}. 

\begin{figure}[ht]
\centering
{\includegraphics[width=6.5cm]{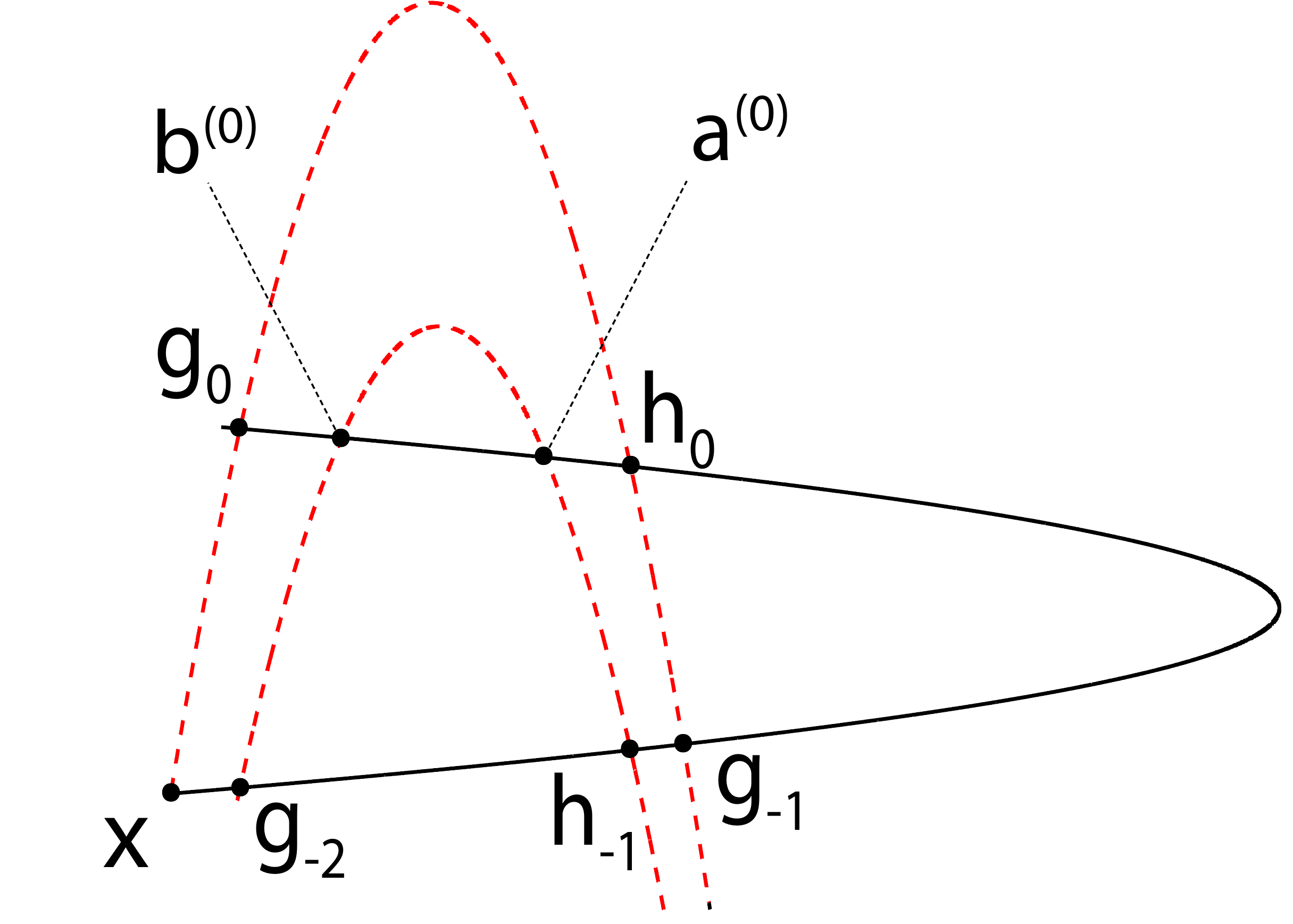}}
 \caption{(Color online) Homoclinic points in $T_{-1,0}$. The symbolic codes are: $h_{-1} \Rightarrow \overline{0}.11\overline{0}$, $g_{-2} \Rightarrow \overline{0}.01\overline{0}$, $a^{(0)} \Rightarrow \overline{0} 1.11  \overline{0}$, and $b^{(0)} \Rightarrow \overline{0} 1.01  \overline{0}$. The hierarchical relations are: $a^{(0)},b^{(0)} \xhookrightarrow[S]{1} g_{-2}$. Notice that the hierarchical relations are indicative for the assignments of symbolic codes: the codes of $a^{(0)}$ and $b^{(0)}$ can be obtained by adding the substrings ``$11$" and ``$10$", respectively, to the left end of the core of $g_{-2}$, while maintaining the position of the decimal point relative to the right end of the core. }
\label{fig:Trellis_0}
\end{figure}

Every homoclinic orbit has one and only one representative point on $S^{\prime}_{-1}$ and labeling the entire set of orbits can be reduced to labeling the homoclinic points on $S^{\prime}_{-1}$.  Starting from $T_{-1,-1}$, in which $S^{\prime}_{-1}$ is not intersected by any unstable fundamental segment, the only homoclinic points are the primaries $h_{-1} \Rightarrow \overline{0}.11\overline{0}$ and $g_{-2} \Rightarrow \overline{0}.01\overline{0}$, both of which are winding-$1$.  Proceeding to the intersections of $S^{\prime}_{-1}$ with $T_{-1,0}$, there are two winding-$2$ points, $a^{(0)}$ and $b^{(0)}$, as shown by Fig.~\ref{fig:Trellis_0},      
which are the leading terms of the two winding-$2$ families $[a^{(n)}]$ and $[b^{(n)}]$ from the future $T_{-1,n}$ that accumulate on $g_{-2}$.  Their symbolic codes are $a^{(0)} \Rightarrow \overline{0} 1.11  \overline{0}$ and $b^{(0)} \Rightarrow \overline{0} 1.01  \overline{0}$, which emerges quickly by following their excursions. The hierarchical relationship at this stage can be denoted alternatively as
\begin{equation}\label{eq:Accumulation on g first terms appendix}
\begin{split}
& ( a^{(0)} \Rightarrow \overline{0} 1.11  \overline{0} ) \xhookrightarrow[S]{1} ( g_{-2} \Rightarrow \overline{0} .01  \overline{0} ) \\
& ( b^{(0)} \Rightarrow \overline{0} 1.01  \overline{0} ) \xhookrightarrow[S]{1} ( g_{-2} \Rightarrow \overline{0} .01  \overline{0} )
\end{split}
\end{equation}
where the notations ``$\xhookrightarrow[S]{1}$" are defined in Eq.~\eqref{eq:Accumulation along stable}. Notice that the hierarchical relations imply the symbolic code assignments: the codes of $a^{(0)}$ and $b^{(0)}$ can be obtained by adding the substrings ``$11$" and ``$10$", respectively, to the left end of the core of $g_{-2}$, while maintaining the position of the decimal point relative to the right end of the core. Also, the transit times of $a^{(0)}$ and $b^{(0)}$ are both unity, and their core lengths are both $3$.  It turns out in general that
\begin{equation}\label{eq:transition time core length relation}
\textit{core length} = \textit{transit time} + 2,
\end{equation} 
which holds true for all non-primary homoclinic points.  Another important observation is, $a^{(0)}$ and $b^{(0)}$ with core lengths $3$, emerged from $S^{\prime}_{-1} \cap U_0$ in trellis $T_{-1,0}$. This leads to the simple fact that any non-primary homoclinic point that emerges from $S^{\prime}_{-1} \cap U_n$ in $T_{-1,n}$ must have core length $n+3$.  

\begin{figure}[ht]
\centering
{\includegraphics[width=6.5cm]{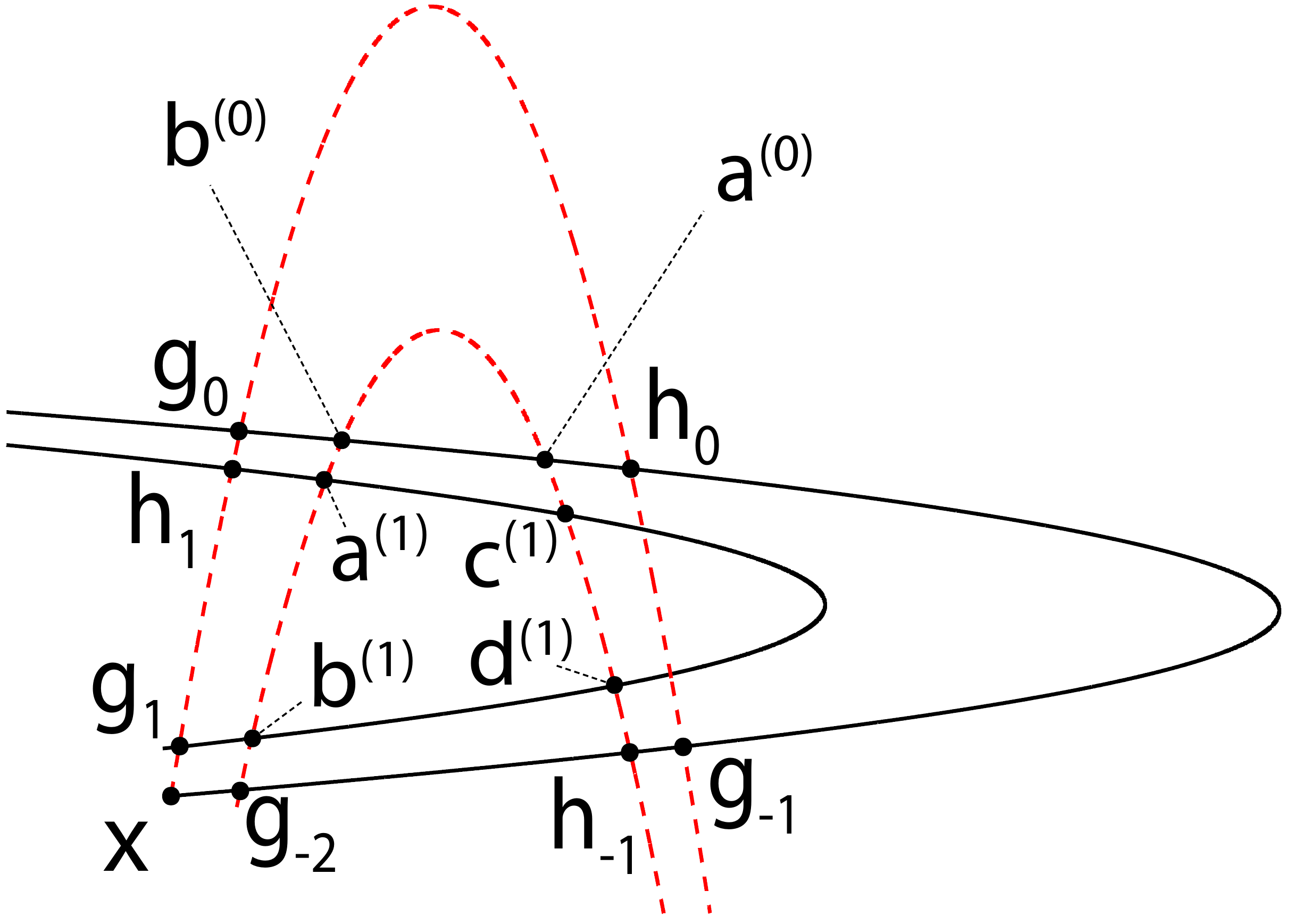}}
 \caption{(Color online) Homoclinic points in $T_{-1,1}$. The symbolic codes are: $a^{(1)} \Rightarrow \overline{0} 11.01  \overline{0}$, $b^{(1)} \Rightarrow \overline{0} 10.01  \overline{0}$, $c^{(1)} \Rightarrow \overline{0} 11.11  \overline{0}$, and $d^{(1)} \Rightarrow \overline{0} 10.11  \overline{0}$. The hierarchical relations are: $a^{(1)},b^{(1)} \xhookrightarrow[S]{2} g_{-2}$ and $c^{(1)},d^{(1)} \xhookrightarrow[S]{1} h_{-1}$. }
\label{fig:Trellis_1}
\end{figure}

There are four new intersections generated by $T_{-1,1}$,i.e., $S^{\prime}_{-1} \cap U_1$.  Figure \ref{fig:Trellis_1} shows the four new winding-$2$ points labeled $a^{(1)}$, $b^{(1)}$, $c^{(1)}$, and $d^{(1)}$. An important 
distinction between them is: $a^{(1)}$ and $b^{(1)}$ are the second realizations of their respective families $[a^{(n)}]$ and $[b^{(n)}]$ ($n \geq 0$) that accumulate on $g_{-2}$, whereas $c^{(1)}$ and $d^{(1)}$ are the first terms of their respective families, $[c^{(n)}]$ and $d^{(n)}$ ($n \geq 1$), that accumulate on $h_{-1}$.  Therefore, following the pattern of Eq.~\eqref{eq:Accumulation on g first terms appendix}, the symbolic codes of $c^{(1)}$ and $d^{(1)}$ should be obtained by adding the substring ``$11$" and ``$10$", respectively, to the left end of the core of $h_{-1}$ (which is ``$11$"), while keeping the position of the decimal point relative to the right end of the core. This leads to the assignments $c^{(1)} \Rightarrow  \overline{0} 11.11 \overline{0}$ and $d^{(1)} \Rightarrow  \overline{0} 10.11 \overline{0}$ according to the hierarchical relations 
\begin{equation}\label{eq:Accumulation on h first terms appendix}
\begin{split}
& ( c^{(1)} \Rightarrow \overline{0} 11.11 \overline{0} ) \xhookrightarrow[S]{1} ( h_{-1} \Rightarrow \overline{0}.11\overline{0} ) \\
& ( d^{(1)} \Rightarrow \overline{0} 10.11 \overline{0} ) \xhookrightarrow[S]{1} ( h_{-1} \Rightarrow \overline{0}.11\overline{0} ).
\end{split}
\end{equation}  

As for the symbolic codes of $a^{(1)}$ and $b^{(1)}$, since they are the second terms in their respective accumulating families, the substrings ``$110$" and ``$100$", instead of ``$11$" and ``$10$", should be added to the left end of the core of $g_{-2}$, respectively, while keeping the position of the decimal point relative to the right end of the core unchanged:
\begin{equation}\label{eq:Accumulation on g second terms appendix}
\begin{split}
& ( a^{(1)} \Rightarrow \overline{0} 11.01  \overline{0} ) \xhookrightarrow[S]{2} ( g_{-2} \Rightarrow \overline{0} .01  \overline{0} ) \\
& ( b^{(1)} \Rightarrow \overline{0} 10.01  \overline{0} ) \xhookrightarrow[S]{2} ( g_{-2} \Rightarrow \overline{0} .01  \overline{0} ).
\end{split}
\end{equation}
Calculating the orbits numerically, one readily verifies that Eqs.~\eqref{eq:Accumulation on h first terms appendix} and \eqref{eq:Accumulation on g second terms appendix} indeed give the correct desired symbolic codes for the orbits. 

Generalization of the above relations gives the general rule for the assignment of symbolic codes.  Given an arbitrary winding-$m$ homoclinic point $y$, and two winding-$(m+1)$ homoclinic points $z$ and $w$ from the two winding-$(m+1)$ families accumulating on $y$, such that $z \xhookrightarrow[S]{k} y$ and $w \xhookrightarrow[S]{k} y$ ($k \geq 1$) and $S[y,w] \subset S[y,z]$, then the symbolic codes of $z$ and $w$ can be obtained by adding the substrings ``$110^{k-1}$" and ``$100^{k-1}$", respectively, to the left end of the core of $y$, keeping the position of the decimal point relative to the right end of the core.  The notation ``$0^{k-1}$" denotes a string composed of $(k-1)$ consecutive ``$0$"s. Or equivalently, let the symbolic code of the orbit $\lbrace y \rbrace$ be $\lbrace y \rbrace \Rightarrow \overline{0} \tilde{s} \overline{0}$, where the string $\tilde{s}$ denotes the core, then the symbolic codes of orbits $\lbrace z \rbrace$ and $\lbrace w \rbrace$ are determined as
\begin{equation}\label{eq:Orbit symbolic codes general rules}
\begin{split}
& \lbrace z \rbrace \Rightarrow \overline{0} 110^{k-1} \tilde{s}  \overline{0}\\
& \lbrace w \rbrace \Rightarrow \overline{0} 100^{k-1} \tilde{s}  \overline{0}
\end{split}
\end{equation}
and the position of the decimal points in the symbolic codes of $z$ and $w$ are identical to that of $y$, when counted from the right ends of their cores.   

Concrete examples of the preceding assignment rules are labeled in Fig.~\ref{fig:Area_Cells_Zoom}.  Choose the winding-$2$ point $a^{(0)} \Rightarrow \overline{0} 1.11 \overline{0}$ as the base, and notice the accumulating points $e^{(k)},f^{(k)}  \xhookrightarrow[S]{k} a^{(0)}$, where the $k=1$ case is explicitly shown in the figure. According to the preceding assignment rules, the symbolic codes of $e^{(k)}$ and $f^{(k)}$ are constructed as $f^{(k)} \Rightarrow \overline{0} 110^{k-1} 1.11 \overline{0}$ and $e^{(k)} \Rightarrow \overline{0} 100^{k-1} 1.11 \overline{0}$, which was verified numerically. 

The proof of Eq.~\eqref{eq:Orbit symbolic codes general rules} involves mapping the base point $y$ simultaneously with $z$ and $w$ forward and inversely, to study the deformation of $S[y,z/w]$ under forward iterations, and the deformation of $U[y,z/w]$ under inverse iterations. Notice that the stable segments $S[y,z/w]$ belong to either $S[g_{-2},b^{(0)}]$ or $S[h_{-1},a^{(0)}]$, which will become even shorter under forward iterations. Therefore, forward iterations of $y$ and $z/w$ are guaranteed to locate on the same side of $S^{\prime}_{-1}$, thus in the same generating partition ($V_0$ or $V_1$). For the inverse mappings, the unstable segments $U[y,z/w]$ are constrained to deform in a specific way such that the images of $y$ and $z/w$ must locate in the same partition along the code segment ``$0^{k-1}\tilde{s}$" first. After that, the backward images of $z$ immediately visit $V_1$ twice, then stay in $V_0$ as they approach $x$; on the contrary, the backward images of $w$ visit $V_0$ and $V_1$ consecutively, and then stay in $V_0$ as they approach $x$. The slight difference in their behaviors give rise to the ``$\overline{0}11$" and ``$\overline{0}10$" in their respective symbolic codes in Eq.~\eqref{eq:Orbit symbolic codes general rules}. The detailed derivation is quite lengthy and skipped here for brevity. 

With Eq.~\eqref{eq:Orbit symbolic codes general rules}, the complete set of symbolic codes is generated based on just the symbolic codes of the two primary orbits.  For a finite trellis $T_{-1,N}$ (presumably with large $N$), the maximum transition time of homoclinic orbits is $N+1$, i.e., those arise from $S^{\prime}_{-1} \cap U_N$. According to Eq.~\eqref{eq:transition time core length relation}, the corresponding maximum core length is $N+3$. Therefore, starting from $\lbrace h_{-1} \rbrace \Rightarrow \overline{0} 11 \overline{0}$ and $\lbrace g_{-2} \rbrace \Rightarrow \overline{0} 1 \overline{0}$, by intersecting $S^{\prime}_{-1}$ with successive $U_i$ where $0\leq i \leq N$ and recursive use of Eq.~\eqref{eq:Orbit symbolic codes general rules} up to core length $N+3$, the symbolic codes of all homoclinic orbits present in $T_{-1,N}$ are generated according to the relative positions of their representative points on $S^{\prime}_{-1}$. This process is equivalent to the $>_s$ ordering in Lemma 7 of \cite{Sterling99}. 

A similar prescription could have been generated for the accumulating homoclinic families along the unstable manifold under inverse mappings.  Given any winding-$n$ homoclinic point $y^{\prime}$, and two winding-$(n+1)$ homoclinic points $z^{\prime}$ and $w^{\prime}$ such that $z^{\prime}  \xhookrightarrow[U]{k} y^{\prime} $, $w^{\prime}  \xhookrightarrow[U]{k} y^{\prime}$ ($k \geq 1$) and $U[y^{\prime},w^{\prime}] \subset U[y^{\prime}, z^{\prime}]$, the symbolic codes of $z^{\prime} $ and $w^{\prime}$ can be constructed by adding the substrings ``$0^{k-1}11$" and ``$0^{k-1}01$", respectively, to the right end of the core of the symbolic code of $y^{\prime}$, while keeping the position of the decimal point relative to the left end of the core unchanged. Or equivalently, if we let the symbolic code of the orbit be $\lbrace y^{\prime} \rbrace \Rightarrow \overline{0} \tilde{s}^{\prime} \overline{0}$ where $\tilde{s}^{\prime}$ denotes the core, then the symbolic codes of orbits $\lbrace z^{\prime} \rbrace$ and $\lbrace w^{\prime} \rbrace$ are constructed as
\begin{equation}\label{eq:Orbit symbolic codes general rules inverse mapping}
\begin{split}
& \lbrace z^{\prime} \rbrace \Rightarrow \overline{0} \tilde{s}^{\prime} 0^{k-1}11 \overline{0} \\
& \lbrace w^{\prime} \rbrace \Rightarrow \overline{0}  \tilde{s}^{\prime} 0^{k-1}01 \overline{0}
\end{split}
\end{equation}
which is in complete analogy to Eq.~\eqref{eq:Orbit symbolic codes general rules}, and equivalent to the $>_u$ ordering in Lemma 7 of \cite{Sterling99}. For example, in Fig.~\ref{fig:Scaling_Inverse} we have $g_0 \Rightarrow \overline{0} 1. \overline{0} $, and $v^{(-k)},w^{(-k)} \xhookrightarrow[U]{k} g_0$, where the $k=1,2$ cases are explicitly shown in the figure. Then according to the preceding rules, the symbolic codes of $v^{(-k)}$ and $w^{(-k)}$ are constructed from the symbolic code of $g_0$ as $v^{(-k)} \Rightarrow  \overline{0} 1. 0^{k-1}11 \overline{0}$ and $w^{(-k)} \Rightarrow  \overline{0} 1. 0^{k-1}01 \overline{0}$, respectively.

\section{Asymptotic accumulation exponent}
\label{ASYMPTOTIC ACCUMULATION EXPONENT}

\begin{figure}[ht]
\centering
{\includegraphics[width=6.5cm]{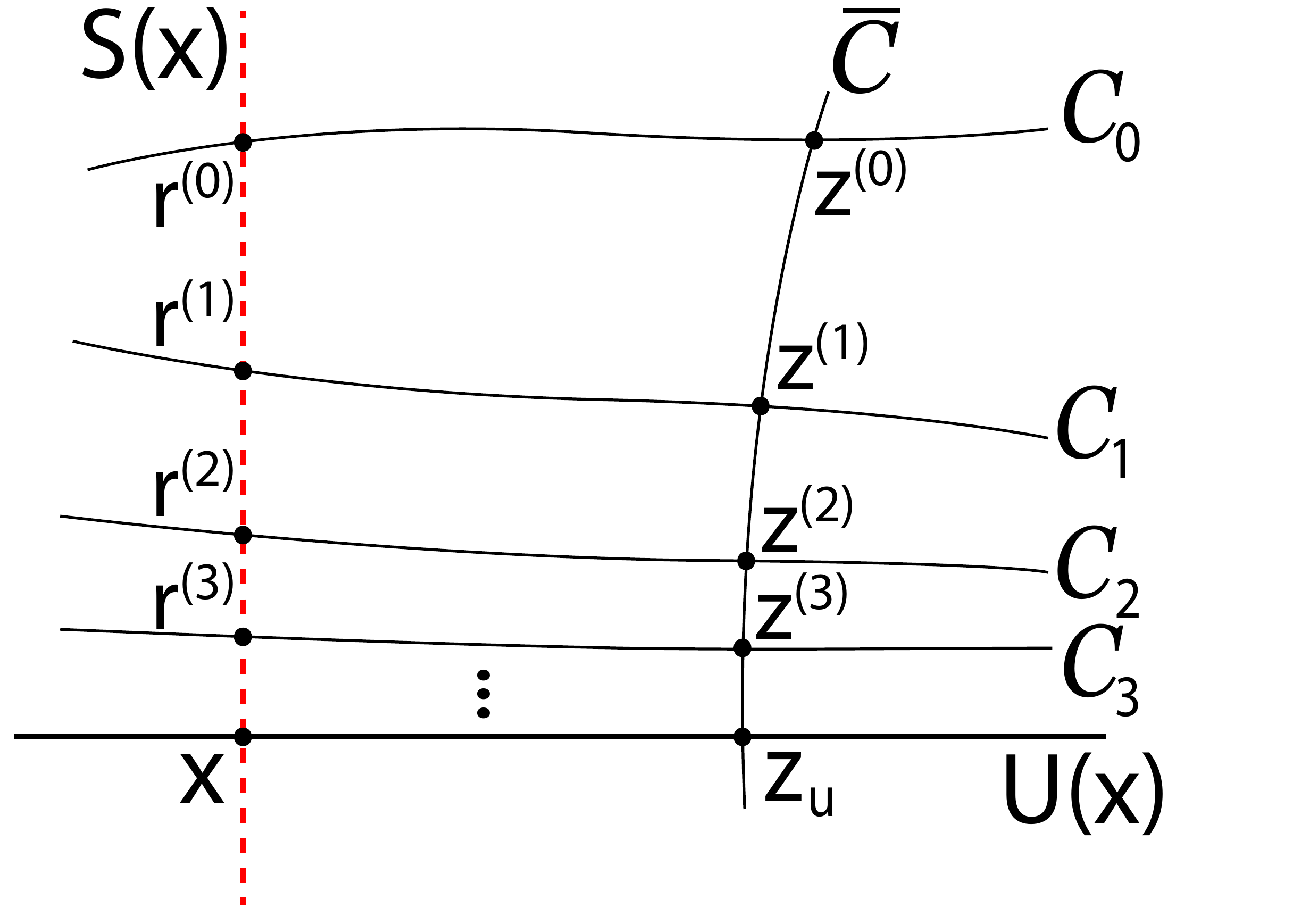}}
 \caption{(Schematic, color online) Iterates of a curve intersecting the stable manifold approach the unstable manifold. Future iterations of the curve ${\cal C}_0$ creates a family of curves $[{\cal C}_n]$, which intersect $\overline{{\cal C}}$ at a family of points $[z^{(n)}]$. $[z^{(n)}]$ accumulates on $z_u$ under the exponent $\mu_x$, as given by Eq.~\eqref{eq:Mitchell scaling relation}. }
\label{fig:Mitchell_Theorem}
\end{figure}  

The foundation of Sec.~\ref{Asymptotic accumulation} is established by Lemma 2 in Appendix.~B.~3 of \cite{Mitchell03a}, and a brief overview of their results is given here. The setting of the lemma is demonstrated schematically by Fig.~\ref{fig:Mitchell_Theorem}. Let $z_u$ be an arbitrary point on $U(x)$, and $\overline{{\cal C}}$ an arbitrary differentiable curve passing transversely through $U(x)$ at $z_u$. Consider another arbitrary differentiable curve, ${\cal C}_0$, which passes through $S(x)$ transversely at $r^{(0)}$, and intersects $\overline{{\cal C}}$ at $z^{(0)}$. Then, its future iterations ${\cal C}_n = M^n ({\cal C}_0)$ ($n \geq 1$) pass through $S(x)$ transversely at $r^{(n)}$, and intersect $\overline{ {\cal C} }$ at $z^{(n)}$, which form a family of points $[z^{(n)}]$ that accumulate asymptotically on the base point $z_u$:
\begin{equation}\label{eq:Mitchell scaling relation}
\begin{split}
& \lim_{n \to \infty} z^{(n)} = z_{u} \\
& \lim_{n \to \infty} | z^{(n)} - z_{u} | e^{n\mu_x} = C(z_u,z^{(0)})
\end{split}
\end{equation}
where $||$ is the standard Euclidean vector norm, $\mu_x$ is the stability exponent of $x$, and $C(z_u,z^{(0)})$ is a positive constant depending on the base point $z_u$ and the leading term $z^{(0)}$ in the asymptotic family. Notice that Eq.~\eqref{eq:Mitchell scaling relation} is just a re-expression of Eqs.~(B5) and (B6) of \cite{Mitchell03a}. What is surprising here is that even though the manifolds explore the vast majority of phase space with clearly non-uniform expansion rates, the asymptotic exponent in the above equation is still that of the hyperbolic fixed point.

\section{Area correspondence relations}
\label{Area correspondence relations}
Given any homoclinic point $y \in ( S^{\prime}_{-1} \cap U_{m} )$, there is an explicit relation that links ${\cal A}^{\circ}(y)={\cal A}^{\circ}_{SUSU[ y, P_S(y), P_S P_U(y), P_U(y) ]}$ with specific linear combinations of cell areas from the type-I and type-II partition trees of $T_{-1,m}$. The transition time of $y$ is $m+1$, so its core length is $m+3$. Let $\tilde{s} = s_1 s_2 \cdots s_{m+2} s_{m+3} $ ($s_i \in \lbrace 0,1 \rbrace$, $s_1=s_{m+3}=1$) be the core of the symbolic code of $y$, then the linear combination of cell areas depends solely on $\tilde{s}$. The correspondence relation is established in the following step.

\begin{enumerate}
\item[1)] Define $\Phi_{B \mapsto A}$ to be a mapping from the cells of the type-II partition trees to the cells of the type-I partition trees, such that for any finite Greek alphabet string $\tilde{\omega}$ composed of $\alpha$, $\gamma$, and $\beta$ ($\tilde{\omega}$ could also be an empty string) we have 
\begin{equation}\label{eq:Definition Phi B to A}
 \begin{cases}
    \Phi_{B \mapsto A} (B_{ \tilde{\omega}  }) = A_{ \tilde{\omega} }  \\
    \Phi_{B \mapsto A} (A_{ \tilde{\omega}  }) = \varnothing \\
    \Phi_{B \mapsto A} ( \varnothing ) = \varnothing
  \end{cases}
\end{equation}
where $\varnothing$ denotes a null cell that gives zero contribution to the action calculations. 

\item[2)] Define $\Phi_{\beta \mapsto \alpha}$ to be a mapping between the cells of the partition trees, such that for any finite Greek alphabet string $\tilde{\omega}$ composed of $\alpha$ and $\gamma$ (but not $\beta$, note also that $\tilde{\omega}$ could be an empty string), we have 
\begin{equation}\label{eq:Definition Phi beta to alpha}
 \begin{cases}
    \Phi_{\beta \mapsto \alpha} (A_{ \tilde{\omega} \beta }) = A_{ \tilde{\omega} \alpha }  \\
    \Phi_{\beta \mapsto \alpha} (B_{ \tilde{\omega} \beta }) = B_{ \tilde{\omega} \alpha }  \\ 
    \Phi_{\beta \mapsto \alpha} (A_{ \tilde{\omega} \alpha }) = \Phi_{\beta \mapsto \alpha} (B_{ \tilde{\omega} \alpha }) = \varnothing  \\
    \Phi_{\beta \mapsto \alpha} (A) =  \Phi_{\beta \mapsto \alpha} (B) = \varnothing  \\
    \Phi_{\beta \mapsto \alpha} ( \varnothing ) = \varnothing 
  \end{cases}
\end{equation}

\item[3)] Define $\Gamma$ to be a mapping from the core $\tilde{s}=s_1 s_2 \cdots s_{m+2} s_{m+3}$ ($s_1=s_{m+3}=1$) of the symbolic code of any non-primary homoclinic point $y \in (S^{\prime}_{-1} \cap U_m)$ to the cells of the partition trees, such that depending on the detailed forms of $\tilde{s}$, the mapping $\Gamma$ takes the forms:
\begin{equation}\label{eq:Definition Gamma}
\Gamma(\tilde{s})=
\begin{cases}
	\ \Gamma(101 )=A  \\
	\\
	\ \Gamma(111)=B   \\
	\\
\begin{array}{cccccccccc}
\Gamma( & \underbracket{10} & \cdots & \underbracket{0} & \cdots & \underbracket{1} & \cdots & \underbracket{01} & ) & = A_{\cdots \gamma \cdots \alpha \cdots \alpha}\\
& \downarrow && \downarrow && \downarrow && \downarrow &&\\
& \alpha & \cdots & \alpha & \cdots & \gamma & \cdots & A && 
\end{array}	
     \\
     \\
     \begin{array}{cccccccccc}
\Gamma( & \underbracket{10} & \cdots & \underbracket{0} & \cdots & \underbracket{1} & \cdots & \underbracket{11} & ) & = B_{\cdots \gamma \cdots \alpha \cdots \alpha}\\
& \downarrow && \downarrow && \downarrow && \downarrow &&\\
& \alpha & \cdots & \alpha & \cdots & \gamma & \cdots & B &&
\end{array}
 	\\
 	\\
 	\begin{array}{cccccccccc}
\Gamma( & \underbracket{11} & \cdots & \underbracket{0} & \cdots & \underbracket{1} & \cdots & \underbracket{01} & ) & = A_{\cdots \gamma \cdots \alpha \cdots \beta}\\
& \downarrow && \downarrow && \downarrow && \downarrow &&\\
& \beta & \cdots & \alpha & \cdots & \gamma & \cdots & A &&
\end{array}
	\\
	\\
	\begin{array}{cccccccccc}
\Gamma( & \underbracket{11} & \cdots & \underbracket{0} & \cdots & \underbracket{1} & \cdots & \underbracket{11} & ) & = B_{\cdots \gamma \cdots \alpha \cdots \beta}\\
& \downarrow && \downarrow && \downarrow && \downarrow &&\\
& \beta & \cdots & \alpha & \cdots & \gamma & \cdots & B &&
\end{array}
\end{cases}
\end{equation} 
in which the $\tilde{s} = 101$ and $\tilde{s}=111$ cases yield cells $A$ and $B$, respectively; and all the rest of cases with core lengths $\geq 4$ (or equivalently $m \geq 1$) are categorized into four cases, $\lbrace s_2 =0 , s_{m+2}=0 \rbrace$, $\lbrace s_2 = 0, s_{m+2}=1 \rbrace$, $\lbrace s_2 = 1, s_{m+2} = 0 \rbrace$, and $\lbrace s_2 = 1, s_{m+2} = 1 \rbrace$, which correspond to the third, fourth, fifth, and sixth line of Eq.~\eqref{eq:Definition Gamma}, respectively. Notice in those four cases, the letters $A$ and $B$ of the cell names are given by the last two digits $s_{m+2}s_{m+3}$ of $\tilde{s}$ with grammar ``$01 \mapsto A$" and ``$11 \mapsto B$". The Greek alphabet string of the cell names are given by the first $m+1$ digits of $\tilde{s}$ in an reversed order: $s_1s_2$ gives the last alphabet in the Greek string, with grammar ``$10 \mapsto \alpha$" and ``$11 \mapsto \beta$"; and $s_{m+1}s_{m}\cdots s_4s_3$ (reversed string of $s_3s_4\cdots s_m s_{m+1}$) gives the first $m-1$ alphabets in the Greek string, with grammar ``$0 \mapsto \alpha$" and ``$1 \mapsto \gamma$". 
 
 \item[4)] Finally, ${\cal A}^{\circ}(y)$ can be calculated as
 \begin{equation}\label{eq:A_SUSU in terms of partition tree cells}
 \begin{split}
& {\cal A}^{\circ}(y)=(-1)^{ n_{\gamma} \left( \Gamma ( \tilde{ s } ) \right) } \cdot \Big[ \Gamma ( \tilde{ s } ) + \Phi_{\beta \mapsto \alpha}( \Gamma ( \tilde{s} ) )\\
& + \Phi_{B \mapsto A}( \Gamma ( \tilde{s} ) ) + \Phi_{B \mapsto A} \big( \Phi_{\beta \mapsto \alpha}( \Gamma ( \tilde{s} ) ) \big)   \Big]
 \end{split}
 \end{equation}
 where $n_{\gamma} ( \Gamma ( \tilde{ s } ))$ is a function that returns the total number of $\gamma$ in the Greek alphabet string of the cell $\Gamma ( \tilde{ s } )$. For example, $n_{\gamma} (A_{\alpha \beta})=0$ and $n_{\gamma} (B_{\gamma\beta})=1$. Again, we emphasize that Eq.~\eqref{eq:A_SUSU in terms of partition tree cells} only applies to non-primary homoclinic points $y$ located on $S^{\prime}_{-1}$. 

\end{enumerate}

Eq.~\eqref{eq:A_SUSU in terms of partition tree cells} gives a systematic way of identifying the $ {\cal A}^{\circ}(y)$ term in the homoclinic action decomposition [Eq.~\eqref{eq:Action formulae once reduction}] in terms of a linear combination of cell areas from the type-I and type-II partition trees. In practice, some of the terms in Eq.~\eqref{eq:A_SUSU in terms of partition tree cells} will vanish due to the presence of null areas ($\varnothing $) in Eqs.~\eqref{eq:Definition Phi B to A} and \eqref{eq:Definition Phi beta to alpha}. Depending on $y$, Eq.~\eqref{eq:A_SUSU in terms of partition tree cells} may take four possible forms, as listed below:
\begin{enumerate}

\item[1)] A single type-I cell area: $A$ (for $y$=$b^{(0)}$ only) or $A_{\tilde{\omega} \alpha}$, where $\tilde{\omega}$ denotes some Greek alphabet string composed by $\alpha$ and $\gamma$. Examples are
\begin{enumerate}
\item[i)] In Fig.~\ref{fig:Trellis_0}, let $y=b^{(0)} \in ( S^{\prime}_{-1}\cap U_0) $, then
\begin{equation}
{\cal A}^{\circ}(y)=A, \nonumber
\end{equation}
which a type-I cell of $T_{-1,0}$. 
\item[ii)] In Fig.~\ref{fig:Trellis_Partition_1}, let $y=b^{(1)} \in (S^{\prime}_{-1} \cap U_1)$, then
\begin{equation}
{\cal A}^{\circ}(y) =A_{\alpha}, \nonumber
\end{equation}
which a type-I cell of $T_{-1,1}$. 
\item[iii)] In Fig.~\ref{fig:Area_Cells_Zoom}, let $y=r^{(1)} \in ( S^{\prime}_{-1}\cap U_2 ) $, then
\begin{equation}
{\cal A}^{\circ}(y)  = -A_{\gamma \alpha}, \nonumber
\end{equation}
which is a type-I cell area of $T_{-1,2}$. 
\end{enumerate} 

\item[2)] Two type-I areas: $A_{\tilde{\omega}\beta} + A_{\tilde{\omega}\alpha}$. Examples are
\begin{enumerate}
\item[i)] In Fig.~\ref{fig:Trellis_Partition_1}, let $y=a^{(1)} \in ( S^{\prime}_{-1}\cap U_1 )$, then
\begin{equation}
{\cal A}^{\circ}(y) =A_{\beta}+A_{\alpha}, \nonumber
\end{equation}
which is the sum of two type-I areas of $T_{-1,1}$. 
\item[ii)] In Fig.~\ref{fig:Area_Cells_Zoom}, let $y=s^{(1)} \in ( S^{\prime}_{-1}\cap U_2 )$, then 
\begin{equation}
{\cal A}^{\circ}(y)=-(A_{\gamma\beta}+A_{\gamma\alpha}), \nonumber
\end{equation}
which is the sum of two type-I areas of $T_{-1,2}$.
\end{enumerate}

\item[3)] A type-I area and a type-II area: $A+B$ (for $y=a^{(0)}$ only) or $A_{\tilde{\omega}\alpha}+B_{\tilde{\omega}\alpha}$. Examples are
\begin{enumerate}
\item[i)] In Fig.~\ref{fig:Trellis_0}, let $y=a^{(0)} \in ( S^{\prime}_{-1} \cap U_0 )$. Recall that only for the special case of $y=a^{(0)}$, we alter Eq.~\eqref{eq:Action formulae once reduction} into Eq.~\eqref{eq:Action formulae once reduction alternative for a0}, whose area term gives  
\begin{equation}
{\cal A}^{\circ}_{SUSU[ a^{(0)}, h_{-1}, x, g_0 ]}=A+B \nonumber
\end{equation}
which is the sum of a type-I and a type-II area of $T_{-1,0}$.
\item[ii)] In Fig.~\ref{fig:Trellis_Partition_1}, let $y=d^{(1)} \in ( S^{\prime}_{-1}\cap U_1 )$, then  
\begin{equation}
{\cal A}^{\circ}(y) = A_{\alpha} + B_{\alpha}, \nonumber
\end{equation}
which is the sum of a type-I and a type-II area. 
\item[iii)] In Fig.~\ref{fig:Area_Cells_Zoom}, let $y=e^{(1)} \in (S^{\prime}_{-1}\cap U_2)$, then
\begin{equation}
{\cal A}^{\circ}(y) = -( A_{\gamma\alpha} + B_{\gamma\alpha} ), \nonumber
\end{equation}
which is the sum of a type-I and a type-II area of $T_{-1,2}$.
\end{enumerate}

\item[4)] Two type-I areas plus two type-II areas: $A_{\tilde{\omega}\alpha} + A_{\tilde{\omega}\beta} + B_{\tilde{\omega}\alpha} + B_{\tilde{\omega}\beta} $. Examples are
\begin{itemize}
\item[i)] In Fig.~\ref{fig:Trellis_Partition_1}, let $y=c^{(1)} \in ( S^{\prime}_{-1}\cap U_1 )$, then 
\begin{equation}
{\cal A}^{\circ}(y)= A_{\alpha} + A_{\beta} + B_{\alpha} + B_{\beta}, \nonumber
\end{equation}
which is the sum of two type-I and two type-II areas of $T_{-1,1}$. 
\item[ii)] In Fig.~\ref{fig:Area_Cells_Zoom}, let $y=f^{(1)} \in ( S^{\prime}_{-1}\cap U_2 )$, then
\begin{equation}
\begin{split}
&{\cal A}^{\circ}(y)=-(A_{\gamma\alpha} + A_{\gamma\beta} + B_{\gamma\alpha} + B_{\gamma\beta}), \nonumber
\end{split}
\end{equation}
which is the sum of two type-I and two type-II areas of $T_{-1,2}$. 
\end{itemize}

\end{enumerate}

\acknowledgments

JL gratefully acknowledges many inspiring discussions with Akira Shudo during several productive visits to Tokyo Metropolitan University.

\bibliography{classicalchaos,quantumchaos,general_ref}

\begin{thebibliography}{66}%
\makeatletter
\providecommand \@ifxundefined [1]{%
 \@ifx{#1\undefined}
}%
\providecommand \@ifnum [1]{%
 \ifnum #1\expandafter \@firstoftwo
 \else \expandafter \@secondoftwo
 \fi
}%
\providecommand \@ifx [1]{%
 \ifx #1\expandafter \@firstoftwo
 \else \expandafter \@secondoftwo
 \fi
}%
\providecommand \natexlab [1]{#1}%
\providecommand \enquote  [1]{``#1''}%
\providecommand \bibnamefont  [1]{#1}%
\providecommand \bibfnamefont [1]{#1}%
\providecommand \citenamefont [1]{#1}%
\providecommand \href@noop [0]{\@secondoftwo}%
\providecommand \href [0]{\begingroup \@sanitize@url \@href}%
\providecommand \@href[1]{\@@startlink{#1}\@@href}%
\providecommand \@@href[1]{\endgroup#1\@@endlink}%
\providecommand \@sanitize@url [0]{\catcode `\\12\catcode `\$12\catcode
  `\&12\catcode `\#12\catcode `\^12\catcode `\_12\catcode `\%12\relax}%
\providecommand \@@startlink[1]{}%
\providecommand \@@endlink[0]{}%
\providecommand \url  [0]{\begingroup\@sanitize@url \@url }%
\providecommand \@url [1]{\endgroup\@href {#1}{\urlprefix }}%
\providecommand \urlprefix  [0]{URL }%
\providecommand \Eprint [0]{\href }%
\providecommand \doibase [0]{http://dx.doi.org/}%
\providecommand \selectlanguage [0]{\@gobble}%
\providecommand \bibinfo  [0]{\@secondoftwo}%
\providecommand \bibfield  [0]{\@secondoftwo}%
\providecommand \translation [1]{[#1]}%
\providecommand \BibitemOpen [0]{}%
\providecommand \bibitemStop [0]{}%
\providecommand \bibitemNoStop [0]{.\EOS\space}%
\providecommand \EOS [0]{\spacefactor3000\relax}%
\providecommand \BibitemShut  [1]{\csname bibitem#1\endcsname}%
\let\auto@bib@innerbib\@empty
\bibitem [{\citenamefont {Cvitanovi{\'c}}\ \emph {et~al.}(2016)\citenamefont
  {Cvitanovi{\'c}}, \citenamefont {Artuso}, \citenamefont {Mainieri},
  \citenamefont {Tanner},\ and\ \citenamefont {Vattay}}]{ChaosBook}%
  \BibitemOpen
  \bibfield  {author} {\bibinfo {author} {\bibfnamefont {P.}~\bibnamefont
  {Cvitanovi{\'c}}}, \bibinfo {author} {\bibfnamefont {R.}~\bibnamefont
  {Artuso}}, \bibinfo {author} {\bibfnamefont {R.}~\bibnamefont {Mainieri}},
  \bibinfo {author} {\bibfnamefont {G.}~\bibnamefont {Tanner}}, \ and\ \bibinfo
  {author} {\bibfnamefont {G.}~\bibnamefont {Vattay}},\ }\href
  {http://ChaosBook.org/} {\emph {\bibinfo {title} {Chaos: Classical and
  Quantum}}}\ (\bibinfo  {publisher} {Niels Bohr Inst.},\ \bibinfo {address}
  {Copenhagen},\ \bibinfo {year} {2016})\BibitemShut {NoStop}%
\bibitem [{\citenamefont {So}(2007)}]{So07}%
  \BibitemOpen
  \bibfield  {author} {\bibinfo {author} {\bibfnamefont {P.}~\bibnamefont
  {So}},\ }\href@noop {} {\bibfield  {journal} {\bibinfo  {journal}
  {Scholarpedia}\ }\textbf {\bibinfo {volume} {2}},\ \bibinfo {pages} {1353}
  (\bibinfo {year} {2007})}\BibitemShut {NoStop}%
\bibitem [{\citenamefont {Gutzwiller}(1971)}]{Gutzwiller71}%
  \BibitemOpen
  \bibfield  {author} {\bibinfo {author} {\bibfnamefont {M.~C.}\ \bibnamefont
  {Gutzwiller}},\ }\href@noop {} {\bibfield  {journal} {\bibinfo  {journal}
  {J.~Math.~Phys.}\ }\textbf {\bibinfo {volume} {12}},\ \bibinfo {pages} {343}
  (\bibinfo {year} {1971})},\ \bibinfo {note} {and references
  therein}\BibitemShut {NoStop}%
\bibitem [{\citenamefont {Du}\ and\ \citenamefont
  {Delos}(1988{\natexlab{a}})}]{Du88a}%
  \BibitemOpen
  \bibfield  {author} {\bibinfo {author} {\bibfnamefont {M.~L.}\ \bibnamefont
  {Du}}\ and\ \bibinfo {author} {\bibfnamefont {J.~B.}\ \bibnamefont {Delos}},\
  }\href@noop {} {\bibfield  {journal} {\bibinfo  {journal} {Phys.~Rev.~A}\
  }\textbf {\bibinfo {volume} {38}},\ \bibinfo {pages} {1896} (\bibinfo {year}
  {1988}{\natexlab{a}})}\BibitemShut {NoStop}%
\bibitem [{\citenamefont {Du}\ and\ \citenamefont
  {Delos}(1988{\natexlab{b}})}]{Du88b}%
  \BibitemOpen
  \bibfield  {author} {\bibinfo {author} {\bibfnamefont {M.~L.}\ \bibnamefont
  {Du}}\ and\ \bibinfo {author} {\bibfnamefont {J.~B.}\ \bibnamefont {Delos}},\
  }\href@noop {} {\bibfield  {journal} {\bibinfo  {journal} {Phys.~Rev.~A}\
  }\textbf {\bibinfo {volume} {38}},\ \bibinfo {pages} {1913} (\bibinfo {year}
  {1988}{\natexlab{b}})}\BibitemShut {NoStop}%
\bibitem [{\citenamefont {Tomsovic}\ and\ \citenamefont
  {Heller}(1993)}]{Tomsovic93}%
  \BibitemOpen
  \bibfield  {author} {\bibinfo {author} {\bibfnamefont {S.}~\bibnamefont
  {Tomsovic}}\ and\ \bibinfo {author} {\bibfnamefont {E.~J.}\ \bibnamefont
  {Heller}},\ }\href@noop {} {\bibfield  {journal} {\bibinfo  {journal}
  {Phys.~Rev.~E}\ }\textbf {\bibinfo {volume} {47}},\ \bibinfo {pages} {282}
  (\bibinfo {year} {1993})}\BibitemShut {NoStop}%
\bibitem [{\citenamefont {Birkhoff}(1927{\natexlab{a}})}]{Birkhoff27}%
  \BibitemOpen
  \bibfield  {author} {\bibinfo {author} {\bibfnamefont {G.~D.}\ \bibnamefont
  {Birkhoff}},\ }\href@noop {} {\bibfield  {journal} {\bibinfo  {journal} {Acta
  Math.}\ }\textbf {\bibinfo {volume} {50}},\ \bibinfo {pages} {359} (\bibinfo
  {year} {1927}{\natexlab{a}})}\BibitemShut {NoStop}%
\bibitem [{\citenamefont {Moser}(1956)}]{Moser56}%
  \BibitemOpen
  \bibfield  {author} {\bibinfo {author} {\bibfnamefont {J.}~\bibnamefont
  {Moser}},\ }\href@noop {} {\bibfield  {journal} {\bibinfo  {journal}
  {Commun.~Pure Appl.~Math.}\ }\textbf {\bibinfo {volume} {9}},\ \bibinfo
  {pages} {673} (\bibinfo {year} {1956})}\BibitemShut {NoStop}%
\bibitem [{\citenamefont {da~Silva~Ritter}\ \emph {et~al.}(1987)\citenamefont
  {da~Silva~Ritter}, \citenamefont {Ozorio~de Almeida},\ and\ \citenamefont
  {Douady}}]{Silva87}%
  \BibitemOpen
  \bibfield  {author} {\bibinfo {author} {\bibfnamefont {G.~L.}\ \bibnamefont
  {da~Silva~Ritter}}, \bibinfo {author} {\bibfnamefont {A.~M.}\ \bibnamefont
  {Ozorio~de Almeida}}, \ and\ \bibinfo {author} {\bibfnamefont
  {R.}~\bibnamefont {Douady}},\ }\href@noop {} {\bibfield  {journal} {\bibinfo
  {journal} {Physica~D}\ }\textbf {\bibinfo {volume} {29}},\ \bibinfo {pages}
  {181} (\bibinfo {year} {1987})}\BibitemShut {NoStop}%
\bibitem [{\citenamefont {Ozorio~de Almeida}(1989)}]{Ozorio89}%
  \BibitemOpen
  \bibfield  {author} {\bibinfo {author} {\bibfnamefont {A.~M.}\ \bibnamefont
  {Ozorio~de Almeida}},\ }\href@noop {} {\bibfield  {journal} {\bibinfo
  {journal} {Nonlinearity}\ }\textbf {\bibinfo {volume} {2}},\ \bibinfo {pages}
  {519} (\bibinfo {year} {1989})}\BibitemShut {NoStop}%
\bibitem [{\citenamefont {Li}\ and\ \citenamefont
  {Tomsovic}(2017{\natexlab{a}})}]{Li17a}%
  \BibitemOpen
  \bibfield  {author} {\bibinfo {author} {\bibfnamefont {J.}~\bibnamefont
  {Li}}\ and\ \bibinfo {author} {\bibfnamefont {S.}~\bibnamefont {Tomsovic}},\
  }\href@noop {} {\bibfield  {journal} {\bibinfo  {journal} {Phys.~Rev.~E}\
  }\textbf {\bibinfo {volume} {95}},\ \bibinfo {pages} {062224} (\bibinfo
  {year} {2017}{\natexlab{a}})},\ \bibinfo {note} {arXiv:1703.07045
  [nlin.CD]}\BibitemShut {NoStop}%
\bibitem [{\citenamefont {Li}\ and\ \citenamefont {Tomsovic}(2018)}]{Li18}%
  \BibitemOpen
  \bibfield  {author} {\bibinfo {author} {\bibfnamefont {J.}~\bibnamefont
  {Li}}\ and\ \bibinfo {author} {\bibfnamefont {S.}~\bibnamefont {Tomsovic}},\
  }\href@noop {} {\bibfield  {journal} {\bibinfo  {journal} {Phys.~Rev.~E}\
  }\textbf {\bibinfo {volume} {97}},\ \bibinfo {pages} {022216} (\bibinfo
  {year} {2018})},\ \bibinfo {note} {arXiv:1712.05568 [nlin.CD]}\BibitemShut
  {NoStop}%
\bibitem [{\citenamefont {MacKay}\ \emph {et~al.}(1984)\citenamefont {MacKay},
  \citenamefont {Meiss},\ and\ \citenamefont {Percival}}]{MacKay84a}%
  \BibitemOpen
  \bibfield  {author} {\bibinfo {author} {\bibfnamefont {R.~S.}\ \bibnamefont
  {MacKay}}, \bibinfo {author} {\bibfnamefont {J.~D.}\ \bibnamefont {Meiss}}, \
  and\ \bibinfo {author} {\bibfnamefont {I.~C.}\ \bibnamefont {Percival}},\
  }\href@noop {} {\bibfield  {journal} {\bibinfo  {journal} {Physica~D}\
  }\textbf {\bibinfo {volume} {13}},\ \bibinfo {pages} {55} (\bibinfo {year}
  {1984})}\BibitemShut {NoStop}%
\bibitem [{\citenamefont {Meiss}(1992)}]{Meiss92}%
  \BibitemOpen
  \bibfield  {author} {\bibinfo {author} {\bibfnamefont {J.~D.}\ \bibnamefont
  {Meiss}},\ }\href@noop {} {\bibfield  {journal} {\bibinfo  {journal}
  {Rev.~Mod.~Phys.}\ }\textbf {\bibinfo {volume} {64}},\ \bibinfo {pages} {795}
  (\bibinfo {year} {1992})}\BibitemShut {NoStop}%
\bibitem [{\citenamefont {Doedel}\ and\ \citenamefont
  {Friedman}(1989)}]{Doedel89}%
  \BibitemOpen
  \bibfield  {author} {\bibinfo {author} {\bibfnamefont {E.~J.}\ \bibnamefont
  {Doedel}}\ and\ \bibinfo {author} {\bibfnamefont {M.~J.}\ \bibnamefont
  {Friedman}},\ }\href@noop {} {\bibfield  {journal} {\bibinfo  {journal}
  {J.~Comput.~Appl.~Math.}\ }\textbf {\bibinfo {volume} {26}},\ \bibinfo
  {pages} {155} (\bibinfo {year} {1989})}\BibitemShut {NoStop}%
\bibitem [{\citenamefont {Beyn}(1990)}]{Beyn90}%
  \BibitemOpen
  \bibfield  {author} {\bibinfo {author} {\bibfnamefont {W.~J.}\ \bibnamefont
  {Beyn}},\ }\href@noop {} {\bibfield  {journal} {\bibinfo  {journal}
  {IMA~J.~Numer.~Anal.}\ }\textbf {\bibinfo {volume} {9}},\ \bibinfo {pages}
  {379} (\bibinfo {year} {1990})}\BibitemShut {NoStop}%
\bibitem [{\citenamefont {Moore}(1995)}]{Moore95b}%
  \BibitemOpen
  \bibfield  {author} {\bibinfo {author} {\bibfnamefont {G.}~\bibnamefont
  {Moore}},\ }\href@noop {} {\bibfield  {journal} {\bibinfo  {journal}
  {IMA~J.~Numer.~Anal.}\ }\textbf {\bibinfo {volume} {15}},\ \bibinfo {pages}
  {245} (\bibinfo {year} {1995})}\BibitemShut {NoStop}%
\bibitem [{\citenamefont {Li}\ and\ \citenamefont
  {Tomsovic}(2017{\natexlab{b}})}]{Li17}%
  \BibitemOpen
  \bibfield  {author} {\bibinfo {author} {\bibfnamefont {J.}~\bibnamefont
  {Li}}\ and\ \bibinfo {author} {\bibfnamefont {S.}~\bibnamefont {Tomsovic}},\
  }\href@noop {} {\bibfield  {journal} {\bibinfo  {journal} {J.~Phys.~A:
  Math.~Theor.}\ }\textbf {\bibinfo {volume} {50}},\ \bibinfo {pages} {135101}
  (\bibinfo {year} {2017}{\natexlab{b}})},\ \bibinfo {note} {arXiv:1507.06455
  [nlin.CD]}\BibitemShut {NoStop}%
\bibitem [{\citenamefont {Brudno}(1978)}]{Brudno78}%
  \BibitemOpen
  \bibfield  {author} {\bibinfo {author} {\bibfnamefont {A.~A.}\ \bibnamefont
  {Brudno}},\ }\href@noop {} {\bibfield  {journal} {\bibinfo  {journal}
  {Russ.~Math.~Surv.}\ }\textbf {\bibinfo {volume} {33}},\ \bibinfo {pages}
  {197} (\bibinfo {year} {1978})}\BibitemShut {NoStop}%
\bibitem [{\citenamefont {Alekseev}\ and\ \citenamefont
  {Yakobson}(1981)}]{Alekseev81}%
  \BibitemOpen
  \bibfield  {author} {\bibinfo {author} {\bibfnamefont {V.~M.}\ \bibnamefont
  {Alekseev}}\ and\ \bibinfo {author} {\bibfnamefont {M.~V.}\ \bibnamefont
  {Yakobson}},\ }\href@noop {} {\bibfield  {journal} {\bibinfo  {journal}
  {Phys.~Rep.}\ }\textbf {\bibinfo {volume} {75}},\ \bibinfo {pages} {287}
  (\bibinfo {year} {1981})}\BibitemShut {NoStop}%
\bibitem [{\citenamefont {Kolmogorov}(1958)}]{Kolmogorov58}%
  \BibitemOpen
  \bibfield  {author} {\bibinfo {author} {\bibfnamefont {A.~N.}\ \bibnamefont
  {Kolmogorov}},\ }\href@noop {} {\bibfield  {journal} {\bibinfo  {journal}
  {Doklady of Russian Academy of Sciences}\ }\textbf {\bibinfo {volume}
  {119}},\ \bibinfo {pages} {861} (\bibinfo {year} {1958})}\BibitemShut
  {NoStop}%
\bibitem [{\citenamefont {Kolmogorov}(1959)}]{Kolmogorov59}%
  \BibitemOpen
  \bibfield  {author} {\bibinfo {author} {\bibfnamefont {A.~N.}\ \bibnamefont
  {Kolmogorov}},\ }\href@noop {} {\bibfield  {journal} {\bibinfo  {journal}
  {Doklady of Russian Academy of Sciences}\ }\textbf {\bibinfo {volume}
  {124}},\ \bibinfo {pages} {754} (\bibinfo {year} {1959})}\BibitemShut
  {NoStop}%
\bibitem [{\citenamefont {Sinai}(1959)}]{Sinai59}%
  \BibitemOpen
  \bibfield  {author} {\bibinfo {author} {\bibfnamefont {Y.~G.}\ \bibnamefont
  {Sinai}},\ }\href@noop {} {\bibfield  {journal} {\bibinfo  {journal} {Doklady
  of Russian Academy of Sciences}\ }\textbf {\bibinfo {volume} {124}},\
  \bibinfo {pages} {768} (\bibinfo {year} {1959})}\BibitemShut {NoStop}%
\bibitem [{\citenamefont {Pesin}(1977)}]{Pesin77}%
  \BibitemOpen
  \bibfield  {author} {\bibinfo {author} {\bibfnamefont {Y.~B.}\ \bibnamefont
  {Pesin}},\ }\href@noop {} {\bibfield  {journal} {\bibinfo  {journal}
  {Russ.~Math.~Surv.}\ }\textbf {\bibinfo {volume} {32}},\ \bibinfo {pages}
  {55} (\bibinfo {year} {1977})}\BibitemShut {NoStop}%
\bibitem [{\citenamefont {Gaspard}\ and\ \citenamefont
  {Nicolis}(1990)}]{Gaspard90}%
  \BibitemOpen
  \bibfield  {author} {\bibinfo {author} {\bibfnamefont {P.}~\bibnamefont
  {Gaspard}}\ and\ \bibinfo {author} {\bibfnamefont {G.}~\bibnamefont
  {Nicolis}},\ }\href@noop {} {\bibfield  {journal} {\bibinfo  {journal}
  {Phys.~Rev.~Lett.}\ }\textbf {\bibinfo {volume} {65}},\ \bibinfo {pages}
  {1693} (\bibinfo {year} {1990})}\BibitemShut {NoStop}%
\bibitem [{\citenamefont {Connes}\ \emph {et~al.}(1987)\citenamefont {Connes},
  \citenamefont {Narnhofer},\ and\ \citenamefont {Thirring}}]{Connes87}%
  \BibitemOpen
  \bibfield  {author} {\bibinfo {author} {\bibfnamefont {A.}~\bibnamefont
  {Connes}}, \bibinfo {author} {\bibfnamefont {H.}~\bibnamefont {Narnhofer}}, \
  and\ \bibinfo {author} {\bibfnamefont {W.}~\bibnamefont {Thirring}},\
  }\href@noop {} {\bibfield  {journal} {\bibinfo  {journal}
  {Commun.~Math.~Phys.}\ }\textbf {\bibinfo {volume} {112}},\ \bibinfo {pages}
  {691} (\bibinfo {year} {1987})}\BibitemShut {NoStop}%
\bibitem [{\citenamefont {Alicki}\ and\ \citenamefont
  {Fannes}(1994)}]{Alicki94}%
  \BibitemOpen
  \bibfield  {author} {\bibinfo {author} {\bibfnamefont {R.}~\bibnamefont
  {Alicki}}\ and\ \bibinfo {author} {\bibfnamefont {M.}~\bibnamefont
  {Fannes}},\ }\href@noop {} {\bibfield  {journal} {\bibinfo  {journal}
  {Math.~Phys.}\ }\textbf {\bibinfo {volume} {32}},\ \bibinfo {pages} {75}
  (\bibinfo {year} {1994})}\BibitemShut {NoStop}%
\bibitem [{\citenamefont {Lindblad}(1988)}]{Lindblad88}%
  \BibitemOpen
  \bibfield  {author} {\bibinfo {author} {\bibfnamefont {G.}~\bibnamefont
  {Lindblad}},\ }in\ \href@noop {} {\emph {\bibinfo {booktitle} {Quantum
  Probability and Applications}}},\ \bibinfo {editor} {edited by\ \bibinfo
  {editor} {\bibfnamefont {L.}~\bibnamefont {Accardi}}\ and\ \bibinfo {editor}
  {\bibfnamefont {W.}~\bibnamefont {von Waldenfels}}}\ (\bibinfo  {publisher}
  {Springer},\ \bibinfo {address} {Berlin},\ \bibinfo {year} {1988})\ pp.\
  \bibinfo {pages} {183--191},\ \bibinfo {note} {vol.~III}\BibitemShut
  {NoStop}%
\bibitem [{\citenamefont {Cvitanovi\'{c}}(1992)}]{Cvitanovic92}%
  \BibitemOpen
  \bibfield  {author} {\bibinfo {author} {\bibfnamefont {P.}~\bibnamefont
  {Cvitanovi\'{c}}},\ }\href@noop {} {\bibfield  {journal} {\bibinfo  {journal}
  {Chaos}\ }\textbf {\bibinfo {volume} {2}},\ \bibinfo {pages} {1} (\bibinfo
  {year} {1992})}\BibitemShut {NoStop}%
\bibitem [{\citenamefont {Cvitanovi\'{c}}(1988)}]{Cvitanovic88}%
  \BibitemOpen
  \bibfield  {author} {\bibinfo {author} {\bibfnamefont {P.}~\bibnamefont
  {Cvitanovi\'{c}}},\ }\href@noop {} {\bibfield  {journal} {\bibinfo  {journal}
  {Phys.~Rev.~Lett.}\ }\textbf {\bibinfo {volume} {61}},\ \bibinfo {pages}
  {2729} (\bibinfo {year} {1988})}\BibitemShut {NoStop}%
\bibitem [{\citenamefont {Cvitanovi\'{c}}\ and\ \citenamefont
  {Eckhardt}(1989)}]{Cvitanovic89}%
  \BibitemOpen
  \bibfield  {author} {\bibinfo {author} {\bibfnamefont {P.}~\bibnamefont
  {Cvitanovi\'{c}}}\ and\ \bibinfo {author} {\bibfnamefont {B.}~\bibnamefont
  {Eckhardt}},\ }\href@noop {} {\bibfield  {journal} {\bibinfo  {journal}
  {Phys.~Rev.~Lett.}\ }\textbf {\bibinfo {volume} {63}},\ \bibinfo {pages}
  {823} (\bibinfo {year} {1989})}\BibitemShut {NoStop}%
\bibitem [{\citenamefont {Bogomolny}(1992)}]{Bogomolny92}%
  \BibitemOpen
  \bibfield  {author} {\bibinfo {author} {\bibfnamefont {E.~B.}\ \bibnamefont
  {Bogomolny}},\ }\href@noop {} {\bibfield  {journal} {\bibinfo  {journal}
  {Chaos}\ }\textbf {\bibinfo {volume} {2}},\ \bibinfo {pages} {5} (\bibinfo
  {year} {1992})}\BibitemShut {NoStop}%
\bibitem [{\citenamefont {Kaplan}(1998)}]{Kaplan98b}%
  \BibitemOpen
  \bibfield  {author} {\bibinfo {author} {\bibfnamefont {L.}~\bibnamefont
  {Kaplan}},\ }\href@noop {} {\bibfield  {journal} {\bibinfo  {journal}
  {Phys.~Rev.~Lett.}\ }\textbf {\bibinfo {volume} {81}},\ \bibinfo {pages}
  {3371} (\bibinfo {year} {1998})}\BibitemShut {NoStop}%
\bibitem [{\citenamefont {Poincar\'e}(1899)}]{Poincare99}%
  \BibitemOpen
  \bibfield  {author} {\bibinfo {author} {\bibfnamefont {H.}~\bibnamefont
  {Poincar\'e}},\ }\href@noop {} {\emph {\bibinfo {title} {Les m\'ethodes
  nouvelles de la m\'ecanique c\'eleste}}},\ Vol.~\bibinfo {volume} {3}\
  (\bibinfo  {publisher} {Gauthier-Villars et fils},\ \bibinfo {address}
  {Paris},\ \bibinfo {year} {1899})\BibitemShut {NoStop}%
\bibitem [{\citenamefont {Easton}(1986)}]{Easton86}%
  \BibitemOpen
  \bibfield  {author} {\bibinfo {author} {\bibfnamefont {R.~W.}\ \bibnamefont
  {Easton}},\ }\href@noop {} {\bibfield  {journal} {\bibinfo  {journal}
  {Trans.~Am.~Math.~Soc.}\ }\textbf {\bibinfo {volume} {294}},\ \bibinfo
  {pages} {719} (\bibinfo {year} {1986})}\BibitemShut {NoStop}%
\bibitem [{\citenamefont {Rom-Kedar}(1990)}]{Rom-Kedar90}%
  \BibitemOpen
  \bibfield  {author} {\bibinfo {author} {\bibfnamefont {V.}~\bibnamefont
  {Rom-Kedar}},\ }\href@noop {} {\bibfield  {journal} {\bibinfo  {journal}
  {Physica~D}\ }\textbf {\bibinfo {volume} {43}},\ \bibinfo {pages} {229}
  (\bibinfo {year} {1990})}\BibitemShut {NoStop}%
\bibitem [{\citenamefont {Smale}(1963)}]{Smale63}%
  \BibitemOpen
  \bibfield  {author} {\bibinfo {author} {\bibfnamefont {S.}~\bibnamefont
  {Smale}},\ }\href@noop {} {\emph {\bibinfo {title} {Differential and
  Combinatorial Topology}}},\ edited by\ \bibinfo {editor} {\bibfnamefont
  {S.~S.}\ \bibnamefont {Cairns}}\ (\bibinfo  {publisher} {Princeton University
  Press},\ \bibinfo {address} {Princeton},\ \bibinfo {year} {1963})\BibitemShut
  {NoStop}%
\bibitem [{\citenamefont {Smale}(1980)}]{Smale80}%
  \BibitemOpen
  \bibfield  {author} {\bibinfo {author} {\bibfnamefont {S.}~\bibnamefont
  {Smale}},\ }\href@noop {} {\emph {\bibinfo {title} {The Mathematics of Time:
  Essays on Dynamical Systems, Economic Processes and Related Topics}}}\
  (\bibinfo  {publisher} {Springer-Verlag},\ \bibinfo {address} {New York,
  Heidelberg, Berlin},\ \bibinfo {year} {1980})\BibitemShut {NoStop}%
\bibitem [{\citenamefont {H\'enon}(1976)}]{Henon76}%
  \BibitemOpen
  \bibfield  {author} {\bibinfo {author} {\bibfnamefont {M.}~\bibnamefont
  {H\'enon}},\ }\href@noop {} {\bibfield  {journal} {\bibinfo  {journal}
  {Comm.~Math.~Phys.}\ }\textbf {\bibinfo {volume} {50}},\ \bibinfo {pages}
  {69} (\bibinfo {year} {1976})}\BibitemShut {NoStop}%
\bibitem [{\citenamefont {Hadamard}(1898)}]{Hadamard1898}%
  \BibitemOpen
  \bibfield  {author} {\bibinfo {author} {\bibfnamefont {J.}~\bibnamefont
  {Hadamard}},\ }\href@noop {} {\bibfield  {journal} {\bibinfo  {journal}
  {J.~Math.~Pures~Appl.~series 5}\ }\textbf {\bibinfo {volume} {4}},\ \bibinfo
  {pages} {27} (\bibinfo {year} {1898})}\BibitemShut {NoStop}%
\bibitem [{\citenamefont {Birkhoff}(1927{\natexlab{b}})}]{Birkhoff27a}%
  \BibitemOpen
  \bibfield  {author} {\bibinfo {author} {\bibfnamefont {G.~D.}\ \bibnamefont
  {Birkhoff}},\ }\href@noop {} {\emph {\bibinfo {title} {A.M.S. Coll.
  Publications, vol. 9}}}\ (\bibinfo  {publisher} {American Mathematical
  Society},\ \bibinfo {address} {Providence},\ \bibinfo {year}
  {1927})\BibitemShut {NoStop}%
\bibitem [{\citenamefont {Birkhoff}(1935)}]{Birkhoff35}%
  \BibitemOpen
  \bibfield  {author} {\bibinfo {author} {\bibfnamefont {G.~D.}\ \bibnamefont
  {Birkhoff}},\ }\href@noop {} {\bibfield  {journal} {\bibinfo  {journal}
  {Mem.~Pont.~Acad.~Sci.~Novi~Lyncaei}\ }\textbf {\bibinfo {volume} {1}},\
  \bibinfo {pages} {85} (\bibinfo {year} {1935})}\BibitemShut {NoStop}%
\bibitem [{\citenamefont {Morse}\ and\ \citenamefont
  {Hedlund}(1938)}]{Morse38}%
  \BibitemOpen
  \bibfield  {author} {\bibinfo {author} {\bibfnamefont {M.}~\bibnamefont
  {Morse}}\ and\ \bibinfo {author} {\bibfnamefont {G.~A.}\ \bibnamefont
  {Hedlund}},\ }\href@noop {} {\bibfield  {journal} {\bibinfo  {journal}
  {Amer.~J.~Math.}\ }\textbf {\bibinfo {volume} {60}},\ \bibinfo {pages} {815}
  (\bibinfo {year} {1938})}\BibitemShut {NoStop}%
\bibitem [{\citenamefont {Wiggins}(1992)}]{Wiggins92}%
  \BibitemOpen
  \bibfield  {author} {\bibinfo {author} {\bibfnamefont {S.}~\bibnamefont
  {Wiggins}},\ }\href@noop {} {\emph {\bibinfo {title} {Chaotic Transport in
  Dynamical Systems}}}\ (\bibinfo  {publisher} {Springer},\ \bibinfo {address}
  {New York},\ \bibinfo {year} {1992})\BibitemShut {NoStop}%
\bibitem [{\citenamefont {Sterling}\ \emph {et~al.}(1999)\citenamefont
  {Sterling}, \citenamefont {Dullin},\ and\ \citenamefont
  {Meiss}}]{Sterling99}%
  \BibitemOpen
  \bibfield  {author} {\bibinfo {author} {\bibfnamefont {D.}~\bibnamefont
  {Sterling}}, \bibinfo {author} {\bibfnamefont {H.~R.}\ \bibnamefont
  {Dullin}}, \ and\ \bibinfo {author} {\bibfnamefont {J.~D.}\ \bibnamefont
  {Meiss}},\ }\href@noop {} {\bibfield  {journal} {\bibinfo  {journal}
  {Physica~D}\ }\textbf {\bibinfo {volume} {134}},\ \bibinfo {pages} {153}
  (\bibinfo {year} {1999})}\BibitemShut {NoStop}%
\bibitem [{\citenamefont {Tabacman}(1995)}]{Tabacman95}%
  \BibitemOpen
  \bibfield  {author} {\bibinfo {author} {\bibfnamefont {E.}~\bibnamefont
  {Tabacman}},\ }\href@noop {} {\bibfield  {journal} {\bibinfo  {journal}
  {Physica~D}\ }\textbf {\bibinfo {volume} {85}},\ \bibinfo {pages} {548}
  (\bibinfo {year} {1995})}\BibitemShut {NoStop}%
\bibitem [{\citenamefont {Hockett}\ and\ \citenamefont
  {Holmes}(1986)}]{Hockett86}%
  \BibitemOpen
  \bibfield  {author} {\bibinfo {author} {\bibfnamefont {K.}~\bibnamefont
  {Hockett}}\ and\ \bibinfo {author} {\bibfnamefont {P.}~\bibnamefont
  {Holmes}},\ }\href@noop {} {\bibfield  {journal} {\bibinfo  {journal}
  {Ergod.~Th.~\&~Dynam.~Sys.}\ }\textbf {\bibinfo {volume} {6}},\ \bibinfo
  {pages} {205} (\bibinfo {year} {1986})}\BibitemShut {NoStop}%
\bibitem [{\citenamefont {Bevilaqua}\ and\ \citenamefont {Bas\'{i}lio~de
  Matos}(2000)}]{Bevilaqua00}%
  \BibitemOpen
  \bibfield  {author} {\bibinfo {author} {\bibfnamefont {D.}~\bibnamefont
  {Bevilaqua}}\ and\ \bibinfo {author} {\bibfnamefont {M.}~\bibnamefont
  {Bas\'{i}lio~de Matos}},\ }\href@noop {} {\bibfield  {journal} {\bibinfo
  {journal} {Physica}\ }\textbf {\bibinfo {volume} {D 145}},\ \bibinfo {pages}
  {13} (\bibinfo {year} {2000})}\BibitemShut {NoStop}%
\bibitem [{\citenamefont {Mitchell}\ \emph
  {et~al.}(2003{\natexlab{a}})\citenamefont {Mitchell}, \citenamefont
  {Handley}, \citenamefont {Tighe}, \citenamefont {Delos},\ and\ \citenamefont
  {Knudson}}]{Mitchell03a}%
  \BibitemOpen
  \bibfield  {author} {\bibinfo {author} {\bibfnamefont {K.~A.}\ \bibnamefont
  {Mitchell}}, \bibinfo {author} {\bibfnamefont {J.~P.}\ \bibnamefont
  {Handley}}, \bibinfo {author} {\bibfnamefont {B.}~\bibnamefont {Tighe}},
  \bibinfo {author} {\bibfnamefont {J.~B.}\ \bibnamefont {Delos}}, \ and\
  \bibinfo {author} {\bibfnamefont {S.~K.}\ \bibnamefont {Knudson}},\
  }\href@noop {} {\bibfield  {journal} {\bibinfo  {journal} {Chaos}\ }\textbf
  {\bibinfo {volume} {13}},\ \bibinfo {pages} {880} (\bibinfo {year}
  {2003}{\natexlab{a}})}\BibitemShut {NoStop}%
\bibitem [{\citenamefont {Li}\ and\ \citenamefont {Tomsovic}(2019)}]{Li19b}%
  \BibitemOpen
  \bibfield  {author} {\bibinfo {author} {\bibfnamefont {J.}~\bibnamefont
  {Li}}\ and\ \bibinfo {author} {\bibfnamefont {S.}~\bibnamefont {Tomsovic}},\
  }\href@noop {} {\  (\bibinfo {year} {2019})}\BibitemShut {NoStop}%
\bibitem [{\citenamefont {Cvitanovi\'{c}}\ \emph {et~al.}(1988)\citenamefont
  {Cvitanovi\'{c}}, \citenamefont {Gunaratne},\ and\ \citenamefont
  {Procaccia}}]{Cvitanovic88a}%
  \BibitemOpen
  \bibfield  {author} {\bibinfo {author} {\bibfnamefont {P.}~\bibnamefont
  {Cvitanovi\'{c}}}, \bibinfo {author} {\bibfnamefont {G.}~\bibnamefont
  {Gunaratne}}, \ and\ \bibinfo {author} {\bibfnamefont {I.}~\bibnamefont
  {Procaccia}},\ }\href@noop {} {\bibfield  {journal} {\bibinfo  {journal}
  {Phys.~Rev.~A}\ }\textbf {\bibinfo {volume} {38}},\ \bibinfo {pages} {1503}
  (\bibinfo {year} {1988})}\BibitemShut {NoStop}%
\bibitem [{\citenamefont {Cvitanovi\'{c}}(1991)}]{Cvitanovic91}%
  \BibitemOpen
  \bibfield  {author} {\bibinfo {author} {\bibfnamefont {P.}~\bibnamefont
  {Cvitanovi\'{c}}},\ }\href@noop {} {\bibfield  {journal} {\bibinfo  {journal}
  {Physica~D}\ }\textbf {\bibinfo {volume} {51}},\ \bibinfo {pages} {138}
  (\bibinfo {year} {1991})}\BibitemShut {NoStop}%
\bibitem [{\citenamefont {Hagiwara}\ and\ \citenamefont
  {Shudo}(2004)}]{Hagiwara04}%
  \BibitemOpen
  \bibfield  {author} {\bibinfo {author} {\bibfnamefont {R.}~\bibnamefont
  {Hagiwara}}\ and\ \bibinfo {author} {\bibfnamefont {A.}~\bibnamefont
  {Shudo}},\ }\href@noop {} {\bibfield  {journal} {\bibinfo  {journal}
  {J.~Phys.~A: Math.~Gen.}\ }\textbf {\bibinfo {volume} {37}},\ \bibinfo
  {pages} {10521–10543} (\bibinfo {year} {2004})}\BibitemShut {NoStop}%
\bibitem [{\citenamefont {Mitchell}\ \emph
  {et~al.}(2003{\natexlab{b}})\citenamefont {Mitchell}, \citenamefont
  {Handley}, \citenamefont {Delos},\ and\ \citenamefont
  {Knudson}}]{Mitchell03b}%
  \BibitemOpen
  \bibfield  {author} {\bibinfo {author} {\bibfnamefont {K.~A.}\ \bibnamefont
  {Mitchell}}, \bibinfo {author} {\bibfnamefont {J.~P.}\ \bibnamefont
  {Handley}}, \bibinfo {author} {\bibfnamefont {J.~B.}\ \bibnamefont {Delos}},
  \ and\ \bibinfo {author} {\bibfnamefont {S.~K.}\ \bibnamefont {Knudson}},\
  }\href@noop {} {\bibfield  {journal} {\bibinfo  {journal} {Chaos}\ }\textbf
  {\bibinfo {volume} {13}},\ \bibinfo {pages} {892} (\bibinfo {year}
  {2003}{\natexlab{b}})}\BibitemShut {NoStop}%
\bibitem [{\citenamefont {Mitchell}\ and\ \citenamefont
  {Delos}(2006)}]{Mitchell06}%
  \BibitemOpen
  \bibfield  {author} {\bibinfo {author} {\bibfnamefont {K.~A.}\ \bibnamefont
  {Mitchell}}\ and\ \bibinfo {author} {\bibfnamefont {J.~B.}\ \bibnamefont
  {Delos}},\ }\href@noop {} {\bibfield  {journal} {\bibinfo  {journal}
  {Physica~D}\ }\textbf {\bibinfo {volume} {221}},\ \bibinfo {pages} {170}
  (\bibinfo {year} {2006})}\BibitemShut {NoStop}%
\bibitem [{\citenamefont {Novick}\ \emph {et~al.}(2012)\citenamefont {Novick},
  \citenamefont {Keeler}, \citenamefont {Giefer},\ and\ \citenamefont
  {Delos}}]{Novick12a}%
  \BibitemOpen
  \bibfield  {author} {\bibinfo {author} {\bibfnamefont {J.}~\bibnamefont
  {Novick}}, \bibinfo {author} {\bibfnamefont {M.~L.}\ \bibnamefont {Keeler}},
  \bibinfo {author} {\bibfnamefont {J.}~\bibnamefont {Giefer}}, \ and\ \bibinfo
  {author} {\bibfnamefont {J.~B.}\ \bibnamefont {Delos}},\ }\href@noop {}
  {\bibfield  {journal} {\bibinfo  {journal} {Phys.~Rev.~E}\ }\textbf {\bibinfo
  {volume} {85}},\ \bibinfo {pages} {016205} (\bibinfo {year}
  {2012})}\BibitemShut {NoStop}%
\bibitem [{\citenamefont {Novick}\ and\ \citenamefont
  {Delos}(2012)}]{Novick12b}%
  \BibitemOpen
  \bibfield  {author} {\bibinfo {author} {\bibfnamefont {J.}~\bibnamefont
  {Novick}}\ and\ \bibinfo {author} {\bibfnamefont {J.~B.}\ \bibnamefont
  {Delos}},\ }\href@noop {} {\bibfield  {journal} {\bibinfo  {journal}
  {Phys.~Rev.~E}\ }\textbf {\bibinfo {volume} {85}},\ \bibinfo {pages} {016206}
  (\bibinfo {year} {2012})}\BibitemShut {NoStop}%
\bibitem [{\citenamefont {Bowen}(1975)}]{Bowen75}%
  \BibitemOpen
  \bibfield  {author} {\bibinfo {author} {\bibfnamefont {R.}~\bibnamefont
  {Bowen}},\ }\href@noop {} {\emph {\bibinfo {title} {Lect. Notes in Math. Vol.
  470.}}}\ (\bibinfo  {publisher} {Springer-Verlag},\ \bibinfo {address}
  {Berlin},\ \bibinfo {year} {1975})\BibitemShut {NoStop}%
\bibitem [{\citenamefont {Gaspard}(1998)}]{Gaspard98}%
  \BibitemOpen
  \bibfield  {author} {\bibinfo {author} {\bibfnamefont {P.}~\bibnamefont
  {Gaspard}},\ }\href@noop {} {\emph {\bibinfo {title} {Chaos, Scattering and
  Statistical Mechanics}}}\ (\bibinfo  {publisher} {Cambridge University
  Press},\ \bibinfo {address} {Cambridge, UK},\ \bibinfo {year}
  {1998})\BibitemShut {NoStop}%
\bibitem [{\citenamefont {Grassberger}\ and\ \citenamefont
  {Kantz}(1985)}]{Grassberger85a}%
  \BibitemOpen
  \bibfield  {author} {\bibinfo {author} {\bibfnamefont {P.}~\bibnamefont
  {Grassberger}}\ and\ \bibinfo {author} {\bibfnamefont {H.}~\bibnamefont
  {Kantz}},\ }\href@noop {} {\bibfield  {journal} {\bibinfo  {journal}
  {Phys.~Lett.}\ }\textbf {\bibinfo {volume} {113A}},\ \bibinfo {pages} {235}
  (\bibinfo {year} {1985})}\BibitemShut {NoStop}%
\bibitem [{\citenamefont {Christiansen}\ and\ \citenamefont
  {Politi}(1995)}]{Christiansen95}%
  \BibitemOpen
  \bibfield  {author} {\bibinfo {author} {\bibfnamefont {F.}~\bibnamefont
  {Christiansen}}\ and\ \bibinfo {author} {\bibfnamefont {A.}~\bibnamefont
  {Politi}},\ }\href@noop {} {\bibfield  {journal} {\bibinfo  {journal}
  {Phys.~Rev.~E}\ }\textbf {\bibinfo {volume} {51}},\ \bibinfo {pages} {R3811}
  (\bibinfo {year} {1995})}\BibitemShut {NoStop}%
\bibitem [{\citenamefont {Christiansen}\ and\ \citenamefont
  {Politi}(1996)}]{Christiansen96}%
  \BibitemOpen
  \bibfield  {author} {\bibinfo {author} {\bibfnamefont {F.}~\bibnamefont
  {Christiansen}}\ and\ \bibinfo {author} {\bibfnamefont {A.}~\bibnamefont
  {Politi}},\ }\href@noop {} {\bibfield  {journal} {\bibinfo  {journal}
  {Nonlinearity}\ }\textbf {\bibinfo {volume} {9}},\ \bibinfo {pages} {1623}
  (\bibinfo {year} {1996})}\BibitemShut {NoStop}%
\bibitem [{\citenamefont {Christiansen}\ and\ \citenamefont
  {Politi}(1997)}]{Christiansen97}%
  \BibitemOpen
  \bibfield  {author} {\bibinfo {author} {\bibfnamefont {F.}~\bibnamefont
  {Christiansen}}\ and\ \bibinfo {author} {\bibfnamefont {A.}~\bibnamefont
  {Politi}},\ }\href@noop {} {\bibfield  {journal} {\bibinfo  {journal}
  {Physica~D}\ }\textbf {\bibinfo {volume} {109}},\ \bibinfo {pages} {32}
  (\bibinfo {year} {1997})}\BibitemShut {NoStop}%
\bibitem [{\citenamefont {Rubido}\ \emph {et~al.}(2018)\citenamefont {Rubido},
  \citenamefont {Grebogi},\ and\ \citenamefont {Baptista}}]{Rubido18}%
  \BibitemOpen
  \bibfield  {author} {\bibinfo {author} {\bibfnamefont {N.}~\bibnamefont
  {Rubido}}, \bibinfo {author} {\bibfnamefont {C.}~\bibnamefont {Grebogi}}, \
  and\ \bibinfo {author} {\bibfnamefont {M.~S.}\ \bibnamefont {Baptista}},\
  }\href@noop {} {\bibfield  {journal} {\bibinfo  {journal} {Chaos}\ }\textbf
  {\bibinfo {volume} {28}},\ \bibinfo {pages} {033611} (\bibinfo {year}
  {2018})}\BibitemShut {NoStop}%
\bibitem [{\citenamefont {Aubry}\ and\ \citenamefont
  {Abramovici}(1990)}]{Aubry90}%
  \BibitemOpen
  \bibfield  {author} {\bibinfo {author} {\bibfnamefont {S.}~\bibnamefont
  {Aubry}}\ and\ \bibinfo {author} {\bibfnamefont {G.}~\bibnamefont
  {Abramovici}},\ }\href@noop {} {\bibfield  {journal} {\bibinfo  {journal}
  {Physica~D}\ }\textbf {\bibinfo {volume} {43}},\ \bibinfo {pages} {199}
  (\bibinfo {year} {1990})}\BibitemShut {NoStop}%
\bibitem [{\citenamefont {Aubry}(1995)}]{Aubry95}%
  \BibitemOpen
  \bibfield  {author} {\bibinfo {author} {\bibfnamefont {S.}~\bibnamefont
  {Aubry}},\ }\href@noop {} {\bibfield  {journal} {\bibinfo  {journal}
  {Physica~D}\ }\textbf {\bibinfo {volume} {86}},\ \bibinfo {pages} {284}
  (\bibinfo {year} {1995})}\BibitemShut {NoStop}%
\end{thebibliography}%

\end{document}